\documentclass[astrosymb,tighten,twocolumn]{aastex631}

\usepackage{amsmath}
\usepackage{graphicx}
\usepackage{longtable}
\usepackage{bm}
\usepackage{romannum}

\newcommand{\mainred}{\texttt{MAIN-RED}}
\newcommand{\mainblue}{\texttt{MAIN-BLUE}}
\newcommand{\mainbroad}{\texttt{MAIN-BROAD}}
\newcommand{\mwsmain}{\texttt{MWS-MAIN}}
\DeclareRobustCommand{\msol}{\mathrm{M_{\sun}}}
\DeclareRobustCommand{\rsol}{\mathrm{R_{\sun}}}

\DeclareRobustCommand{\zsol}{\mathrm{Z_{\sun}}}

\DeclareRobustCommand{\kms}{\mathrm{km\,s^{-1}}}

\newcommand{\hdbscan}{\texttt{HDBSCAN}}

\usepackage[T1]{fontenc}

\begin{document}

\title{Disentangling the Distant Stellar Halo Using K-Giants in the DESI Year 3 Data}

\author[0000-0003-0853-8887]{Namitha Kizhuprakkat}
\affiliation{Institute of Astronomy and Department of Physics, National Tsing Hua University, Hsinchu 30013, Taiwan}
\affiliation{Center for Informatics and Computation in Astronomy, National Tsing Hua University, Hsinchu 30013, Taiwan}
\email{namitha96@gapp.nthu.edu.tw}

\author[0000-0001-8274-158X]{Andrew P. Cooper}
\affiliation{Institute of Astronomy and Department of Physics, National Tsing Hua University, Hsinchu 30013, Taiwan}
\affiliation{Center for Informatics and Computation in Astronomy, National Tsing Hua University, Hsinchu 30013, Taiwan}
\email{apcooper@gapp.nthu.edu.tw}
\author[0000-0002-5762-7571]{Wenting Wang}
\affiliation{Department of Astronomy, School of Physics and Astronomy, Shanghai Jiao Tong University, Shanghai 200240, China}
\author[0000-0002-5689-8791]{Amanda~Bystr\"om}
\affiliation{Institute for Astronomy, University of Edinburgh, Royal Observatory, Blackford Hill, Edinburgh EH9 3HJ, UK}
\author[0000-0002-5758-150X]{Joan~Najita}
\affiliation{NSF NOIRLab, 950 N. Cherry Ave., Tucson, AZ 85719, USA}
\author[0000-0003-2644-135X]{Sergey~E.~Koposov}
\affiliation{Institute for Astronomy, University of Edinburgh, Royal Observatory, Blackford Hill, Edinburgh EH9 3HJ, UK}
\affiliation{Institute of Astronomy, University of Cambridge, Madingley Road, Cambridge CB3 0HA, UK}
\author{Guillaume~Thomas}
\affiliation{Instituto de Astrof\'{\i}sica de Canarias, C/ V\'{\i}a L\'{a}ctea, s/n, E-38205 La Laguna, Tenerife, Spain}
\author[0000-0002-6469-8263]{Songting Li}
\affiliation{Department of Astronomy, School of Physics and Astronomy, Shanghai Jiao Tong University, Shanghai 200240, China}
\author[0009-0008-1319-1084]{Ruizhi Zhang}
\affiliation{Instituto de Astrof\'{\i}sica de Canarias, C/ V\'{\i}a L\'{a}ctea, s/n, E-38205 La Laguna, Tenerife, Spain}
\affiliation{Departamento de Astrof\'{\i}sica, Universidad de La Laguna (ULL), E-38206, La Laguna, Tenerife, Spain}
\author[0000-0002-0084-572X]{Carlos~Allende~Prieto}
\affiliation{Instituto de Astrof\'{\i}sica de Canarias, C/ V\'{\i}a L\'{a}ctea, s/n, E-38205 La Laguna, Tenerife, Spain}
\affiliation{Departamento de Astrof\'{\i}sica, Universidad de La Laguna (ULL), E-38206, La Laguna, Tenerife, Spain}
\author{Mika~Lambert}
\affiliation{Department of Astronomy and Astrophysics, UCO/Lick Observatory, University of California, 1156 High Street, Santa Cruz, CA 95064, USA}
\author[0000-0001-9852-9954]{Oleg~Y.~Gnedin}
\affiliation{Department of Astronomy, University of Michigan, Ann Arbor, MI 48109, USA}
\affiliation{University of Michigan, 500 S. State Street, Ann Arbor, MI 48109, USA}
\author[0000-0001-5805-5766]{Alexander~H.~Riley}
\affiliation{Lund Observatory, Division of Astrophysics, Department of Physics, Lund University, SE-221 00 Lund, Sweden}
\author{Jessica N.~Aguilar}
\affiliation{Lawrence Berkeley National Laboratory, 1 Cyclotron Road, Berkeley, CA 94720, USA}
\author[0000-0001-6098-7247]{Steven~Ahlen}
\affiliation{Department of Physics, Boston University, 590 Commonwealth Avenue, Boston, MA 02215 USA}
\author[0000-0001-9712-0006]{Davide~Bianchi}
\affiliation{Dipartimento di Fisica ``Aldo Pontremoli'', Universit\`a degli Studi di Milano, Via Celoria 16, I-20133 Milano, Italy}
\affiliation{Dipartimento di Fisica ``Aldo Pontremoli'', Universit\`a degli Studi di Milano, Via Celoria 16, I-20133 Milano, Italy}
\author{David~Brooks}
\affiliation{Department of Physics \& Astronomy, University College London, Gower Street, London, WC1E 6BT, UK}
\author{Todd~Claybaugh}
\affiliation{Lawrence Berkeley National Laboratory, 1 Cyclotron Road, Berkeley, CA 94720, USA}
\author[0000-0002-1769-1640]{Axel~de la Macorra}
\affiliation{Instituto de F\'{\i}sica, Universidad Nacional Aut\'{o}noma de M\'{e}xico, Circuito de la Investigaci\'{o}n Cient\'{\i}fica, Ciudad Universitaria, Cd. de M\'{e}xico C.~P.~04510, M\'{e}xico}
\author{Peter~Doel}
\affiliation{Department of Physics \& Astronomy, University College London, Gower Street, London, WC1E 6BT, UK}
\author[0000-0001-9632-0815]{Enrique~Gaztañaga}
\affiliation{Institut d'Estudis Espacials de Catalunya (IEEC), c/ Esteve Terradas 1, Edifici RDIT, Campus PMT-UPC, 08860 Castelldefels, Spain}
\affiliation{Institute of Cosmology and Gravitation, University of Portsmouth, Dennis Sciama Building, Portsmouth, PO1 3FX, UK}
\affiliation{Institute of Space Sciences, ICE-CSIC, Campus UAB, Carrer de Can Magrans s/n, 08913 Bellaterra, Barcelona, Spain}
\author[0000-0003-3142-233X]{Satya~Gontcho A Gontcho}
\affiliation{University of Virginia, Department of Astronomy, Charlottesville, VA 22904, USA}
\author{Gaston~Gutierrez}
\affiliation{Fermi National Accelerator Laboratory, PO Box 500, Batavia, IL 60510, USA}
\author[0000-0003-0201-5241]{Dick~Joyce}
\affiliation{NSF NOIRLab, 950 N. Cherry Ave., Tucson, AZ 85719, USA}
\author{Robert~Kehoe}
\affiliation{Department of Physics, Southern Methodist University, 3215 Daniel Avenue, Dallas, TX 75275, USA}
\author[0000-0003-3510-7134]{Theodore~Kisner}
\affiliation{Lawrence Berkeley National Laboratory, 1 Cyclotron Road, Berkeley, CA 94720, USA}
\author[0000-0001-6356-7424]{Anthony~Kremin}
\affiliation{Lawrence Berkeley National Laboratory, 1 Cyclotron Road, Berkeley, CA 94720, USA}
\author[0000-0003-1838-8528]{Martin~Landriau}
\affiliation{Lawrence Berkeley National Laboratory, 1 Cyclotron Road, Berkeley, CA 94720, USA}
\author[0000-0001-7178-8868]{Laurent~Le~Guillou}
\affiliation{Sorbonne Universit\'{e}, CNRS/IN2P3, Laboratoire de Physique Nucl\'{e}aire et de Hautes Energies (LPNHE), FR-75005 Paris, France}
\author[0000-0003-4962-8934]{Marc~Manera}
\affiliation{Departament de F\'{i}sica, Serra H\'{u}nter, Universitat Aut\`{o}noma de Barcelona, 08193 Bellaterra (Barcelona), Spain}
\affiliation{Institut de F\'{i}sica d’Altes Energies (IFAE), The Barcelona Institute of Science and Technology, Edifici Cn, Campus UAB, 08193, Bellaterra (Barcelona), Spain}
\author[0000-0002-1125-7384]{Aaron~Meisner}
\affiliation{NSF NOIRLab, 950 N. Cherry Ave., Tucson, AZ 85719, USA}
\author{Ramon~Miquel}
\affiliation{Instituci\'{o} Catalana de Recerca i Estudis Avan\c{c}ats, Passeig de Llu\'{\i}s Companys, 23, 08010 Barcelona, Spain}
\affiliation{Institut de F\'{i}sica d’Altes Energies (IFAE), The Barcelona Institute of Science and Technology, Edifici Cn, Campus UAB, 08193, Bellaterra (Barcelona), Spain}
\author[0000-0002-0644-5727]{Will~Percival}
\affiliation{Department of Physics and Astronomy, University of Waterloo, 200 University Ave W, Waterloo, ON N2L 3G1, Canada}
\affiliation{Perimeter Institute for Theoretical Physics, 31 Caroline St. North, Waterloo, ON N2L 2Y5, Canada}
\affiliation{Waterloo Centre for Astrophysics, University of Waterloo, 200 University Ave W, Waterloo, ON N2L 3G1, Canada}
\author[0000-0001-7145-8674]{Francisco~Prada}
\affiliation{Instituto de Astrof\'{i}sica de Andaluc\'{i}a (CSIC), Glorieta de la Astronom\'{i}a, s/n, E-18008 Granada, Spain}
\author[0000-0001-6979-0125]{Ignasi~P\'erez-R\`afols}
\affiliation{Departament de F\'isica, EEBE, Universitat Polit\`ecnica de Catalunya, c/Eduard Maristany 10, 08930 Barcelona, Spain}
\author{Graziano~Rossi}
\affiliation{Department of Physics and Astronomy, Sejong University, 209 Neungdong-ro, Gwangjin-gu, Seoul 05006, Republic of Korea}
\author[0000-0002-9646-8198]{Eusebio~Sanchez}
\affiliation{CIEMAT, Avenida Complutense 40, E-28040 Madrid, Spain}
\author{David~Schlegel}
\affiliation{Lawrence Berkeley National Laboratory, 1 Cyclotron Road, Berkeley, CA 94720, USA}
\author{Michael~Schubnell}
\affiliation{Department of Physics, University of Michigan, 450 Church Street, Ann Arbor, MI 48109, USA}
\affiliation{University of Michigan, 500 S. State Street, Ann Arbor, MI 48109, USA}
\author[0000-0003-3449-8583]{Ray~Sharples}
\affiliation{Centre for Advanced Instrumentation, Department of Physics, Durham University, South Road, Durham DH1 3LE, UK}
\affiliation{Institute for Computational Cosmology, Department of Physics, Durham University, South Road, Durham DH1 3LE, UK}
\author[0000-0002-3461-0320]{Joseph~H.~Silber}
\affiliation{Lawrence Berkeley National Laboratory, 1 Cyclotron Road, Berkeley, CA 94720, USA}
\author{David~Sprayberry}
\affiliation{NSF NOIRLab, 950 N. Cherry Ave., Tucson, AZ 85719, USA}
\author[0000-0003-1704-0781]{Gregory~Tarl\'{e}}
\affiliation{University of Michigan, 500 S. State Street, Ann Arbor, MI 48109, USA}
\author{Benjamin~A.~Weaver}
\affiliation{NSF NOIRLab, 950 N. Cherry Ave., Tucson, AZ 85719, USA}
\author[0000-0002-6684-3997]{Hu~Zou}
\affiliation{National Astronomical Observatories, Chinese Academy of Sciences, A20 Datun Road, Chaoyang District, Beijing, 100101, P.~R.~China}

\begin{abstract}
We present a sample of 88,959 K-giants from DESI Milky Way Survey Year 3 data, which we use to characterize the chemo-dynamical properties of the stellar halo at Galactocentric distances of 12 to $\sim$100 kpc. Using \hdbscan{}, we identify five prominent stellar halo substructures: Aleph, the Sagittarius stream, Gaia Sausage-Enceladus (GSE), Cetus-Palca and the Orphan-Chenab stream. We present the properties of each of these structures as they appear in our catalog, and examine how uncertainties on distance affect the characterization of substructure with this approach. 
We also examine regions associated with previously reported overdensities (such as the Virgo Overdensities and the Sagittarius spur) that we do not recover with \hdbscan{}.
The size and distance range of our catalog allows us to explore in detail the residual stellar halo, comprising stars that we do not associate with any substructure. We find that samples of $\sim2000$ outer halo stars with both highly prograde and highly retrograde angular momenta have similar metallicity distribution functions (MDFs), which do not resemble the MDFs of either GSE or Sagittarius. Both the prograde and retrograde residual halo MDFs are bimodal, with a metal-poor peak at [Fe/H]$\sim-2$ and a metal-rich peak at [Fe/H]$\sim-1.3$ (prograde) or $-1.5$ (retrograde). The MDF for lower angular momentum residual halo K-giants does not show clear evidence for a metal-poor peak, and broadly resembles the MDF of GSE, even at much lower binding energies than GSE itself. We discuss possible interpretations of these findings for GSE accretion scenarios.
\end{abstract}


\section{Introduction} \label{sec:intro}

In recent decades, wide-field spectroscopic and photometric surveys have transformed our view of the Milky Way's stellar halo. Notable milestones include SDSS \citep{sdss_2000,Gunn2006_SDSS,SDSSiii,SDSSIV,SDSSV}, its extensions SEGUE/SEGUE-2 \citep{Rockosi2009_SEGUE,Rockosi22_SEGUE2} and APOGEE \citep{apogee2013,apogee2017}, the ESA Gaia satellite mission \citep{Gaia_2016, GaiaDR2_2018,GaiaEDR3_21}, LAMOST \citep{lamost2012,lamost_dr1}, RAVE \citep{Rave2003,Rave2020a,Rave2020b}, GALAH \citep{Galah2015,Galah2021} and H3 \citep{Conroy2019_H3}. These surveys have characterized the overall mass and shape of the stellar halo \citep[e.g.][Li et al. in prep.]{Bell:2008aa,Juric:2008aa,Ivezic:2008aa,Deason:2019aa,Han2022} and revealed a plethora of substructure, which in most cases is thought be debris from tidally disrupted satellite galaxies \citep{Searle:1978aa,Johnston:1996aa, Chiba_beers2000,Helmi_zeeuw2000, Bullock:2001aa}. 
\citet{Deason_belokurov2024} review the major substructures known to date; we also discuss several of these in detail in later sections of this paper. 

Simulations \citep[e.g.][]{Bullock:2001aa, Bullock:2005aa, Cooper2010} predict that only a handful of mergers dominate the stellar halo. Reconstructing the most significant events in the assembly history of the Milky Way is possible, in principle, by identifying those remnants and tracing properties back to those of their progenitors. In practice, the ease with which accreted debris can be identified and interpreted depends strongly on the mass and arrival time of its progenitor; those parameters in turn determine (on average) the range of present-day Galactocentric radii over which the debris is deposited and the degree of phase mixing \citep[e.g.][]{Johnston:2008aa,Amorisco:2017aa}. 

In the inner Galaxy, the timescale for phase mixing is short, particularly for the most massive progenitors. A variety of methods have been proposed to identify chemo-dynamical substructure in this region, in particular by exploiting densely-sampled, high-precision Gaia astrometry \citep{streamfinder1,Lovdal22_hdbscan,Kim2025,Berni2025,GFT2025}. Stars from progenitors that have phase-mixed heavily in configuration space may nevertheless retain similar integrals of motion or orbital actions \citep[][]{Helmi:1999ab, Helmi_zeeuw2000}, but even coherence in these quantities may be lost over time in realistic evolving potentials \citep[e.g.][]{Jean-baptiste2017,Dillamore_sanders2025}. A plethora of stellar halo substructures have been uncovered in recent years in the $\sim10$~kpc volume around the Sun \citep[e.g.][]{Myeong2018_shards, Malhan24, Kim2025}. Many of these features are associated with very small numbers of stars: they may be evidence of distinct, intrinsically low-mass progenitors, or else may be fragments of extensive debris systems associated with the complex orbits of more massive progenitors \citep[e.g.][]{Khoperskov2023, Kizhuprakkat2024, Thomas:2025aa}. As well as discovering these features, Gaia has confirmed earlier hints that the majority of the inner halo is dominated by a single well-mixed feature, now known as Gaia Sausage/Enceladus \citep[GSE;][and see below]{Belokurov2018,Helmi2018}.
 
We are currently faced with the challenge of determining whether the census of the most significant accretion events is complete, and interpreting massive structures, such as GSE, in terms of the overall assembly history of the Milky Way \citep[e.g.][and the references therein]{Belokurov2018,Helmi2018,Myeong:2019aa,
Evans:2020aa,Helmi2020,Naidu2020, Naidu:2021aa,Horta2023,Horta2024_GA}. It also remains unclear what fraction of the stellar halo is contributed by the more diffuse debris of currently identified structures, what fraction can be attributed to sparse structure not yet identified due to the limited sampling density of current surveys \citep[e.g][]{Carlin:2016ac}, and what fraction, if any, is contributed by an accreted component that is truly `smooth' (i.e.\ effectively uniform in phase space). Complicating this issue further, cosmological hydrodynamical simulations typically predict a non-negligible fraction of halo stars form `in situ', either by scattering from the Galactic disk or through cooling instabilities in the circumgalactic gas \citep{Abadi:2006aa,Cooper:2015aa}. Currently there is little theoretical consensus on the expected chemistry of in situ halo stars \cite[see e.g.][]{Belokurov:2023aa} or how they distribute in phase space, and hence no obvious procedure to distinguish them from the `smooth' accreted halo.\footnote{It is not necessarily obvious that an in situ halo would appear smooth in the spaces used to search for substructure -- this likely depends on the dominant mechanism for producing in situ halo stars \citep[e.g.][]{Cooper:2015aa}. The nature of known substructures with similarities to the disk (Aleph, ACS and the Monoceros ring, in Section~\ref{aleph}) is a case in point.}

In the outer halo, debris systems remain coherent over a much longer timescale. Access to these distant stars is potentially extremely valuable for reconstructing the Galaxy's assembly history \citep{Searle:1978aa,Bullock:2005aa,Helmi2018,Mackereth2019,Myeong2019,MWS_Cooper23,Kim2025}. Structure in this region may include features related to the early stages of disruption of the most significant progenitors \citep[e.g.][]{Naidu:2021aa,Vasiliev:2022aa,Amarante2022} and may therefore help to disentangle the complex picture painted by recent discoveries in the inner halo -- for example, to test hypotheses regarding the timing and potential multiplicity of events that contributed to the GSE feature \citep[e.g][]{Donlon2022, Chandra:2023aa, Donlon2024, Folsom2025}.

Until recently, however, the sample of stellar spectra from the outer Galaxy has been relatively small, a limitation that next-generation spectroscopic surveys, such as DESI, are now able to overcome \citep{AllendePrieto2020,MWS_Cooper23,MWS_dr1}. Nevertheless, at the depth of current all-sky surveys, the structure of the outer halo can still only be probed by sampling relatively bright `tracer' populations for which accurate distances can be obtained. The most widely-used tracers are RR Lyrae variables \citep[e.g.][]{Medina:2025aa}, blue horizontal branch (BHB) stars \citep[e.g][]{Bystrom:2025aa}, and K-giants \citep[][and the references therein]{Deason_belokurov2024,Ding2025}. Of these three tracers, K-giants have the greatest density and the highest intrinsic luminosities, although they also have substantially greater distance uncertainties than RR Lyrae stars or BHBs, which need to be taken into account.

K-giants are evolved stars with effective temperatures of approximately $3500-5500$~K, masses of $1.1 - 1.2\,\msol$, and absolute magnitudes $2 < M_r < -2$, corresponding to luminosities $60 - 300\, \mathrm{L_\odot}$ \citep{Yanny2009}. They can be identified using straightforward selections on effective temperature and surface gravity measured from low-resolution spectra, with the principle source of contamination being less luminous K- and M-dwarfs of similar temperature. Although the luminosity of K-giants is a steep function of their effective temperature (or color index), their distance can be estimated to an accuracy of $\sim 16 \%$ \citep{Xue2014, Bird2022}. Li et al. (in prep.) found distance uncertainties for DESI Year 3 K-giants as low as $\sim 11\%$ when using estimates from the SpecDis code \citep{Specdis2025}. A sufficiently large sample of K-giants over a large distance range can therefore be used to explore the structure of the stellar halo and hence the accretion history of the Galaxy \citep{Chiba1998,Dohm-Palmer2001,Else_starkenburg_2009,Carollo2010,Janesh2016,yang2019,Bird2019,Naidu2020,Lopez2024}. As dynamical tracers, K-giants have also been used to constrain the total mass of the Galaxy \citep{Kafle2014,Huang2016,Zhai2018,Bird2019} and the shape of the dark matter halo \citep{Vera_carlos_helmi2013,Posti_Helmi2018,Han2022,Zhang2025}.

In this paper we select K-giants in the Dark Energy Spectroscopic Instrument (DESI) Milky Way Survey \citep[MWS,][described below]{MWS_Cooper23,Koposov_2024_EDR} Year 3 stellar catalog and use them to characterize the major structures in the distant stellar halo. To exploit the size and uniformity of the DESI MWS sample, we use a (largely) automated approach to the association of individual stars with structures. We also consider in detail the properties of stars that are not assigned to any structure.

The outline of the paper is as follows. In Section \ref{data and methods}, we provide a brief overview of the DESI survey, MWS, and the Y3 stellar catalog and our criteria for selecting K-giants. In Section \ref{method} we describe our method to identify halo substructure, based on sequential application of the \hdbscan{} clustering algorithm. In Section \ref{results}, we present our results and a discussion of the effect of distance errors, in particular with regard to the recovery of the Sagittarius stream. Section \ref{sec:bulk} explores the properties of the `residual' (apparently unstructured) stellar halo, after removing the structures identified in previous sections. A summary is given in Section \ref{conclusion}. We give further details of our sequential clustering process in Appendix \ref{sequential clustering}. A comparison of our K-giant catalog with SEGUE data is included in Appendix \ref{appendix:segue}.

\section{Data}
\label{data and methods}

In this section, we describe the DESI Milky Way Survey. We introduce the DESI Year 3 stellar catalog, and the quality cuts and selection criteria we use to identify halo K-giants. We then describe our methods to find substructures in the distant stellar halo.

\subsection{The DESI Survey} \label{desi_survey}

DESI is a optical multi-object spectrograph, with a wavelength range of 3600 -- 9800 \AA{} and average spectral resolution ($R=\lambda/\Delta \lambda$) of 2000 -- 5500, installed on the Mayall 4-m telescope at Kitt Peak National Observatory, Arizona \citep{DESI_instrumentation_2022,Poppet2024_fibersys,Miller2024_optical_corr}.\footnote{See \url{https://www.desi.lbl.gov} for general information on the DESI survey and \url{https://data.desi.lbl.gov/doc/} for information on the data releases.} The instrument is capable of collecting spectra for $\sim$5000 objects simultaneously across a $\sim 3^\circ$ field of view, leading to 63 million spectroscopically confirmed galaxies and quasars over an eight-year period \citep{Schlafly2023,DESI_Guy2023_SP}. Since 2021, the instrument has been used for a cosmological redshift survey over a 17,000 square degree footprint, aimed at improving constraints on cosmological parameters, primarily through high-fidelity measurements of the baryon acoustic oscillation feature in the large-scale galaxy distribution
\citep{Levi2013_DESI,DESI2016a,DESI2016b,DESI2024.I.DR1,DESI2024.VII.KP7B,DESI_DR2_BAO}. Targets for DESI were selected from the DESI Legacy Imaging Survey (LS) data release 9 \citep{LS_Dey_2019,AllendePrieto2020,Myers2023}.

\subsubsection{DESI Milky Way Survey} \label{MWS}

Although DESI was developed for a cosmological survey, the instrument is well suited to observing large numbers of stars. Concurrent with the 8-year cosmological survey \footnote{This is an update over the 5-year plan mentioned in \citet{DESI2016a,MWS_Cooper23}.}, the DESI Milky Way Survey (MWS) aims to observe $\gtrsim10$ million unique stars\footnote{The public release of 3 years of data (DR2 or Y3 stellar catalog) will include 12 million stars across all MWS survey programs \citep{MWS_dr1}. See \citet{MWS_Cooper23,MWS_dr1,Dey2025} for more information on the MWS survey programs (Bright, Dark and Backup).} within a magnitude range of $16 < r < 19$, at Galactic latitudes\footnote{Although the main MWS footprint is limited to latitudes $\geq 20^{\circ}$, the DESI Backup program \citep{Dey2025} observes down to latitudes as low as 10$^{\circ}$.} $b\geq 20^{\circ}$. The data from MWS will be used to better constrain the mass and assembly history of the Galaxy, to measure the shape and orientation of its dark matter halo, and to explore statistical stellar astrophysics by gathering large samples of stars at particular evolutionary stages, such as white dwarfs \citep{MWS_Cooper23,Manser2024}.

MWS operates under bright sky conditions, as defined by the overall DESI observing strategy \citep{AllendePrieto2020,MWS_Cooper23,Schlafly2023}. In the bright time observing program, MWS shares the DESI focal plane with the DESI Bright Galaxy Survey \citep[BGS;][]{DESI_BGS_2023}. The bright program (hence MWS) footprint includes most of the north Galactic cap and a fraction of the south Galactic cap. The footprint includes many well-known substructures in the stellar halo, such as the Sagittarius stream, the GD 1 and Orphan streams, the disrupting globular cluster Pal 5, and many dwarf galaxies and intact globular clusters. MWS targets are selected from the LS dataset using a set of straightforward criteria based on color and Gaia astrometry. MWS observes three main classes of stellar target, \mainblue{}, \mainred{} and \mainbroad{}, distinguished by color and proper motion. Essentially all stars in the magnitude range $16 < r < 19$ (in the survey footprint) are potential targets in one of these categories, which we refer to collectively as \mwsmain{}. MWS also observes fainter stars ($19 < r < 20$) in the color-selected classes \texttt{FAINT-BLUE} and \texttt{FAINT-RED}, at a lower priority (and hence lower completeness) than the \mwsmain{} targets. Although targets are selected with an emphasis on large-scale observations of the thick disk and stellar halo, the survey also includes samples of rare stellar types such as Blue Horizontal Branch stars (BHBs), RR-Lyrae stars, white dwarfs and stars within a 100 pc of the Sun. These samples are given high priority for observation, because they are very sparsely distributed on the sky. 

The DESI survey as a whole also defines a backup observing program \citep{Dey2025}, for conditions that are too poor to observe the regular bright-time program. Unlike the bright time program, \textit{all} backup program targets are stars. The categorization of backup program targets is different to that of the MWS survey; in general, backup targets are observed over a much broader magnitude range (in particular, as bright as $G\sim12$), redder stars are favored, and they are observed to a lower limiting Galactic latitude (i.e.\ probe further into the Galactic plane). 

\subsubsection{MWS Year 3 stellar catalog}

The basis for this work is the catalog of all stellar observations up to and including the third year of data taken by the DESI survey. This catalog includes parameters measured by RVS for all stars observed `by design' in the MWS and the backup survey. It also includes stars observed during the dark time programs or BGS, either as flux standards, as part of small `secondary' target programs (an example is a program targeting BHB stars in dark time) or as stellar contaminants in the various DESI galactic target selections. 73\% of the bright time survey was completed as of April 9th, 2024, the last night of observations included in the Year 3 data set.

Our work is based on the MWS RVS pipeline\footnote{MWS has 2 pipelines: (1) RVS, based on \cite{Koposov2011} and (2) SP, based on the \texttt{FERRE} \citep{Allende_Prieto_2006}, a \texttt{FORTRAN} code \citep[see][for more details on the MWS pipelines]{MWS_Cooper23}. The Y3 catalog from the SP pipeline was unavailable at the time of writing this paper.} (specifically its python implementation, RVSpecfit), a post-processing extension to the primary DESI spectral fitting pipeline, Redrock \citep[S.J. Bailey et al., in prep]{DESI_Guy2023_SP}.
RVS fits spectral templates from the PHOENIX library, using an algorithm presented in \cite{Koposov2011}. From the RVS fit to each spectrum we obtain a radial velocity and other atmospheric stellar parameters including [Fe/H], [$\alpha$/Fe], $T_\mathrm{eff}$, log g, and $V \sin i$, along with their uncertainties. More information on the MWS science, target selection, observation strategy and the RVS pipeline is given in \cite{MWS_Cooper23,Koposov_2024_EDR}; Koposov et al\ (\textit{in prep}).

Prior to applying our K-giant selection criteria, we apply the data quality cuts given in \cite{Koposov_2024_EDR}, in order to ensure that the targets are indeed stars. We add a lower limit of 5 on the signal-to-noise ratio in all three wavelength bands. This ensures a cleaner sample where features in the data can be identified robustly and have lower noise-induced biases. Our data quality cuts can be summarized as follows:
\begin{itemize}
    \item \texttt{RR\_SPECTYPE = STAR}
    \item \texttt{RVS\_WARN\footnote{\texttt{RVS\_WARN} is a bitmask integer that is set to zero if there are no warnings during the spectral fitting. See \citet{Koposov_2024_EDR}.}} = 0
    \item \texttt{DESI\_TARGET \& MWS\_ANY > 0}
    \item \texttt{PRIMARY = True} 
    \item\texttt{SURVEY = main}
    \item \texttt{SN\_B/R/Z > 5}
    \item \texttt{Gaia SOURCE\_ID} is valid (\texttt{!= 999999})
    and unique\footnote{If a star is observed multiple times as a result of being targeted by different observing programs,the resulting catalog entries for each program will have different DESI \texttt{TARGETID}s, but the same Gaia \texttt{SOURCE\_ID}. Only a very small number of MWS stars have such duplicate entries, and the recovered stellar parameters are almost always in good agreement. We use the \texttt{PRIMARY} flag to select the unique observation in these cases. The \texttt{PRIMARY} flag is set for the observation with the largest signal-to-noise ratio in the $r$ band \citep{Koposov_2024_EDR}. }.
    \item \texttt{[Fe/H] > -3.9 dex}
\end{itemize}
Application of these criteria to the full Y3 stellar catalog yields 8,718,902 stars\footnote{Comprising 4,579,456 targets from the bright program, 405,235 targets from the dark program and 3,732,758 targets from the backup program.}.
 
Distance information is essential to obtain six-dimensional phase space coordinates for each star. MWS does not provide a single distance measurement, but is instead pursuing several methods of measuring spectro-photometric distances.
In this work, we use distance measurements obtained with the \texttt{rvsdistnn} method, summarized by \citet{Aganze2025} and to be described in full by Koposov et al.\ (\textit{in prep.}).\footnote{In parallel to the \texttt{rvsdistnn} method, \citet{Specdis2025} have constructed an alternative catalog of distances for the DESI stellar sample, SpecDis. Although \texttt{rvsdistnn} and SpecDis both use neural networks to estimate distances, the former trains on stellar parameters and color, while the latter trains on complete DESI spectra.} In Appendix \ref{distance calibration} we provide further details of \texttt{rvsdistnn} and compare distance estimates for our K giant candidates with literature values for stars in known globular clusters and dwarf galaxies.

We restrict our sample to stars with estimated \texttt{rvsdistnn} heliocentric distances below 150 kpc. The mean distance uncertainty (also obtained from \texttt{rvsdistnn}) is 15\%, with $\sim$ 80\% of the sample having uncertainties lower than 20\%.\footnote{5.4\% of stars have distance uncertainties larger than 30\%. Retaining these stars does not change our results in any significant way.} This additional criterion reduces our cleaned parent catalog to 8,717,449 stars. 

\subsubsection{Computing kinematic and orbital quantities}
\label{computing tools}

In order to calculate the spatial, kinematic and orbital parameters in a Galactocentric frame, we use Astropy v6.0.1 \citep{astropy:2018} with the following parameters: Galactocentric location of the Sun, $\rsol = 8.277$ kpc \citep{Gravity2022}, vertical height of the Sun above the Galactic plane, $\zsol = 20.8\,\mathrm{pc}$ \citep{Bennett_Bovy_2019}, and solar velocity relative to the Galactic center $[V_{x,\sun}, V_{y,\sun}, V_{z,\sun}] = [12.9, 245.6, 7.78] \, \kms$ \citep{Drimmel_poggio_2018}. The orbital and kinematic properties including total energy, eccentricities, actions, pericenter and apocenter radius are calculated using a \citet{McMillan2017} potential model in the \texttt{AGAMA} framework \citep{Agama2019}. This potential is static and axisymmetric, and comprises a bulge, thin and thick disks, and an NFW dark matter halo. We use a right-handed convention, whereby prograde stars have $L_z < 0$ and retrograde stars have $L_z > 0$. Throughout the paper, we quote absolute values of energy in units of $10^5\,\mathrm{km^2 s^{-2}}$, absolute values of angular momentum in $10^3\,\mathrm{kpc\,km\,s^{-1}}$, and metallicity in dex.

\subsection{The K-giant sample} \label{kgiant selection}

To select K-giants, we apply the following criteria to the cleaned Y3 stellar catalog described in the previous section: 

\begin{itemize}
    \item $3900 < T_\mathrm{eff} < 5400$;
    \item $0.5 < \log g < 3.0$;
    \item Total proper motion $|\mathrm{\mu}|< 11$ mas/yr.\footnote{The intention of this proper motion cut is to remove nearby dwarf star contaminants. We adopt same value used as in the SEGUE survey \citep{Xue2014}.}
\end{itemize}

This K-giant selection yields a total of 143,635 stars. In order to concentrate on the distant stellar halo we impose additional restrictions on Galactocentric distance and vertical height, to exclude both very nearby halo stars and the Galactic disk:

\begin{itemize}
\item Galactocentric radial distance $r_\mathrm{gal} > 12\,\mathrm{kpc}$;
\item $|z_\mathrm{gal}| > 3\,\mathrm{kpc}$.
\end{itemize}

This halo selection reduces the size of the K-giant sample to 89,806 stars \footnote{This selection may still include some contamination from the disk, including structures such as the TriAnd, the Anticenter stream and the Monoceros ring (see Section \ref{aleph}), which are believed to originate from perturbations in the outer disk. These structures span distances $|z_\mathrm{gal}| \lesssim 10$ kpc.}. 

For this sample of K-giant candidates, we calculate Galactocentric positions and velocities, and orbital properties using the tools described in section \ref{computing tools}. For 847 stars, the total energy was found to be greater than zero, indicating that the stars are unbound. A detailed study of these stars is beyond the scope of this paper; we omit them from further consideration. The final number of K-giants is therefore 88,959\footnote{For comparison, previous substructure detection works using H3 Survey by \citet{Naidu2020} used 5684 giant stars and using LAMOST survey by \citet{yang2019} used 13,000 giant stars.}. This includes 53,934 stars from the bright program, 35,006 stars from the backup program and 19 stars from the dark program.

In appendix \ref{appendix:segue} we compare our K-giant catalog with that of \citet{Xue2014}, comprising $\sim6000$ stars from the Sloan Extension for Galactic Understanding and Exploration (SEGUE). The atmospheric parameter, velocity and distance distributions of the two catalogs are broadly similar, although with systematic offsets that may reflect differences in the spectral fitting pipelines.

\subsubsection{Properties of the K-giant sample}
\begin{figure}
    \centering
    \includegraphics[width=\linewidth]{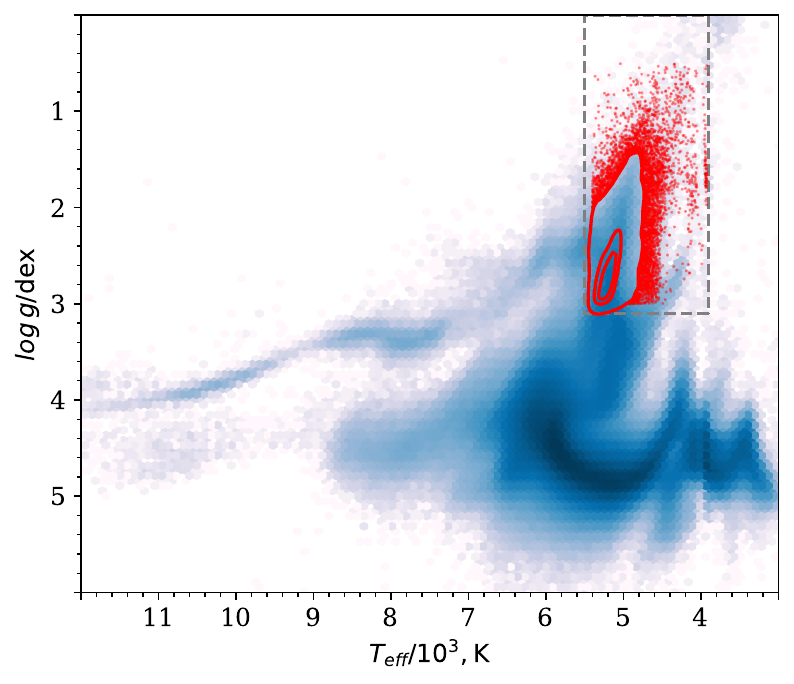}
    \caption{Distribution of stars in the cleaned DESI MWS Y3 catalog in $\log g - T_\mathrm{eff}$ space. The red contours enclose 25\%, 50\% and 95\% of our K-giants. Red points mark K-giants outside the 95\%  contour. The gray dashed rectangle indicates the selection box described in section \ref{kgiant selection}.}
    \label{fig:fig1_logg_teff}
\end{figure}

\begin{figure}
    \centering
    \includegraphics[width=\linewidth]{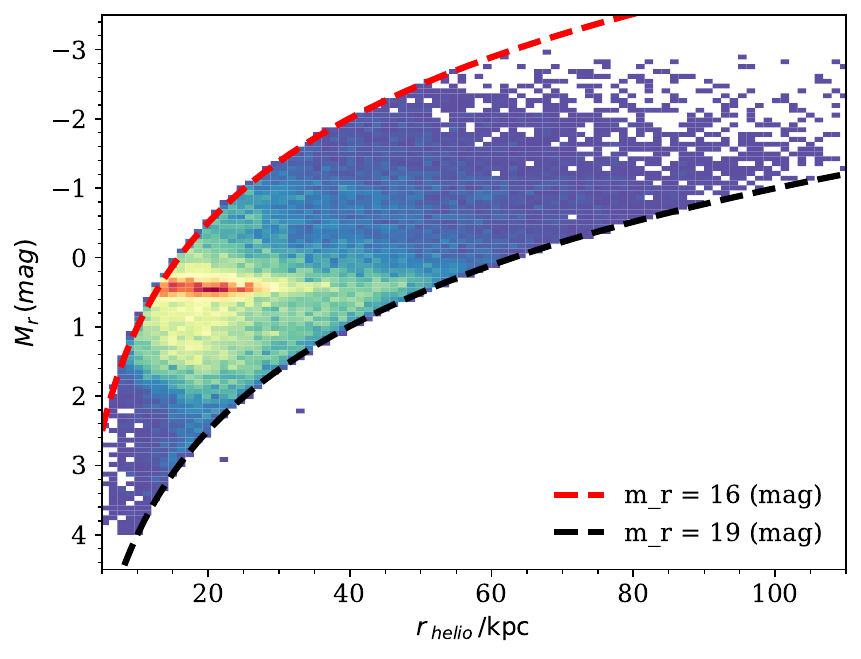}
    \caption{Distribution of our K-giant catalog in absolute magnitude ($M_r$) with heliocentric distance. The red and black dashed lines indicate the bright ($r=16$) and faint ($r=19$) magnitude limits of the MWS main survey. A handful of stars observed by the dark-time DESI programs fall outside these limits.}
    \label{fig:fig2_absr_rhelio}
\end{figure}

\begin{figure*}[h!]
    \centering   
    \includegraphics[width=\linewidth]{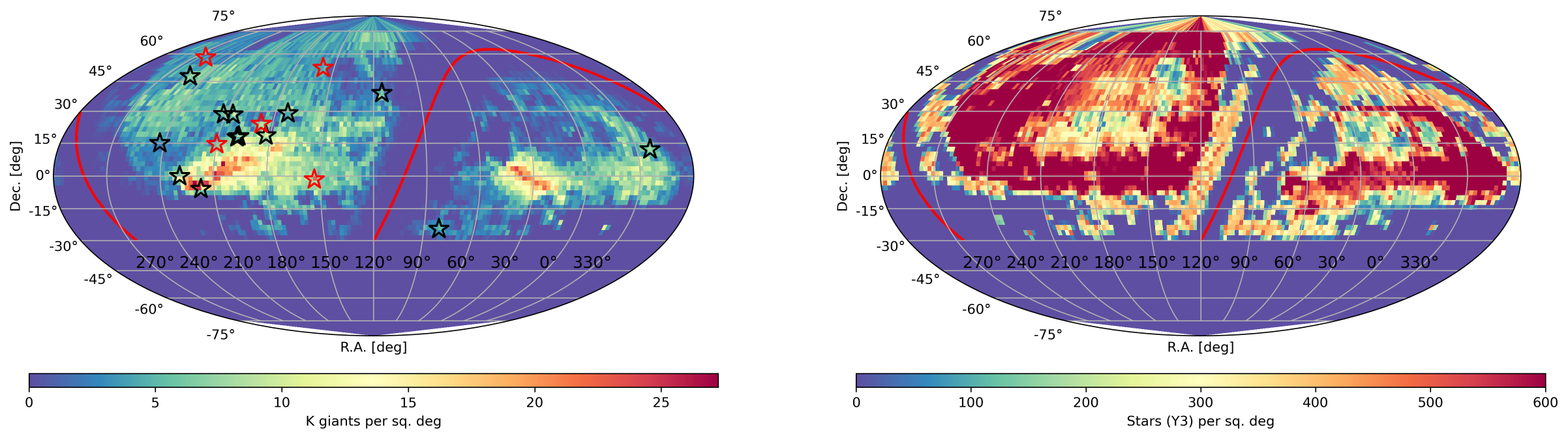}
    \caption{Distribution of stars in the DESI footprint. The Galactic plane is shown by a solid red line. Left panel: distribution of our K-giant sample. The track of the Sagittarius stream is clearly visible. The red star symbols represent the dwarf galaxies and the black star symbols represent globular clusters; these correspond to hotspots in the map. Right panel: distribution of all stars in the DESI Y3 catalog. We use 600 stars per square degree as the saturation point of the color scale to highlight (approximately) the areas that do not have three survey passes in the Y3 catalog (roughly, the region $120^\circ<\alpha<210^\circ$, $15^\circ<\delta<60^\circ$ in the northern hemisphere, and $15^\circ<\delta<30^\circ$ in the southern hemisphere). Regions outside the main DESI footprint are covered by a single pass of the Backup Program.}
    \label{fig:fig3_skyplot}
\end{figure*}

\begin{figure*}
    \centering
    \includegraphics[width=\linewidth]{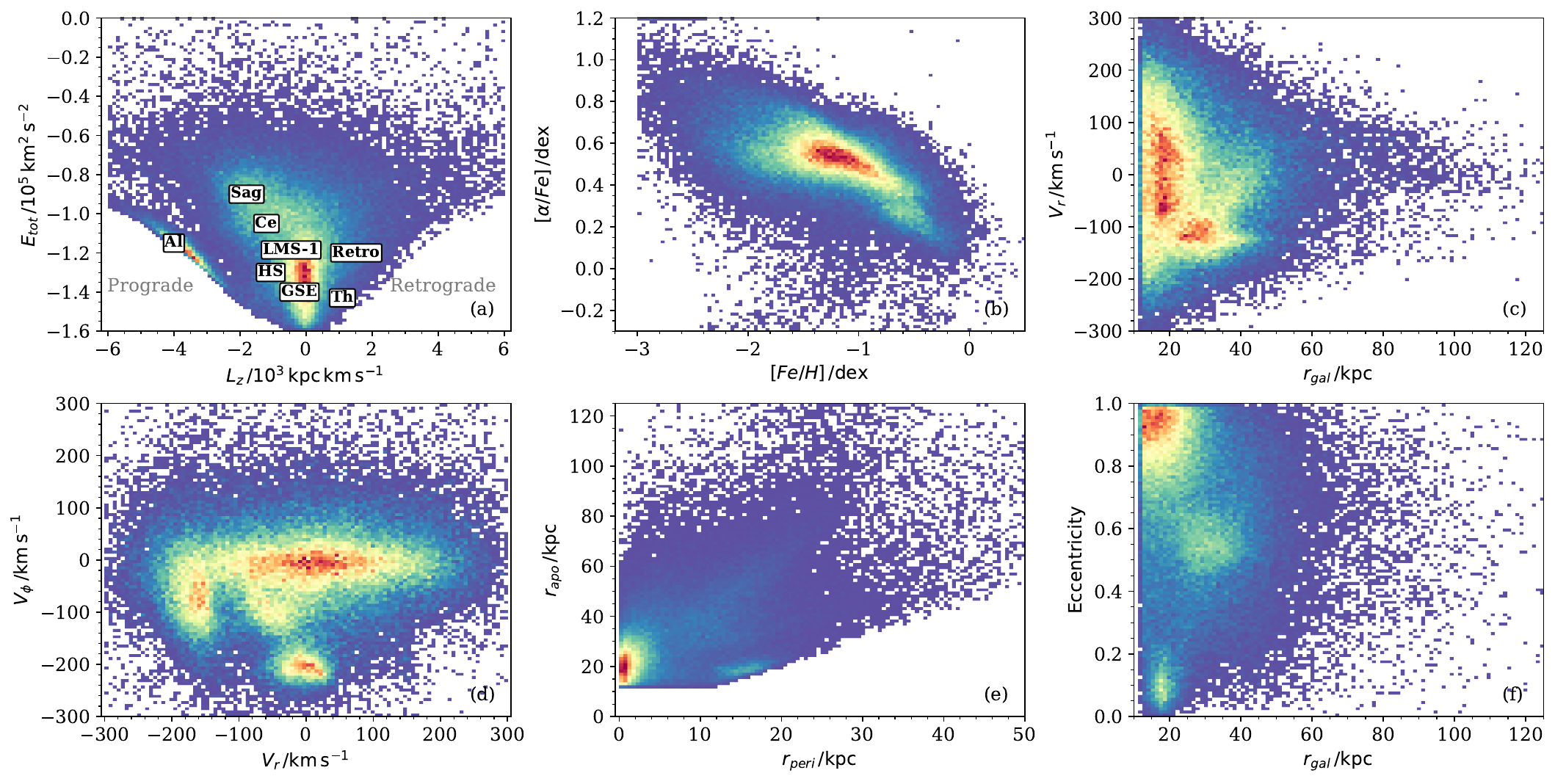}
    \caption{
    An overview of the K-giant sample selected from the DESI Y3 catalog. The sequence of panels runs from top left to bottom right. Panel (a) shows the distribution in $E_{tot} - L_z$ space. The labels indicate the approximate peak positions of previously reported structures: Sag (Sagittarius), Ce (Cetus), Al (Aleph), LMS-1/Wukong, HS (Helmi streams), GSE, Th (Thamnos) and Retro (retrograde structures -- Arjuna, Sequoia and I'itoi). Panel (b) shows the distribution in chemical space. Most stars have metallicities $-0.5<\mathrm{[Fe/H]} < -1.5$ with high [$\alpha$/Fe]. The median [Fe/H] is $-1.25$, with a low metallicity tail extending to [Fe/H] $\approx -3$. Panel (c) shows the distribution of Galactocentric radial velocity with distance. Panel (d) shows the join distribution of velocity components $V_r$ and $V_{\phi}$ in spherical coordinates. Panel (e) shows pericenter and apocenter distributions. Panel (f) shows eccentricity and Galactocentric distance. Overdensities in Panels (c), (d), (e) and (f) correspond to distinct dynamical structures as in Panel (a).}
    \label{fig:fig4_summary}
\end{figure*}

Fig.~\ref{fig:fig1_logg_teff} presents the distribution of the cleaned Y3 MWS stellar catalog in $\log g$ - $T_\mathrm{eff}$ space. The red contours indicate the regions occupied by K-giants at 25\%, 50\% and 95\% density levels and the red points show K-giants outside the 95\% threshold. The dashed gray rectangle illustrates the K-giant selection range boundaries: $0.5 < \log g < 3$ and $3900 < T_\mathrm{eff} < 5400$. This figure shows the turn-off, red giant branch, and the horizontal branch approximately at their expected positions. It also shows the main sequence in the low-temperature, and high gravity region. However, this population exhibits an unusual shape and surprising oscillations, that are thought to be the result of systematic errors in modeling the spectra of these cool stars. Similar systematics have been observed previously in analyses, e.g. in APOGEE or RAVE, and a conclusive explanation is still pending \citep{Holtzman_apogee_2015,Rave2020b}. 

Fig.~\ref{fig:fig2_absr_rhelio} shows the distribution of the K-giant sample in absolute magnitude ($M_r$) and heliocentric distance ($r_\mathrm{helio}$). We mark the bright and faint apparent magnitude limits of the survey, at $r=16$ and $r=19$ mag respectively. 
The figure shows that the majority of the sample has an absolute magnitude in the range $2 < M_r < -2$, corresponding to distances $15\lesssim r_\mathrm{helio} \lesssim 90$~kpc. The sample is dominated by fainter giants/subgiants ($1 < M_r < 4$) at lower distances, 
and brighter giants ($M_r > -2$) at larger distances. 

\begin{figure*}
    \centering
    \includegraphics[width=\linewidth]{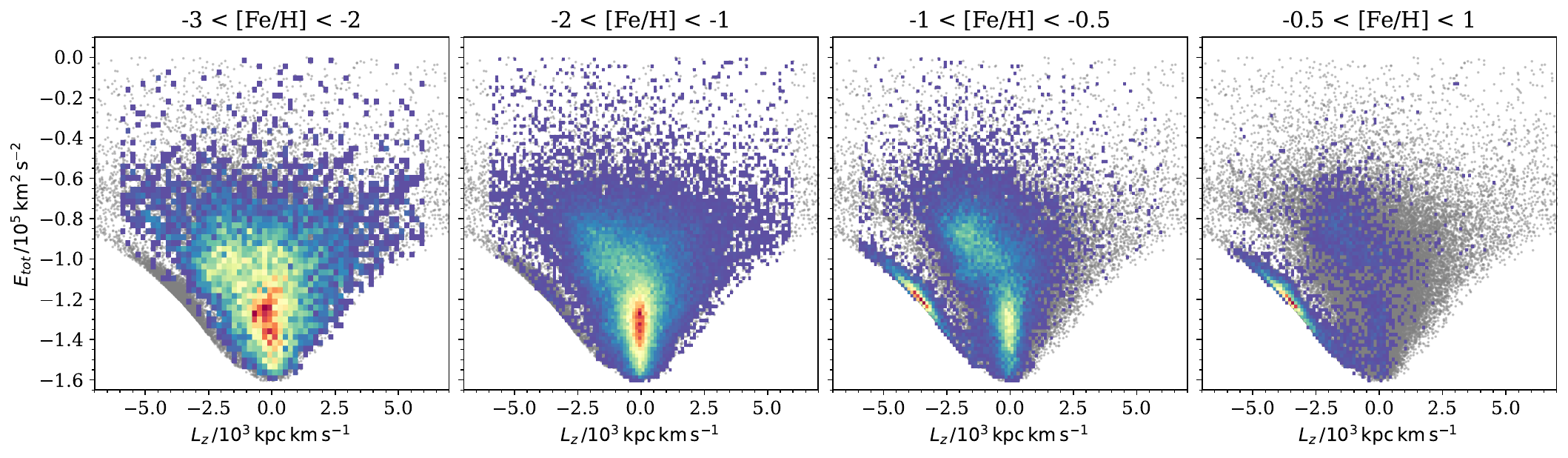}
    \caption{$E_\mathrm{tot} - L_z$ distribution of our K-giants in different metallicity bins (linear scale). Metallicity increases from left to right.}
    \label{fig:fig5_E_Lz_met}
\end{figure*}

Fig.~\ref{fig:fig3_skyplot} shows the distribution of the stars in the MWS footprint. The first panel shows the distribution of our K-giant sample while the second panel shows the coverage of the entire Y3 stellar sample. The Sagittarius stream is clearly visible in the K-giant distribution. 
The five red star-shaped symbols represent the dwarf galaxies Coma Berenices, Bootes \Romannum{1}, Draco, Sextans and Ursa Major, while the thirteen black star-shaped symbols represent globular clusters spanning a distance range of 12 to $
\sim 100$ kpc, with known members appearing in the catalog. A further nine dwarf galaxies (Ursa Minor, Triangulum \Romannum{2}, Segue \Romannum{1}, Segue \Romannum{2}, Leo \Romannum{2}, Hercules, Canes Venatici \Romannum{1}, Bootes \Romannum{3} and Aquarius \Romannum{2}) and six globular clusters contribute a total of 22 stars to the catalog. We also indicate the regions of the sky that do not yet have full survey coverage in the Y3 catalog. In particular, in the northern sky, the portion of the Sagittarius stream at right ascension $\alpha<210^\circ$ has large patches of incomplete coverage. 

Fig.~\ref{fig:fig4_summary} illustrates the distribution of the K-giant sample in different chemical, kinematic and orbital spaces. The $E_{tot} - L_z$ space shown in the first panel is widely used to differentiate the remnants of distinct accretion events. The total energy and the $z$-component of angular momentum are conserved under idealized conditions, which in theory enables each peak in this space to be identified with an individual satellite accretion events, even if the associated stars are dispersed in configuration space \citep{Helmi_zeeuw2000}. In practice, orbits evolve over time due to fluctuations and asymmetries in the potential. Stars from a single progenitor may spread out across this diagram, and one progenitor may generate more than one peak; the interpretation of this diagram is further complicated by survey selection effects \citep[these complications are illustrated by the mock realizations of DESI MWS from cosmological simulations in][and discussed further below; see also \citealt{Jean-baptiste2017}]{Kizhuprakkat2024}. In spite of these caveats, this space has been particularly productive for the discovery of halo substructure \citep{Deason_belokurov2024}. The labels indicate the relative (approximate) positions of significant structures that have been identified in previous studies of the stellar halo. These include GSE, the Sagittarius stream \citep{Ibata94_Sgr}, the Cetus-Palca stream \citep{Newberg2009_Cetus,GFT22_Cetus_Palca}, Aleph \citep{Naidu2020}, the Helmi streams \citep{Helmi1999,Koppelman2019_hs}, LMS-1/Wukong \citep{Naidu2020,Yuan2020a_LMS1}, Thamnos \citep{Koppelman2019a} and other retrograde structures such as Arjuna, Sequoia and I'itoi \citep{Myeong2019,Naidu2020}. In subsequent sections, we will use \hdbscan{} to identify corresponding structures in the DESI MWS dataset, and provide a more comprehensive discussion of those we find.

The second panel shows the distribution in [Fe/H] and [$\alpha$/Fe]. The majority of our K-giants have $-1.5 \lesssim \mathrm{[Fe/H]} \lesssim -0.5$ and $\mathrm{[\alpha/Fe]} \gtrsim0.25$, consistent with the expectations for a metal-poor stellar halo.

The remaining panels show combinations of Galactocentric position (radius $r_\mathrm{gal}$) and velocity (radial $V_{r}$ and azimuthal $V_\phi$), and orbital pericenters, apocenters and eccentricities. The prominent overdensities in these spaces correspond to those in the $E_{tot} - L_z$ space, in particular GSE and Sagittarius.

Fig.~\ref{fig:fig5_E_Lz_met} again shows the distribution of our K-giant sample in $E_{tot} - L_z$ space, here separated into different metallicity bins. This gives a straightforward illustration of the important role of chemical measurements in distinguishing different structures. In the higher metallicity bins, $-1 < \mathrm{[Fe/H]} < 1$, the majority of stars are found on prograde orbits ($L_z < 0$). The most prominent structure in these bins are Aleph, GSE and Sagittarius. The latter two structures are even more prominent in the $-2 < \mathrm{[Fe/H]} < -1$ bin, to which Aleph contributes very little, and in which the proportion of retrograde stars increases. The distribution in the lowest metallicity bin ($-3 < \mathrm{[Fe/H]} < -2$) is more diffuse. The small number of stars in this bin with disk-like orbits may be related to a metal-weak thick disk \citep[e.g.,][]{Carollo2019,Sestito2021,Hong2024}.

\section{Substructure Finding Methods } \label{method}

Figures ~\ref{fig:fig4_summary} and \ref{fig:fig5_E_Lz_met} show that the stellar halo of the Galaxy, as seen in our DESI K-giant sample, is highly structured. One of the major ongoing challenges in the field of Galactic archaeology is to identify and interpret overdensities in the multidimensional space of the properties shown in Fig.~\ref{fig:fig4_summary}. Here, our first aim is to characterize the structures seen in our K-giant sample. Although the figures above lead us to expect that we will recover many well-known structures, the DESI dataset is significantly larger than previous studies and extends to larger distances , potentially offering better characterization of these features and the possibility of discovering additional features. 

The currently known structures in the Milky Way's halo, including coherent tidal streams and more amorphous overdensities, have been identified using different combinations of manual selection (`cuts') in a variety of parameter spaces \citep[among others]{Helmi2018,Naidu2020,Malhan24} and automated clustering algorithms \citep[{and the references therein}]{Lovdal22_hdbscan,Tiwari2023,Gupta2024}. In this work, we employ an unsupervised machine learning algorithm, \hdbscan, which is widely used in the astronomy community for identifying clusters from large datasets \citep{Koppelman2019a,Ou2023_hdbscan,tang24_hdbscan}. \citet{Kim2025} used \hdbscan{} to identify substructures at heliocentric distances $\le 5~\mathrm{kpc}$ in the DESI Year 1 stellar catalog. Using integrals of motion ($E_{tot}, L_z, \log_{10}\, J_r, \log_{10}\, J_z$) and Galactocentric velocities ($V_r, V_{\phi}, V_z$) as input features, they identified 5 structures in the local stellar halo and associated them to Helmi streams, M18-Cand10/MMH-1, Sequoia, Antaeus and ED-2. Although both \citet{Kim2025} and this work use MWS data and \hdbscan{} for structure identification, our dataset differs from theirs in terms of size, the targeted stellar population, as well as the distance range we are exploring. Therefore, we do not transfer any of the parameters, or use a similar process as in the previous study.
The following sections describe how we use \hdbscan{} to identify structures.

\begin{deluxetable*}{|l|c|c|c|c|c|c|c|}
\tabletypesize{\scriptsize}
\tablewidth{0pt} 
\tablecaption{The table provides all the properties of the clusters identified by \hdbscan{}. Each column corresponds to the cluster number associated with the \hdbscan{} pass, and the properties are represented along the rows. From top to bottom; row (1) gives the total number of stars in a given cluster. The value in parenthesis shows the number of stars added to the cluster in the combined fiducial second and metal-poor second passes -- details of this procedure are given in the text. (2) the most likely association of the structure with a known feature, (3) [Fe/H] value, (4) heliocentric radial velocity, (5) [$\alpha$/Fe] value, (6)[Mg/Fe] value, (7) Galactocentric distance in kpc, (8) Eccentricity, (9) maximum vertical excursion (z$_{max}$) in kpc, (10) pericentric distance in kpc and (11) apocentric distance in kpc. All the numbers provided are their respective median values, with uncertainties corresponding to the $16^\mathrm{th}$ and $84^\mathrm{th}$ percentiles.}
\label{tab:cluster_details1}
\tablehead{
\colhead{Properties} & \colhead{Cluster 1} & \colhead{Cluster 2} & \colhead{Cluster 3} & \colhead{Cluster 4} & \colhead{Cluster A} & \colhead{Cluster B}}
\startdata
    $N_{\star}$ & 7,270 (720) & 5,566 (0) & 18,242 (2,124) & 2,252 (655) & 137 & 95 \\
    Association & Sgr (S) & Aleph & GSE & Sgr (N) & Cetus-Palca & Orphan-Chenab \\
    $\mathrm{[Fe/H]/dex}$ & $-0.93^{+0.34}_{-0.49}$ & $-0.53^{+0.24}_{-0.22}$ & $-1.27^{+0.36}_{-0.40}$ & $-1.02^{+0.38}_{-0.47}$ & $-2.13^{+0.08}_{-0.19}$ & $-1.89^{+0.32}_{-0.42}$ \\
    $V_\mathrm{helio}/\mathrm{km s^{-1}}$ & $-139.32^{+25.77}_{-32.62}$ & $-41.62^{+138.57}_{-79.45}$ & $-37.44^{+174.67}_{-144.15}$ & $0.07^{+38.25}_{-34.62}$ & $-90.22^{+36.95}_{-40.85}$ & $200.70^{+24.44}_{-33.12}$ \\
    $\mathrm{[\alpha/Fe]/dex}$ & $0.44^{+0.12}_{-0.09}$ & $0.24^{+0.07}_{-0.08}$ & $0.53^{+0.09}_{-0.09}$ & $0.49^{+0.12}_{-0.11}$ & $0.72^{+0.12}_{-0.15}$ & $0.54^{+0.15}_{-0.14}$ \\
    $\mathrm{[Mg/Fe]/dex}$ & $-0.0045^{+0.21}_{-0.14}$ & $0.14^{+0.08}_{-0.07}$ & $0.19^{+0.16}_{-0.14}$ & $0.05^{+0.23}_{-0.17}$ & $0.40^{+0.22}_{-0.23}$ & $0.004^{+0.30}_{-0.17}$ \\
    Distance$/\mathrm{kpc}$ & $29.56^{+6.62}_{-5.71}$ & $18.23^{+3.54}_{-2.71}$ & $18.88^{+5.38}_{-4.17}$ & $34.15^{+5.75}_{-7.66}$ & $33.04^{+5.39}_{-3.81}$ & $31.72^{+7.12}_{-10.89}$ \\
    Eccentricity & $0.56^{+0.07}_{-0.06}$ & $0.12^{+0.09}_{-0.06}$ & $0.92^{+0.05}_{-0.08}$ & $0.51^{+0.13}_{-0.16}$ & $0.46^{+0.09}_{-0.08}$ & $0.49^{+0.02}_{-0.04}$ \\
    $z_\mathrm{max}/\mathrm{kpc}$ & $51.63^{+19.26}_{-12.55}$ & $5.37^{+2.86}_{-1.48}$ & $16.46^{+5.74}_{-4.86}$ & $33.58^{+5.84}_{-5.85}$ & $28.69^{+5.25}_{-4.76}$ & $35.14^{+5.95}_{-9.68}$ \\ 
    $r_\mathrm{peri}$/kpc & $15.83^{+5.44}_{-4.24}$ & $15.14^{+2.65}_{-2.61}$ & $0.87^{+0.92}_{-0.54}$ & $11.50^{+3.34}_{-3.29}$ & $12.66^{+1.57}_{-1.75}$ & $17.60^{+2.36}_{-3.20}$ \\
    $r_\mathrm{apo}$/kpc & $54.68^{+18.28}_{-11.06}$ & $19.38^{+3.50}_{-2.63}$ & $20.91^{+5.12}_{-4.69}$ & $34.97^{+5.31}_{-5.12}$ & $34.07^{+4.86}_{-3.81}$ & $52.63^{+7.05}_{-14.79}$ \\
\enddata
\end{deluxetable*}

\subsection{HDBSCAN} \label{hdbscan_intro}

\hdbscan, an open-source unsupervised machine learning algorithm \citep{hdbscan_2013,hdbscan_McInnes2017}, is widely used to identify structures in multidimensional parameter spaces \citep{Hunt2021_hdbscan,Brauer2022,Shank2022_hdbscan,Shank2022_b,Shank2023_hdbscan}. The main benefit of an unsupervised algorithm such as \hdbscan{} is that it does not require any apriori information regarding structure in the underlying dataset, such as the number, location, shape or other characteristics of the clusters. The identification of clusters is based solely on the density of points\footnote{Further information about \hdbscan{}, including comparisons to other supervised and unsupervised clustering algorithms, can be found at the following URL: \url{https://hdbscan.readthedocs.io/en/latest/advanced_hdbscan.html}}.

\hdbscan{} measures the density of points in a given dataset and creates a density-based clustering hierarchy tree. This tree comprises clusters with different shapes and sizes. \hdbscan{} has two primary adjustable parameters, \texttt{min\_cluster\_size} and \texttt{min\_samples}, and one potentially important secondary parameter for our purposes, \texttt{cluster\_selection\_method}.

The most significant parameter is \texttt{min\_cluster\_size}, which defines the minimum number of points that will be considered as a cluster. The \texttt{min\_samples} parameter determines the density of the clusters: it specifies the minimum number of neighboring points around a core point required for that point to be considered part of the cluster. A larger value of \texttt{min\_samples} corresponds to a more conservative algorithm, because a greater proportion of points are considered `noise' rather than being associated to a cluster. In order to identify smaller clusters, it is recommended that \texttt{min\_cluster\_size} and \texttt{min\_samples} parameters be set to a small number \citep{hdbscan_2013}. 

In the context of the search for substructure in the stellar halo, it is not immediately obvious how to select the combination of these two parameters for best effect. Empirically, we find (as expected from previous works) that our sample is overwhelmingly dominated by a small number of large clusters. However, it may also contain smaller structures which are also of significant interest. Parameters that favor the recovery of structures with very few members may also artificially fragment the larger structures. This suggests that the optimum approach may involve a `hierarchical' or iterative aspect, significantly more complicated than a single application of the \hdbscan{} algorithm \citep{Lovdal22_hdbscan,Ou2023_hdbscan,Kim2025}.
 
Since our primary aim is to quantify the overall structure of the halo, we first attempt to identify the most prominent clusters (e.g. Sagittarius, GSE, Aleph) without fragmenting them, and also without over-merging. Here we are guided by existing knowledge of the major substructures as in Fig.~\ref{fig:fig4_summary}(a). For this purpose we set \texttt{min\_cluster\_size = 500} and \texttt{min\_samples = 100}. We refer to this as the \textit{first pass} of \hdbscan{}. Although a large \texttt{min\_cluster\_size} prevents the fragmentation of large clusters, it also precludes the identification of real features with only a small number of members in our catalog. Therefore, we apply \hdbscan{} for a second time to the stars that are not assigned to any clusters in this first pass, adjusting the values of these parameters in order to recover smaller clusters. Our final assignment of stars to clusters is obtained by combining these successive passes. We describe this procedure in more detail in Section~\ref{clustering_process}.

The \texttt{cluster\_selection\_method} parameter determines the manner in which the algorithm selects clusters from the clustering hierarchy tree. The default is the Excess of Mass (\texttt{eom}) method, which identifies the most stable and condensed clusters within the tree. To identify finer, more homogeneous clusters, the algorithm offers an option called \texttt{leaf}, which selects clusters from the leaf nodes of the tree. Since our objective is to identify as many structures as possible, we use the \texttt{leaf} method for all passes, rather than the default \texttt{eom} method.

\begin{figure*}[ht!]
    \centering
    \includegraphics[width=\linewidth]{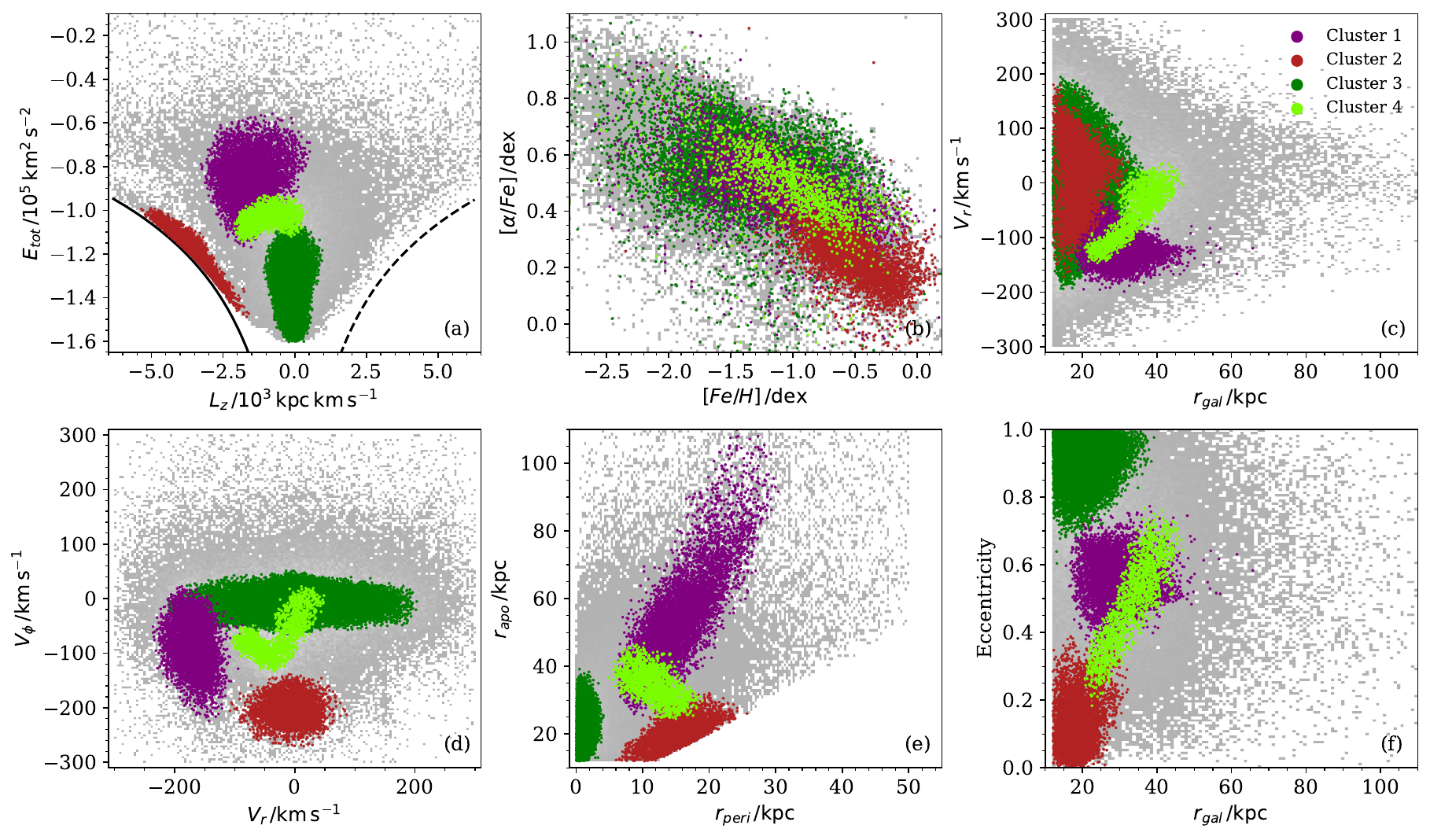}
    \caption{Results of the first pass of \hdbscan{} algorithm. The gray background points show the whole halo K-giant sample and the markers of different colors show the four clusters identified by \hdbscan. From top left to bottom right, panels show the distribution of stars in spaces defined by (a) energy and angular momentum (IoM space). The black solid line at $L_z < 0$ represents the track of Milky Way's disk while the black dashed line represents the retrograde track; (b) [Fe/H] and [$\alpha$/Fe] (chemical space); (c) Galactocentric radial velocity and distance; (d) $V_r$ and $V_\phi$ (velocity in Galactocentric spherical coordinates); (e) orbital pericenter and apocenter; and (f) eccentricity and Galactocentric distance.} 
    \label{fig:fig6_hdbscan_summary1}
\end{figure*}

The feature space in which we apply \hdbscan{} has 6 dimensions, comprising `integrals of motion' (IoM; $E_\mathrm{tot}, L_z$, and $L_{\perp}$\footnote{$L_\perp$ = $\sqrt{L_{x}^2 + L_{y}^2}$, i.e. the angular momentum in the plane of the Galactic disk.}) and velocity components in spherical coordinates ($V_r, V_{\phi}, V_z$). Previous works \citep[e.g.][]{Lovdal22_hdbscan, Ou2023_hdbscan,Kim2025} have applied \hdbscan{} to velocity and IoM spaces separately, in the context of the local volume accessible to dense sampling with Gaia. Empirically, we find combining the velocity dimensions  with the IoM dimensions in a single clustering pass improves the recovery of large-scale clusters in our sample, relative to clustering in IoM dimensions alone.

Note that we do not use $\mathrm{[Fe/H]}$ (or $\mathrm{[\alpha/Fe]}$) as clustering features. This is because, in principle, a single halo progenitor can have a wide range of metallicity, and the metallicity distributions of different progenitors are likely to have significant overlap. In other words, for progenitors with extended star formation histories, we do not expect strong clustering in metallicity. Including metallicity as a clustering feature has the effect of concentrating the resulting clusters around their respective metallicity distribution peaks. At least in the first instance, this is not desirable. We will consider the role of metallicity further below.

Prior to running \hdbscan{}, we standardize each of the 6 features with the \texttt{RobustScaler} function from the \texttt{scikit\_learn} library \citep{scikit_learn}. This pre-processing step is essential due to the significantly different numerical ranges of the features. \texttt{RobustScaler} standardizes the data by subtracting the median and scaling to the inter-quartile range (25th to 75th percentiles), thereby mitigating the impact of outliers. 

\subsection{Successive HDBSCAN passes} \label{clustering_process}

Fig.~\ref{fig:fig6_hdbscan_summary1} presents the results of the first pass of \hdbscan{} on the DESI Y3 dataset, using the procedure and parameters described above. We show the distribution of stars in IoM, chemical and kinematic spaces\footnote{Note that there are no new independent dimensions shown in this figure. The Galactocentric distance was used to obtain the stellar orbits, and eccentricity and peri/apocenter distances are determined by the integrals of motion.}. In each panel of this figure, points in gray show the distribution of all stars in our K-giant sample, and points of different colors indicate members of each \hdbscan{} cluster. In the first pass we find four major clusters. We refer to these as Clusters 1 to 4. Each can be readily identified with a well-known structure, as described in Section \ref{hdbscan results}. The distinction between these major clusters is clearest in energy -- angular momentum ($E-L_z$) space. 

Although there are many retrograde stars in our sample, we find no retrograde clusters in this first \hdbscan{} pass. To explore the apparent absence of retrograde clusters, and to attempt to recover smaller clusters, we ran a second pass of \hdbscan{}, applied only to the \textit{unclustered} data from the first pass. In this second pass, we reduced each parameter to 20\% of its value in the first pass: \texttt{[min\_cluster\_size, min\_samples]} $= [100,20]$. In the second pass, we identified nine clusters. By construction, these are smaller than those identified in the first pass. A large fraction of stars in the new clusters are within regions of feature space dominated by the first pass clusters. As described in Section~\ref{sec:combining_passes}, we consider all of these clusters to correspond to less dense regions of the first pass clusters.

Even in the second pass no cluster was identified in the retrograde halo. One plausible reason for this could be the lack of coherence of these structures in velocity spaces. Previous claims of retrograde structures have identified them predominantly in integral-of-motion (or angle-action) spaces \citep{Koppelman2019a,Myeong2019,Naidu2020}.

The second pass also did not identify any structure among the most metal-poor stars in our catalog ($\mathrm{[Fe/H] \lesssim -2}$). To examine this further, we ran an alternative \textit{metal-poor second pass}, again starting with the unclustered data from the first pass, but this time restricted to stars with $\mathrm{[Fe/H] < -2}$. We use the same parameters as the fiducial second pass (\texttt{[min\_cluster\_size, min\_samples]} $= [100,20]$). \hdbscan{} identified two additional clusters in this pass, of which we consider one to be unrelated to any cluster identified in the first pass (see Section~\ref{sec:combining_passes}). We provide further details of the fiducial and metal-poor second passes in Appendix \ref{sequential clustering}. 

Finally, to test the limits of this approach for the recovery of very faint structures, we ran \hdbscan{} a fourth time, this time on stars that were not assigned to clusters in either the first or second passes. We further reduced the parameters to 20\% of their values in the second pass: \texttt{[min\_cluster\_size, min\_samples]} $= [50,10]$. This pass identified 15 clusters, most of which are concentrated in the low-density regions of first pass clusters We did not combine these very small clusters with those identified in the second and metal-poor second passes, although we did identify one previously known structure in this fourth pass. We refer to this as ``Cluster B''. Further details are given in section \ref{additional structures}.

\subsection{Combining passes}
\label{sec:combining_passes}

In order to construct our final list of \hdbscan{} clusters and their K-giant members, we combine the smaller clusters identified in the second and metal-poor second passes with the major clusters identified in the first pass. To guide our decisions regarding the association between these clusters, we first make a quantitative estimate of the most likely first-pass `parent'. We compute the centroid of each cluster in feature space, and, for each second pass cluster, find the nearest centroid of a first-pass cluster. We then adjust these associations by hand, based on our visual assessment of the morphology of the feature in each space. Although this manual method is effective for our approach to sequential clustering, it is subjective and does not scale to large numbers of features. An automated approach to sequential clustering with \hdbscan{} would be worthwhile \citep[similar issues are explored e.g.\ by][]{Lovdal22_hdbscan}.

Through this procedure, we associate all but one second pass cluster with a first pass cluster. The exception is one of the two metal-poor second pass clusters, which we believe corresponds to a previously known feature, Cetus-Palca. The details of this cluster are given in section~\ref{hdbscan results}. We have confirmed this association using data from a separate study of likely Cetus-Palca stars in DESI \citep[G. Thomas, private communication;][]{GFT22_Cetus_Palca,Zhen2022_cetus}. We refer to this as ``Cluster A'' in the remainder of the paper.

\section{Results} \label{results}

In this section we describe the final outcome of the multiple-pass clustering process described in the previous section.

\subsection{Structures identified by \hdbscan{}} \label{hdbscan results}

\begin{figure*}
    \centering
    \includegraphics[width=\linewidth]{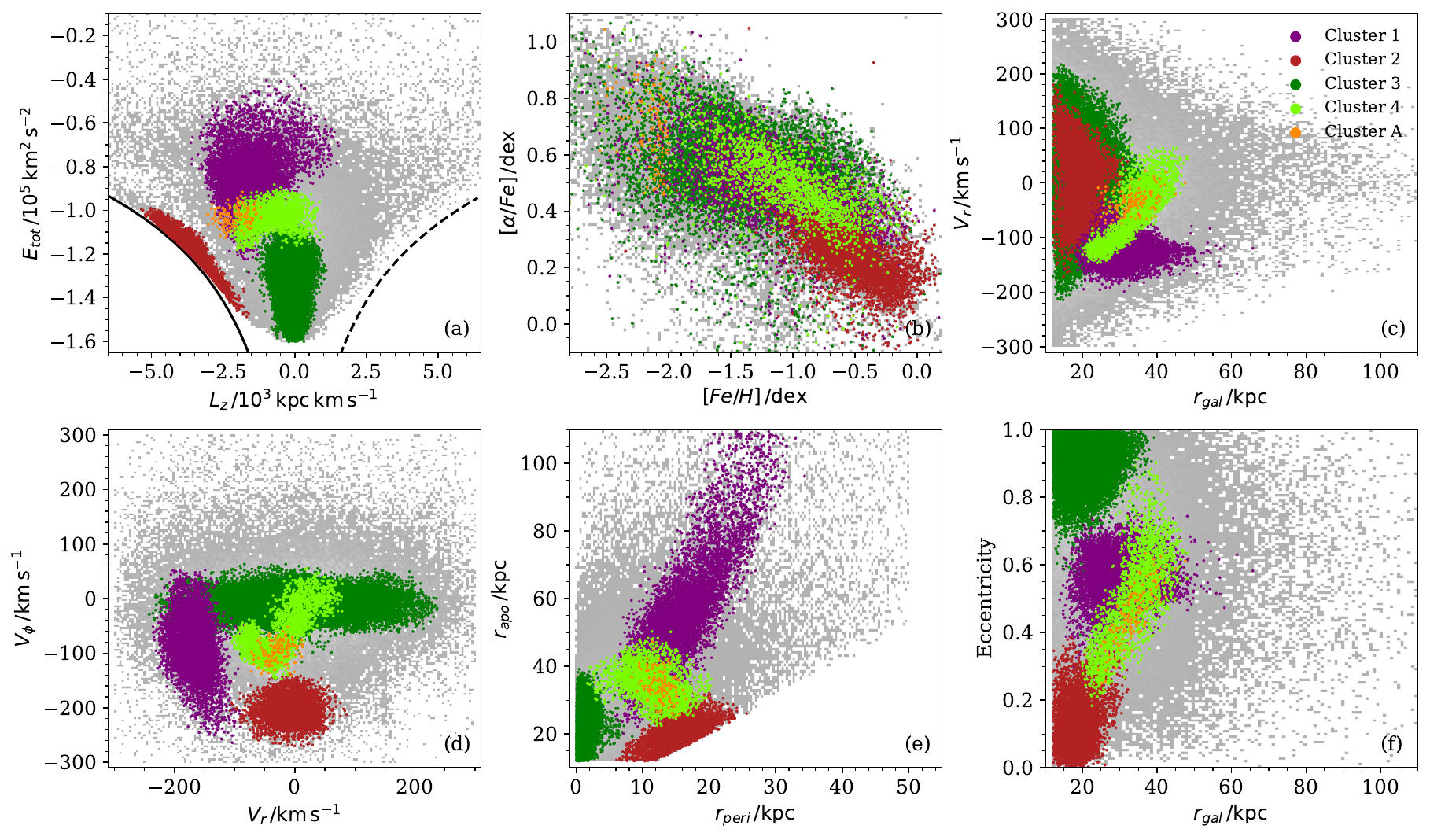}
    \caption{Summary of the \hdbscan{} passes. The gray background points show the whole halo K-giant sample and the markers of different colors show the five clusters identified by \hdbscan. From top left to bottom right, panels show the distribution of stars in spaces defined by (a) energy and angular momentum (IoM space). The black solid (dashed) line at $L_z < 0$ represents the track of prograde (retrograde) circular orbits; (b) [Fe/H] and [$\alpha$/Fe] (chemical space); (c) Galactocentric radial velocity and distance; (d) $V_r$ and $V_\phi$ (velocity in Galactocentric spherical coordinates); (e) orbital pericenter and apocenter; and (f) eccentricity and Galactocentric distance. The legend in Panel (c) shows the color scheme for the clusters used in all the panels and throughout the paper.}
    \label{fig:fig7_hdbscan_summary_a1}
\end{figure*}

Fig.~\ref{fig:fig7_hdbscan_summary_a1} updates Fig.~\ref{fig:fig6_hdbscan_summary1} to include all stars associated with clusters identified in the second and metal-poor second passes. It can be seen that Cluster A, the only likely new cluster found in the sequential passes, is visible as metal-poor cluster in close proximity to Cluster 4 in all spaces, with the exception of chemical space. 

As noted in section \ref{clustering_process}, all the clusters exhibit distinct characteristics in the energy-angular momentum ($E_{tot} - L_z$) space. We first label and describe the clusters independently of their possible associations, and then consider evidence for their relationship to known features in separate subsections.

In [$\alpha$/Fe] - [Fe/H] space (which, we do not use as features for the \hdbscan{} clustering) it is evident that our sample overall has low metallicity and high alpha abundance. Clusters 1, 4 and A overlap each other in a broad region around the ridgeline in this space; they also overlap with Cluster 3, which has a very broad range of metallicity. Cluster 2 occupies a distinct low-$\alpha$ region at the high metallicity end of the distribution.

In the distribution of Galactocentric radial velocity with distance, we see a stratification in distance corresponding approximately to that in total energy. All clusters are distributed between 12 and 65 kpc, although our full sample extends considerably further.

Panel (d) presents the distribution in Galactocentric $V_r - V_\phi$ space. Cluster 2 stars with near-circular orbits occupy the region in this plot around $V_\phi \sim -200$ and low $V_r$. As evident in previous figures, this cluster accounts for a large fraction of such stars in our sample. Conversely, Cluster 3 is characterized by highly radial orbits. The other clusters fall somewhere in between. There are signs of substructures in this space that are not associated with clusters, for example at $(V_r,V_\phi) \sim (-150\,\kms, -200\,\kms)$.   

In the distribution of pericenters and apocenters, most clusters exhibit large apocenters, suggesting an accreted origin. These clusters have narrow distributions in eccentricity, thereby distinguishing them from one another.

\begin{figure*}
    \centering    
    \includegraphics[width=0.8\linewidth]{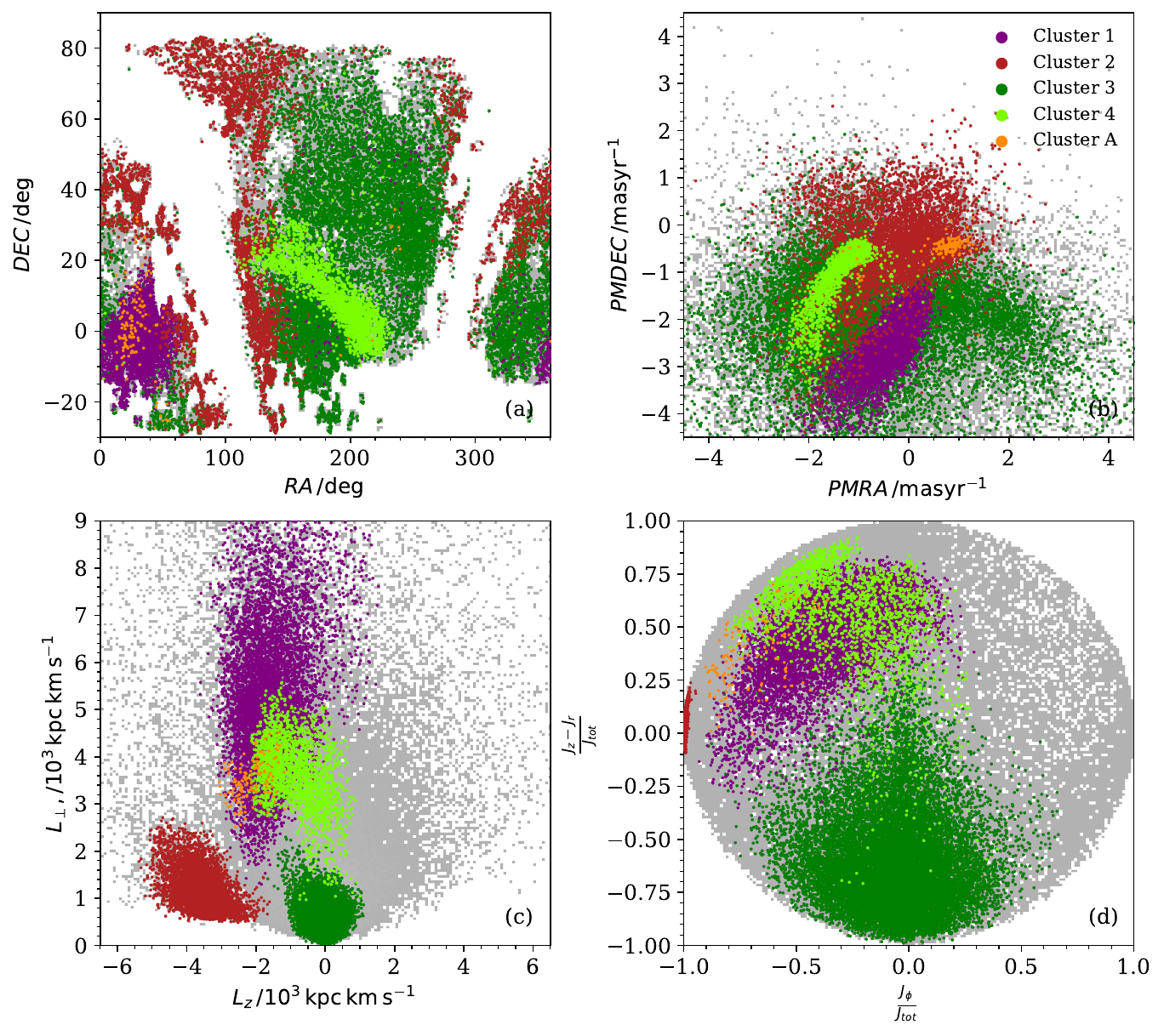}
    \caption{Second summary of the \hdbscan{} results. From top left to bottom right, panels show the distribution of stars: (a) in equatorial coordinates (coherent structures and concentrations around the Galactic plane and Galactic center are readily apparent) (b) in proper motions; (c) in an alternative angular momentum space of $L_z$ and $L_\perp$; (d) in action space. The legend in panel (b) gives the color scheme for the clusters used in all the panels and throughout the paper.}
    \label{fig:fig8_hdbscan_summary_a2}
\end{figure*}

Fig.~\ref{fig:fig8_hdbscan_summary_a2} shows four more views of the \hdbscan{} clusters. Panel (a) shows the distribution of stars in the DESI footprint. This is particularly important because clustering in the plane of the sky is not implied in any trivial way by clustering in the spaces used for our \hdbscan{} analysis. It is evident that some \hdbscan{} clusters are spread across the sky (e.g.\ Cluster 3), others  are clearly concentrated into spatially coherent features. From this projection of the data we can readily identify the association between most of the clusters and known features, which we will elaborate on below.

Panel (b) shows the distribution of stars in proper motion space. These are the fundamental kinematic observables (together with the DESI heliocentric radial velocities) that underlies the derived dynamical quantities such as energy and angular momentum.

The remaining spaces are effectively permutations of those already discussed. Panel (c) shows $L_{z}$ vs.\ $L_{\perp}$. $L_\perp$, the angular momentum perpendicular to the axis of rotation, helps to differentiate between structures with radial and polar orbits. Panel (d) shows a space of actions scaled by $J_{tot}^2 = J_r^2 + J_{\phi}^2 + J_z^2$. The horizontal axis indicates the contribution of angular momentum to the total orbital motion; stars on prograde orbits have negative values of $J_{\phi}/J_{tot}$. The vertical axis, $(J_z - J_r)/J_{tot}$, quantifies the significance of vertical oscillations in comparison to radial oscillations. Stars exhibiting a greater degree of radial oscillation have negative values of this quantity, while those with more pronounced vertical oscillation have positive values. 

The following subsections consider the properties of the clusters in more detail and their association with known features. For this discussion we group clusters with related properties. Since the clusters provide only a partial view of the progenitor systems that we associate with them, and because those associations are themselves not always clear-cut, we continue to refer to the \hdbscan{} clusters using the simple designations in Figs~\ref{fig:fig7_hdbscan_summary_a1} and \ref{fig:fig8_hdbscan_summary_a2}. Table \ref{tab:cluster_details1} gives a summary of the properties of clusters identified in this work.

\subsubsection{Cluster 2} \label{aleph}

\begin{figure*}
    \centering
    \includegraphics[width=\textwidth]{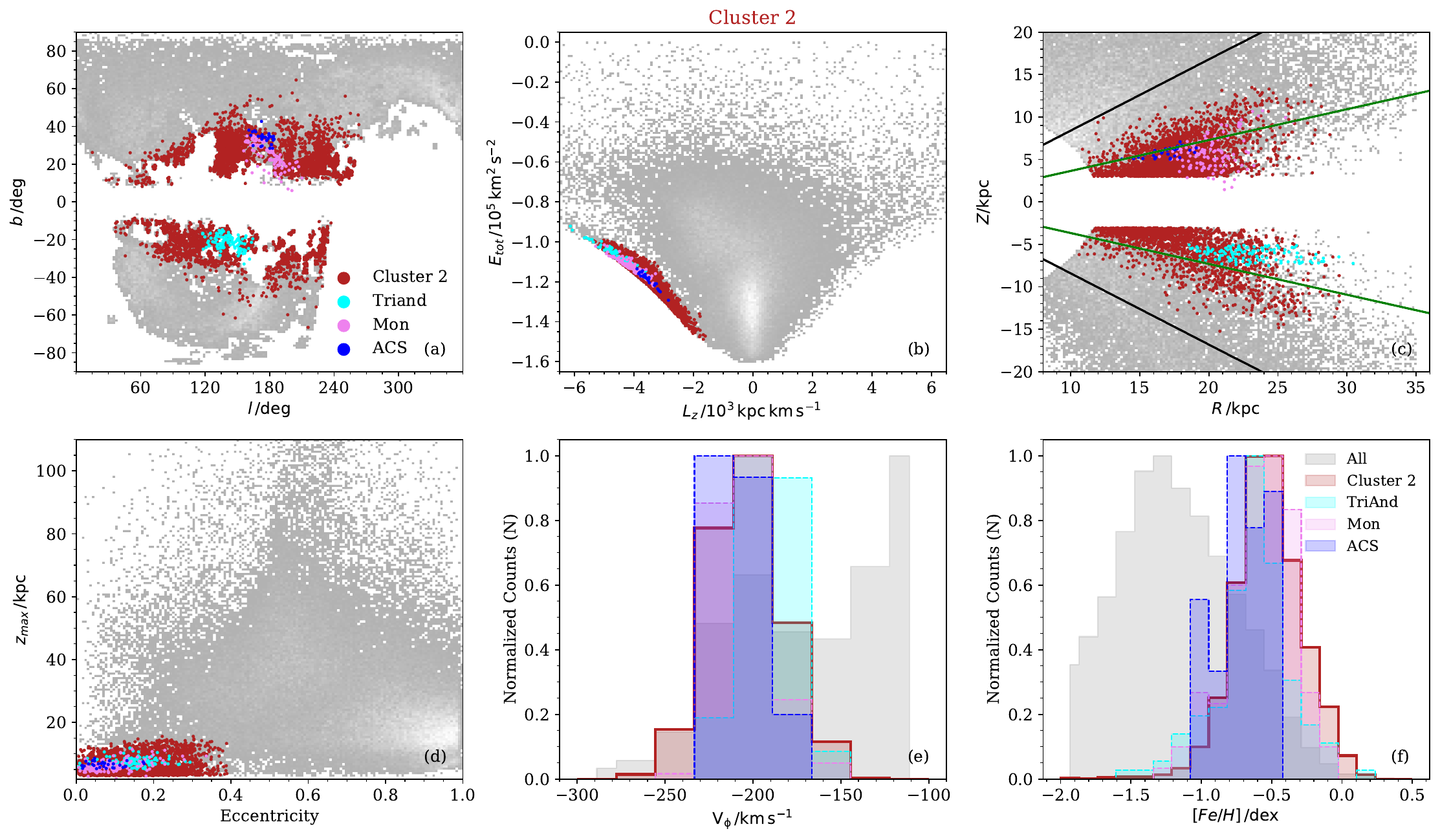}
    \caption{Summary plots for Cluster 2, associated with Aleph. Panels show distributions in (a) Galactic coordinates $(l,b)$; (b) $E_{tot} - L_z$; (c) cylindrical coordinates $(R,Z)$; (d) eccentricity and vertical excursion, $z_{max}$; (e) tangential velocity, $V_\phi$; and (f) [Fe/H]. The grayscale distribution in the background of each panel shows the whole halo K-giant sample and the maroon points show stars in Cluster 2. Lines in panel (c) mark Galactic latitudes $|b|=20$~deg (green) and $b=40$~deg (black). The cyan, pink and blue points on all the panels show the positions of possible TriAnd, Monoceros ring and Anticenter stream members, respectively, taken from \protect\citet{Zhang2022_Acs_mr_tri}.}
    \label{fig:fig9_cluster_2}
\end{figure*}

In Figs.~\ref{fig:fig7_hdbscan_summary_a1} and \ref{fig:fig8_hdbscan_summary_a2}, Cluster 2 (marked with maroon points) is clearly distinguished from the other clusters, primarily by its near-circular orbits and concentration towards the Galactic disk. The properties of this cluster correspond very closely in most respects to the \textit{Aleph} feature identified in the H3 survey by \citet{Naidu2020}. Fig.~\ref{fig:fig9_cluster_2} presents additional information on these stars. We find a total of 5,566 stars in this \hdbscan{} cluster (combining all three passes). They are concentrated in galactic longitude towards the direction of the Galactic anticenter (Fig.~\ref{fig:fig8_hdbscan_summary_a2}a). The apparent asymmetry about the plane (more stars at $b>0$) may be due in part to the DESI footprint. In $(E,L_z)$ space (Fig.~\ref{fig:fig7_hdbscan_summary_a1}a, \ref{fig:fig9_cluster_2}b), these stars extend to high energy (comparable to the highest energy GSE stars) and have high prograde rotation. They follow a narrow track resembling that of an eccentric disk. Fig.~\ref{fig:fig7_hdbscan_summary_a1}f shows a low median eccentricity of $0.12$. 
The disk-like nature of this feature is also apparent in $V_r - V_{\phi}$ space (Fig.~\ref{fig:fig7_hdbscan_summary_a1}d), with $V_{\phi}$ peaking at $\sim -210\,\kms$ and low $V_r$.

The cluster exhibits higher metallicity (median $\mathrm{[Fe/H]}=-0.53$) and much lower alpha enhancement ($\mathrm{[\alpha/Fe]}\simeq0.24$) compared to other clusters and the stellar halo overall (Fig.~\ref{fig:fig7_hdbscan_summary_a1}b). These values are very close to those reported for Aleph by \citet{Naidu2020}. We find a clear negative metallicity gradient for this structure, from $-0.47$ dex at 10 kpc to $-0.66$ dex at 30 kpc.

The Galactocentric location and extent of this structure in our K-giant sample differ somewhat from those reported by \citet{Naidu2020}, who found Aleph members at distances of $9.6\lesssim r_\mathrm{gal} \lesssim 16.8$~kpc, with a maximum extent of $r_\mathrm{gal}\sim25\,\mathrm{kpc}$ (the lower limit was imposed on their sample). Our Cluster 2 spans $12\lesssim r_\mathrm{gal} \lesssim 32$~kpc. In \citet{Naidu2020}, Aleph members are shown to lie along the nominal lower latitude boundary of the H3 survey at $|b|>40\deg$.\footnote{ Figs.~\ref{fig:fig7_hdbscan_summary_a1}a and ~\ref{fig:fig7_hdbscan_summary_a1}c show that we find no Aleph stars at all at $|b|>40\deg$. The discrepancy is likely because the H3 data used by \citet{Naidu2020} included a handful of tiles at lower latitude. Later H3 papers quote $|b|>30\deg$ as the survey limit, consistent with `grazing the edge' of the \hdbscan{} cluster we associated with their Aleph feature.} Our catalog has good coverage over a wide range of longitude for $|b|>20\deg$, and patchy coverage at even lower latitude.

This cluster (in common with Aleph) overlaps several well-studied kinematic substructures around the anticenter: the Triangulum-Andromeda stream (TriAnd), the Monoceros (Mon) stream/ring and the Anticenter stream (ACS). These metal-rich, low-latitude, co-rotating structures are believed to be excitations of the outer disk, possibly induced by the passage of the Sagittarius dwarf \citep{Purcell2011,Laporte2019,Borbolato2024,Qiao2024,Mika26_acs_mr}. \citet{Horta2023} suggest Aleph may be a mixture of "high-$\alpha$" and "low-$\alpha$" disk stars, with qualitative similarity in chemical abundances to ACS. In Fig.~\ref{fig:fig9_cluster_2} we highlight candidate members of these features based on SDSS and LAMOST data from \citet{Zhang2022_Acs_mr_tri}. For this comparison, we take [Fe/H], $r_\mathrm{helio}$ and $V_\mathrm{helio}$ directly from their catalog, and recompute the orbital information using the same solar position, standard of rest and gravitational potential as our catalog. These three structures overlap with our \hdbscan{} cluster, and have similar energies, suggesting they may be overdensities within the more diffuse feature we have identified.

\subsubsection{Clusters 1, 4}
\label{sec:sgr}

\begin{figure*}
    \centering
    \includegraphics[width=\linewidth]{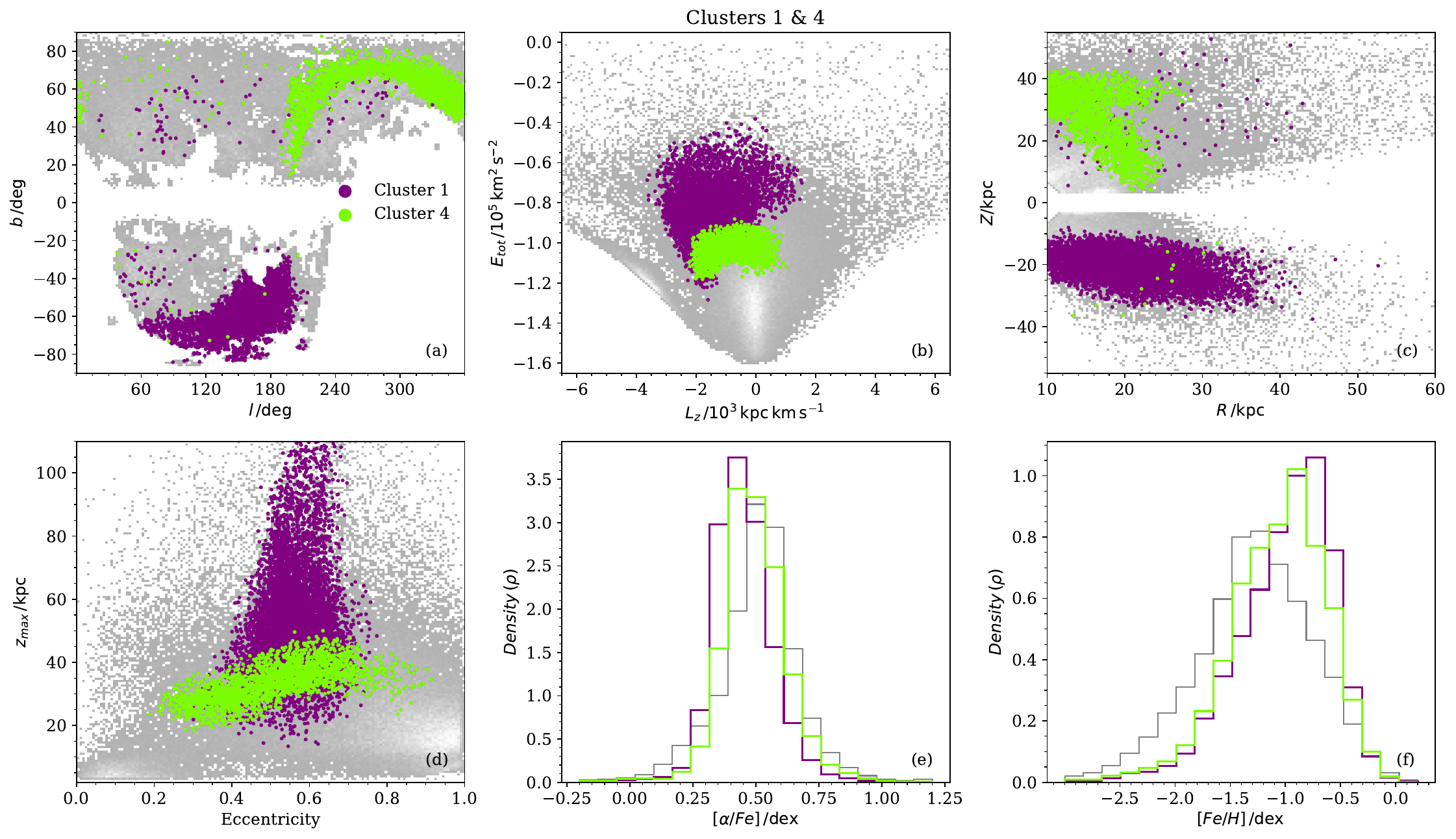}
    \caption{Summary plots for Clusters 1 and 4, associated respectively with the southern and northern Galactic hemisphere sections of the Sagittarius stream. Panels show distributions in (a) Galactic coordinates $(l,b)$; (b) $E_{tot} - L_z$; (c) cylindrical coordinates $(R,Z)$; (d) eccentricity and vertical excursion, $z_{max}$; (e) alpha abundance, [$\alpha/Fe$]; and (f) [Fe/H]. The grayscale distribution in the background of each panel shows the whole halo K-giant sample, the purple points show stars in Cluster 1, and the light green points show stars in Cluster 4.}
    \label{fig:fig10_clusters_1_4}
\end{figure*}

Clusters 1 and 4 identified by \hdbscan{} are visibly components of the Sagittarius stream \citep{Ibata94_Sgr,Law_majewski_2005,Law_Majewski2010,Koposov12_Sgr,Belokurov14_Sgr,Ramos22_Sgr,Cunningham24_Sgr}. As shown clearly by their location on the sky (Fig.~\ref{fig:fig8_hdbscan_summary_a2}a), Cluster 1 (purple points) corresponds to the section of the stream in the southern Galactic hemisphere (Sgr-S) and Cluster 4 (light green points) to the section in the northern hemisphere (Sgr-N). Although we do not use sky position as one of the clustering dimensions for \hdbscan{}, we do use velocity features ($V_r$, $V_\phi$, $V_z$). The gap in our velocity space due to the galactic plane leads \hdbscan{} to identify the northern (leading arm) and southern (trailing arm) sections of the stream as two separate clusters.

We identify 7,270 stars as members of Cluster 1, and 2,252 stars as members of Cluster 4. In line with expectations for Sagittarius, both clusters have prograde rotation at relatively high energy, and a wide range of metallicity ($\mathrm{-3<[Fe/H]} < 0.6$, with a median of $-0.98$). Both clusters are $\alpha$-enhanced, with median $\mathrm{[\alpha/Fe]}= 0.46$. 

The two clusters have different velocity and distance distributions, reflecting the complex orbit of the Sagittarius debris. The stars in Cluster 4 (Sgr-N) are deeper in the Galactic potential, have a wide range of eccentricities, and follow a (partial) `chevron' track in $(r_\mathrm{gal}, V_{r})$ that extends from $\sim20$~kpc to their approximate apocenter at $\sim45$~kpc. In contrast, the stars in cluster 1 are more weakly bound, and have approximately constant $V_{r}$ and eccentricity $\sim0.6$. Their apocenters extend beyond $\sim100$~kpc.
 
Fig.~\ref{fig:fig10_clusters_1_4} shows additional distributions for these two clusters. Although the Sagittarius stream itself has no gaps, it is apparent from the distribution of the two clusters in Galactic $(l,b)$ coordinates (Fig.~\ref{fig:fig10_clusters_1_4}a) and Galactocentric cylindrical coordinates (Fig.~\ref{fig:fig10_clusters_1_4}c) that some parts of the stream are not detected by \hdbscan{} even though they are within our footprint. We now consider why this is the case.

\subsubsection{Comparison with Sagittarius model} \label{sag model}

\begin{figure*}
    \centering
    \includegraphics[width=\linewidth]{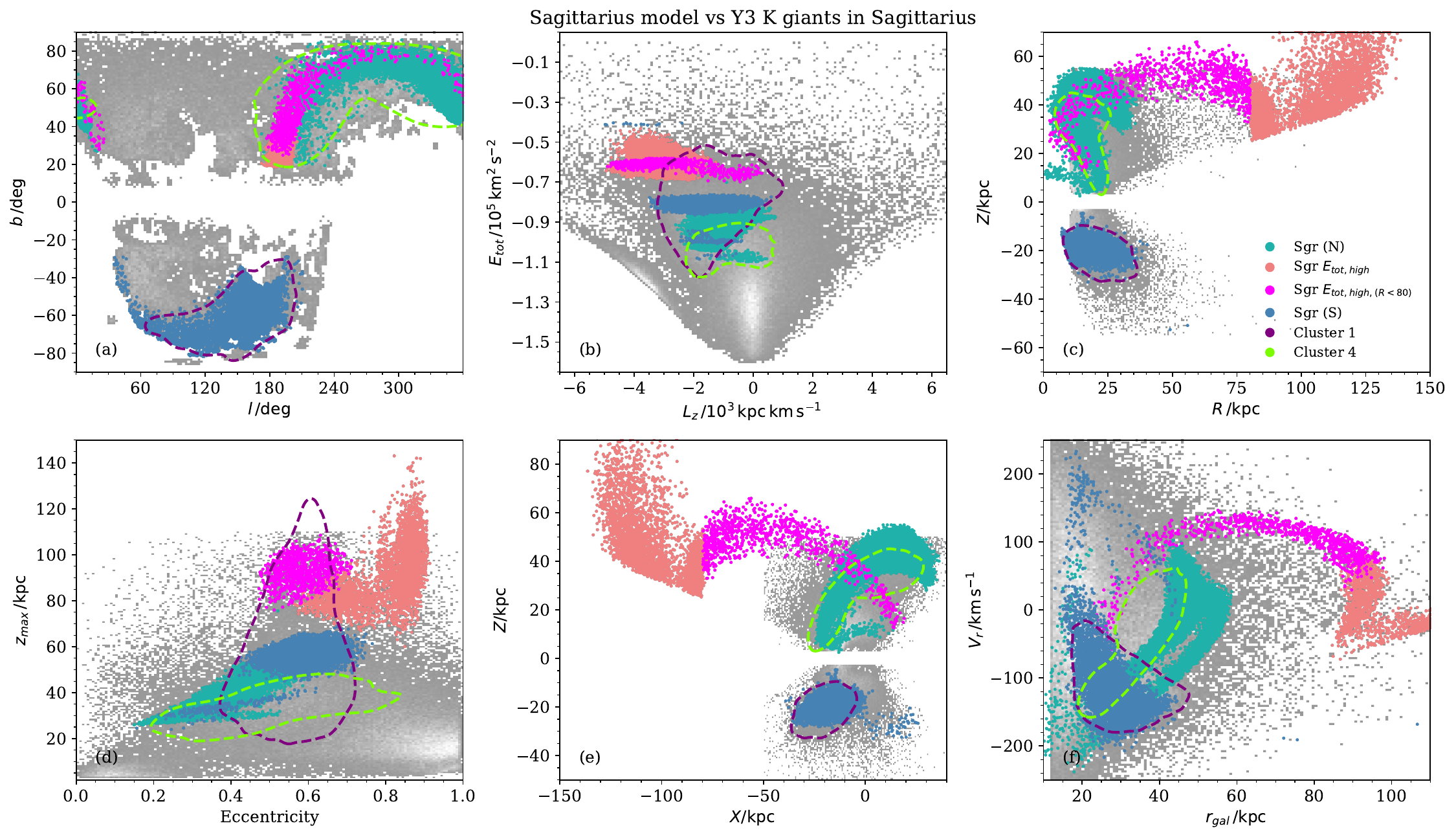}
    \caption{Clusters identified as Sagittarius by \hdbscan{} compared to the Sagittarius model from \protect\cite{Vasiliev21_sag}. Grey points show the whole halo K-giant sample. Our \hdbscan{} clusters 1 and 4 are shown by purple and green contours. The Sagittarius model stars within the DESI footprint are shown by sea green points in the northern hemisphere and dark blue points in the southern hemisphere. We divide the higher-energy wrap (which is only visible in the north) at a cylindrical radius $R>80$~kpc: less distant stars are shown in magenta, and more distant stars in coral pink. The individual panels show (a) Galactic coordinates $(l,b)$; (b) $E_{tot} - L_z$; (c) cylindrical coordinates $(R,Z)$; (d) eccentricity and vertical excursion $z_{max}$; (e) Galactocentric Cartesian coordinates $(X,Z)$; and (f) Galactocentric radial velocity and distance. }
    \label{fig:fig11_sagmodel}
\end{figure*}

Fig.~\ref{fig:fig11_sagmodel} shows a comparison between a Sagittarius model from \citet[][their model including a $1.5 \times 10^ {11} \msol$ LMC]{Vasiliev21_sag} and the clusters identified by \hdbscan{}. We show only stars from the model within the MWS footprint. The model serves only to highlight the regions of these diagrams we expect to be occupied by stars from Sagittarius. We do not assume a distribution of model magnitudes, so the apparent density of model points in these figures cannot be compared with the density of K-giants in our data, and the fact that our selection recovers fewer stars at larger distances is not accounted for.

Fig.~\ref{fig:fig11_sagmodel}a compares the distribution of model stars in Galactic $l$ and $b$ coordinates to the clusters identified by \hdbscan{}. The distribution of Clusters 1 and 4 clearly follow the track of the stream in the model. We also see clearly those regions where the model appears in the footprint but no cluster members are identified by \hdbscan{}. The most obvious omissions are in the southern hemisphere at $l<60^\circ$ (dark blue model points), in the northern hemisphere around $l\sim180^\circ$ (coral pink and magenta model points) and in the northern hemisphere at $l < 60^\circ$ (sea green model points).

Fig.~\ref{fig:fig11_sagmodel}b shows the distribution of the model in $(E,L_z)$ space. In the northern hemisphere, two different wraps of the model stream appear in the DESI footprint. The lower energy\footnote{In this context, all total energies have negative values. ``Lower'' energies ($E_{tot} \rightarrow -\infty$) correspond to more tightly bound stars. Conversely, ``higher'' energies ($E_{tot} \rightarrow 0$) correspond to more weakly bound stars.} component (sea green) corresponds approximately to our Cluster 4, with some notable differences. The higher-energy component (coral pink and magenta) is not recovered at all by \hdbscan{}. Fig.~\ref{fig:fig11_sagmodel}c, an extended view of the cylindrical $(R,Z)$ plane, shows that this higher energy wrap extends to radii $>50$~kpc, well beyond the two \hdbscan{} clusters that we associate with Sagittarius.

\begin{figure}
    \centering
    \includegraphics[width=\linewidth]{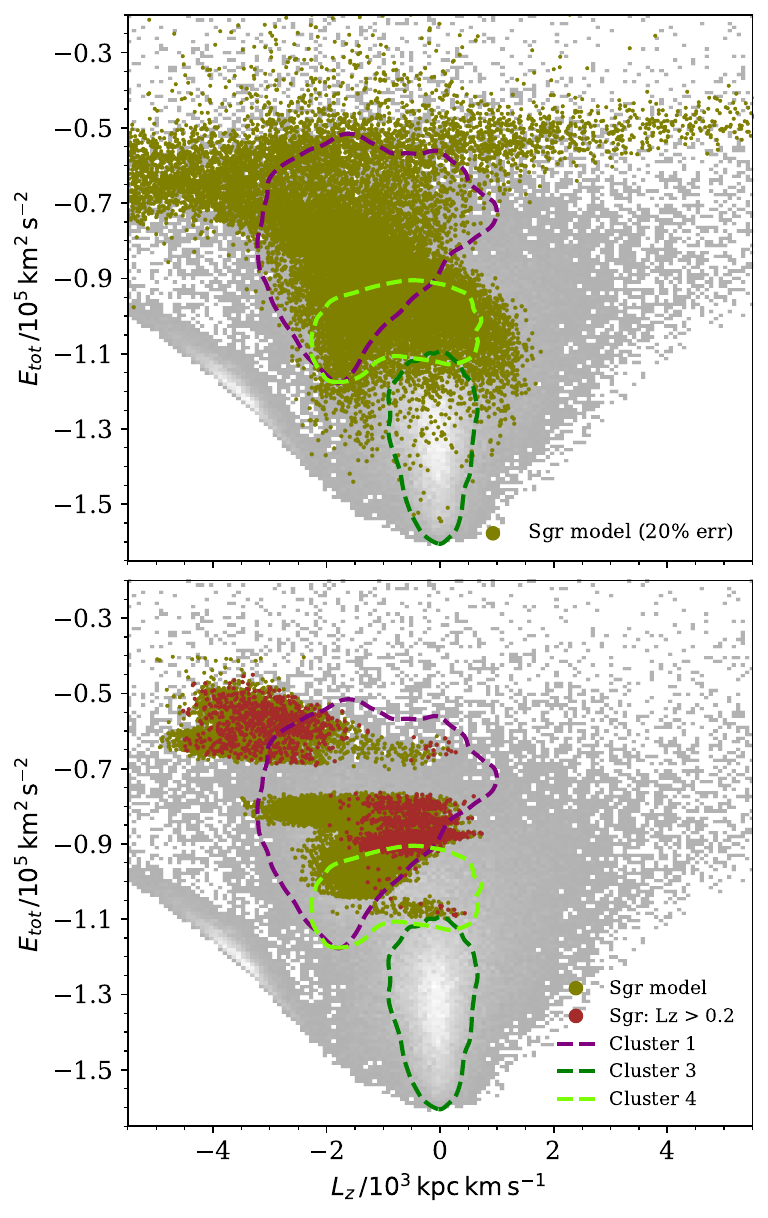}
    \caption{Distribution of Sagittarius model stars in $E_{tot} - L_z$ space. Y3 K-giant sample is shown in the gray background. The olive-colored markers shows the Sagittarius model stars. The purple and lawn green markers shows the \hdbscan{} clusters 1 and 4, associated with Sagittarius. Top panel: A fiducial Gaussian error of 20\% is added to the distances before calculating their $E_{tot}$ and $L_z$, making the model stars to scatter widely on the IoM space, even extending to the retrograde region. Bottom panel: Sagittarius model stars without error are shown with olive colored markers. The brown points shows the stars in the model that were scattered to the retrograde region ($L_z > 0.2$) when model distances were convolved with a Gaussian relative error of 20\%.}
    \label{fig:fig12_Sgr_dist_err}
\end{figure}

\begin{figure*}
    \centering
    \includegraphics[width=\linewidth]{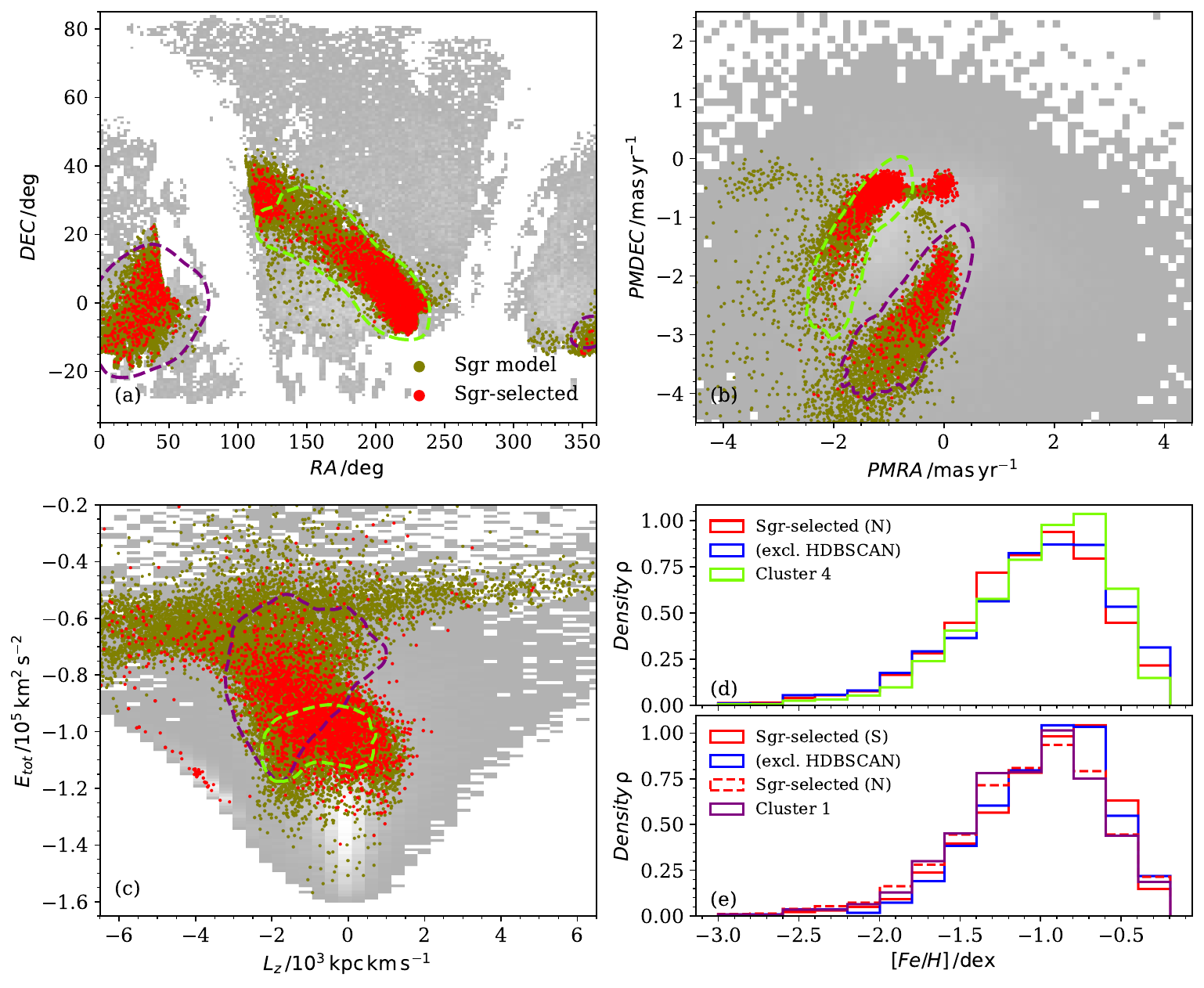}
    \caption{Distribution of K-giants in chemo-dynamic spaces compared to the Sagittarius model stars. Distribution of the whole K-giant sample is shown in the gray background. The olive colored points and purple and light green contours shows the distribution of Sagittarius model stars, K-giants from \hdbscan{} clusters 1 and 4, respectively. The red colored points represent the K-giants matched with Sagittarius model stars in spatial $(\alpha,\delta)$ and kinematic $(\mu_\alpha, \mu_\delta)$ spaces. Panels show distributions of these stars (a) on the sky, (b) in proper motion, and (c) in $E_{tot} - L_z$. The remaining panels show their metallicity distributions in the north (d) and south (e) galactic hemispheres; we compare these to the distributions for the respective \hdbscan{} Sagittarius clusters. In blue, we show the distributions of stars in the model-based selection that are \textit{not} in the corresponding \hdbscan{} cluster. The dashed red line in (e) is the same as the red line in (d), which shows the model selected stars in the northern hemisphere.}
    \label{fig:fig13_sgr_all}
\end{figure*}

We can use the model to understand how distance uncertainties contribute to the incomplete recovery of the lower-energy stream wrap by \hdbscan{}.
The upper panel of Fig.~\ref{fig:fig12_Sgr_dist_err} demonstrates how convolution with a fiducial Gaussian error of 20\% (a conservative estimate of the error on the distances we use for our catalog) redistributes the model stars in $E_{tot} - L_z$ space. The peaks of the `scattered' distribution agree well with the two Sagittarius clusters recovered by \hdbscan{}. A significant number of model stars are scattered to the region of low energy and angular momentum dominated by GSE. Moreover, $\sim$ 11\% of model stars are inferred to be on retrograde orbits after convolution with our fiducial distance error. On this evidence, it seems very likely that a simple selection of GSE stars from our catalog, based on low energy and low angular momentum, would be significantly contaminated by Sagittarius members.

The lower panel of Fig.~\ref{fig:fig12_Sgr_dist_err} shows the original distribution of Sagittarius model stars in the DESI footprint, without any distance error. The model distribution in this space is clearly very different from the \hdbscan{} clusters. We mark (in brown) those stars that are scattered by distance errors into the retrograde region ($L_z > 0.2$). For the high energy wrap, the marked stars are effectively a random sample. For the lower energy clusters, the marked stars are concentrated at lower angular momentum.\footnote{We note that the model distribution in the lower panel, without error convolution, shows an isolated peak at low energy and near-zero angular momentum. This may contribute to our Cluster 4.}

To overcome the limitations of the \hdbscan{} selection in this case, we use the \citet{Vasiliev21_sag} model itself as a prior prediction of the phase space of Sagittarius. In Fig.~\ref{fig:fig13_sgr_all}, we select K-giants neighboring model stars to within $< 1$~degree in $(\alpha,\delta)$ and $< 0.2$~mas/yr in $(\mu_\alpha,\mu_\delta)$.\footnote{An alternative method to isolate Sagittarius stars uses the $L_z$ and $L_y$ angular momentum components \citep[e.g.][]{Johnson2020_sgr,Pennarrubia2021,Chandra2026}. This method selects 21,191 stars in our catalog and accounts for a greater fraction of stars along the locus of the stream, compared to our model-based approach. However, we find the $(L_z,L_y)$ selection also selects a significant fraction of possible interlopers, far from the locus of the stream, likely because our sample extends to large distances and includes many stars with significant distance uncertainties. We prefer the model-based approach as a conservative, higher-purity selection of probable members.} These values were determined by requiring that 90\% of the model points have at least other model point as a neighbor, ensuring that we trace the stream without leaving large gaps. The top row of the figure shows that, as expected, the K-giants we select in this way trace the track of Sagittarius on the sky and in proper motion, although some regions are still not well-sampled due to incomplete sky coverage, as shown in Fig.~\ref{fig:fig3_skyplot}. This approach selects 4,876 stars (denoted as \textit{Sgr selected} in the figure). Of these, 78 were assigned to \hdbscan{} cluster 2 (Aleph); we retain the \hdbscan{} classification for these stars. We treat the remaining 4,798 stars as an alternative `model-based' Sagittarius selection; of these, 1,367 were previously assigned to one of the two \hdbscan{} Sagittarius clusters, and 3,431 were not assigned to any \hdbscan{} cluster. The lower left panel shows that this model-based Sagittarius selection corresponds well to the expected distribution of Sagittarius in $(E,L_z)$ after convolution of the model with distance errors. 

Fig.~\ref{fig:fig13_sgr_all}d compares the metallicity distribution of the model-based selection (in red) to those of the \hdbscan{} clusters. The blue histogram shows the distribution of stars in the model-based selection that are \textit{not} members of clusters 1 or 4. In the south, this distribution is very similar to that of the stars in cluster 1. This suggests that the model-based selection is indeed dominated by Sagittarius members that were not recovered by \hdbscan{}. In the north there is a noticeable difference between the \hdbscan{} and model-based MDFs, with more metal poor stars in the former. The metal-poor tail of the model-based selection is the same in the north and the south; this suggests that the difference between the \hdbscan{} and model selections in the north is the result of lower purity in the \hdbscan{} selection (i.e. a larger fraction of stars that are not true Sagittarius members). The model-based selection itself has a smaller difference from north to south: the peak of the MDF is relatively higher and sharper in the south (around [Fe/H]$\sim-0.7$). Overall, we conclude that the model-based selection of Sagittarius has both higher completeness and higher purity than the \hdbscan{} selection. In the remainder of the paper, we use the model-based selection when subtracting the contribution of Sagittarius stars to other distributions.

We note that \citet{Naidu2020} also explored the effect of distance errors on a Sagittarius-like feature (their appendix A), using a similar stream from the \citet{Bullock:2005aa} simulations rather than a model of Sagittarius specifically. Perhaps because of the specific configuration of Sagittarius, we find a larger and potentially more significant effect -- the spillover of error-convolved Sagittarius stars to the retrograde periphery of GSE could be an unexpected source of contamination for kinematically-selected GSE samples. Since Sagittarius stars have a higher metallicity than GSE, this effect may contribute to the relatively high GSE median metallicity found by \citet{Naidu2020} (see also below). 

\begin{figure*}
    \centering
    \includegraphics[width=\linewidth]{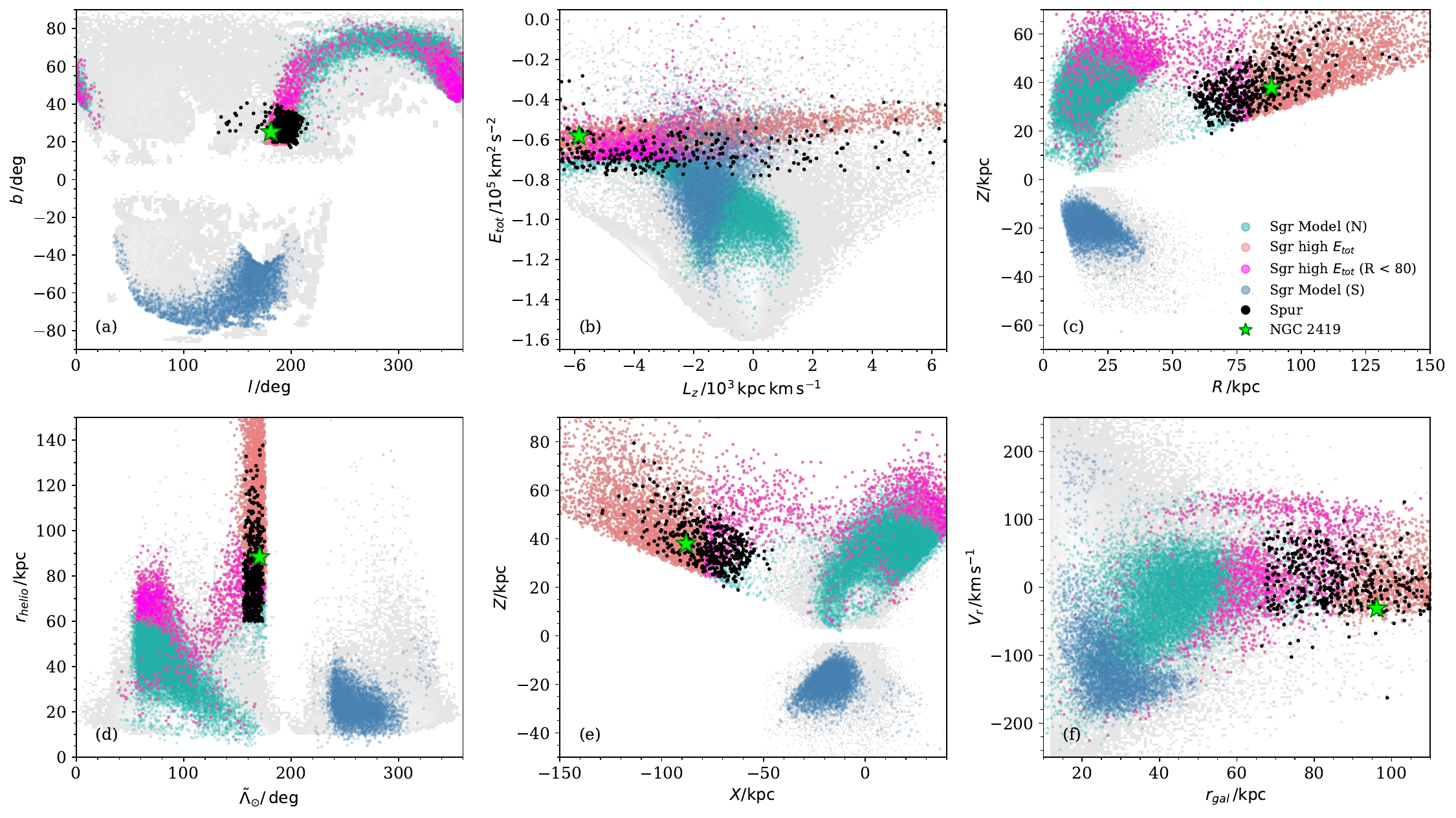}
    \caption{The distribution of stars we associate with the Sagittarius spur (black points) compared to the \citet{Vasiliev21_sag} Sagittarius model (convolved with a fiducial error of 20\% on distance as descried in the text). Other colors have the same meaning as those in Fig.~\ref{fig:fig11_sagmodel}. The green star symbol represents the globular cluster NGC 2419. Panels show distribution of stars in (a) Galactic $(l,b)$ coordinates, (b) $E_{tot} - L_z$ (c) cylindrical coordinates $(R,Z)$, (d) Sagittarius plane angle $\tilde{\Lambda}_{\odot}$ \citep{Vasiliev21_sag} and $r_\mathrm{helio}$ (the parameters we use to define the spur), (e) Galactocentric Cartesian coordinates $(X, Z)$; and (f) Galactocentric radial velocity and distance.}
    \label{fig:fig14_sgr_spur}
\end{figure*}

\subsubsection{Sagittarius Spur} \label{spur}
Using an RR Lyrae catalog, \citet{Sesar:2017ab} identified a bifurcation in the Sagittarius stream near the apocenter of the trailing arm, where one branch turns back towards the Galactic center while the other extends out to $r_\mathrm{helio} > 120\,\mathrm{kpc}$, which they call the ``spur''. Early evidence for structure in this region was found by \citet{Newberg:2003aa}, and recently \citet{Bayer2025} studied the kinematics of this feature using a sample of spectroscopically confirmed BHBs. Although our \hdbscan{} algorithm does not recover this structure (likely because distance errors smear out its distribution in energy and angular momentum, as described in section~\ref{sag model}), it appears as a clear visual overdensity of K-giants in the $(\tilde{\Lambda}_\odot,r_\mathrm{helio})$ plane\footnote{$\tilde{\Lambda}_{\odot}$ is the angular distance along the stream in Sagittarius coordinates, defined as in \citet{Vasiliev21_sag}} consistent with the position of the spur in this space reported by \citet{Sesar:2017ab}. It is also very apparent the sky distribution of our K-giants at large distances ($r_\mathrm{gal}\gtrsim 60$~kpc), which we show below in Fig.~\ref{fig:vod_hac_ovo}. Indeed, it stands out much more clearly in our catalog than in the DESI BHB catalog of \citet{Bystrom:2025aa}, suggesting a higher ratio of K-giants to BHBs than many other halo overdensities.

Fig.~\ref{fig:fig14_sgr_spur} illustrates the relationship between the spur and other components of Sagittarius. We show the same Sagittarius model as in Fig.~\ref{fig:fig11_sagmodel}, but now convolved with a fiducial distance error of 20\% before we compute kinematic and orbital properties (as in Fig.~\ref{fig:fig12_Sgr_dist_err}). The black points on this figure are candidate spur member stars defined as follows:
\begin{equation}
    (r_\mathrm{helio} > 60\, \mathrm{kpc}\,) \cap (155^\circ < \tilde{\Lambda}_{\odot} < 175^\circ)
\end{equation}
This selection space is shown on Fig.~\ref{fig:fig14_sgr_spur}d. In the remaining panels, we see these stars occupy the high energy portion of the stream in the northern hemisphere. In most cases we find good agreement between these stars and the spur in the \citep{Vasiliev21_sag} model.

The extreme distances and low density makes the spur feature a sensitive tracer of the Milky Way's outer halo potential, and the tidal stripping history of the Sagittarius dwarf \citep{Bayer2025}. We also mark on Fig.~\ref{fig:fig14_sgr_spur} the location of the globular cluster NGC 2419 \citep[$r_\mathrm{helio} \sim 90\,kpc$,][]{Baumgardt2021}. This distant and unusual cluster is located near the spur region, spatially and dynamically. A connection between NGC 2419 and Sagittarius has long been suspected, but still remains unclear \citep[e.g.][]{Zhao:1998aa,Newberg:2003aa, Drake:2013aa, Davies2_2024, Bayer2025}.

\subsubsection{Cluster 3}

\begin{figure*}
    \centering
    \includegraphics[width=\linewidth]{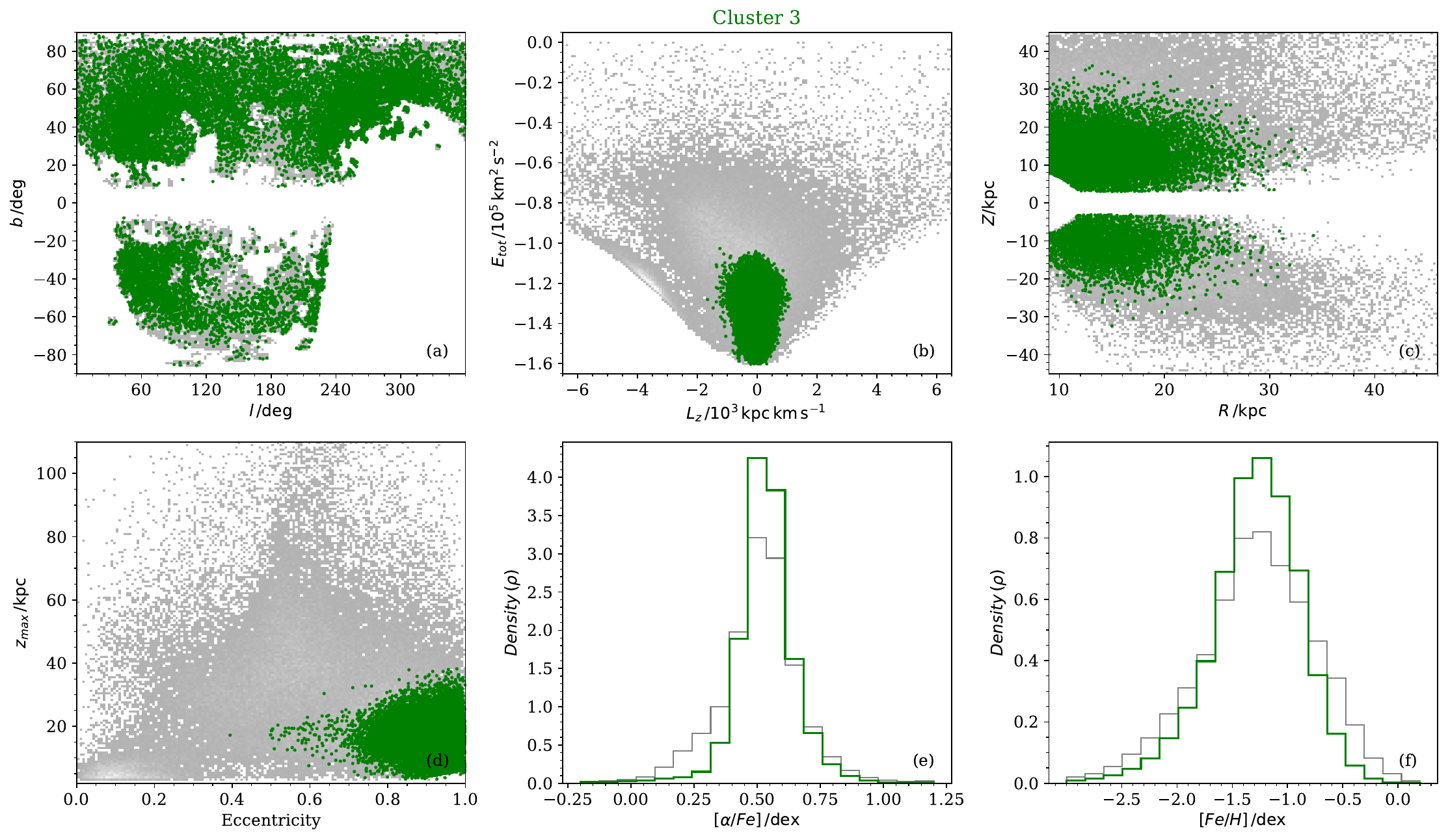}
    \caption{Summary plots for Cluster 3, associated to the known structure GSE. Panel (a): Distribution of GSE on Galactic coordinates l and b. Panel (b): Distribution in $E_{tot} - L_z$ space. Panel (c): Distribution in R - Z plane. Panel (d): Distribution of Eccentricity vs vertical excursion $z_{max}$. Panel (e): Density histogram of alpha abundance, [$\alpha$/Fe]. Panel (f): Density histogram of [Fe/H]. The gray color distribution in the background shows the whole halo K-giant sample and the green markers shows the distribution of Cluster 3.}
    \label{fig:fig15_cluster_3}
\end{figure*}

The largest and most prominent cluster identified by \hdbscan{} is Cluster 3, which is apparently associated with the GSE feature \citep{Belokurov2018,Helmi2018,Myeong2018_GSE}. After combining the first and second pass \hdbscan{} detections, we find a total of 18,242 stars associated with this cluster. Cluster 3/GSE is distinct from other structures in all spatial, kinematic and orbital distributions, and is particularly pronounced in IoM space, with low energy and a narrow distribution around $L_{z} \sim 0$. These highly eccentric orbits give rise to the elongated `sausage' distribution in $(V_r,V_\phi)$. GSE is an apparently relaxed, heavily phase-mixed structure, detectable across the DESI footprint and extending to $\sim 30$~kpc. The sharp `edge' of GSE and corresponding break in the total density of halo stars is well known \citep{Deason:2018aa_break, Naidu2020}. 

Our GSE stars have a wide range of metallicity, with median $\mathrm{[Fe/H]}\sim-1.27$, and $\mathrm{[\alpha/Fe]}\sim0.53$. These properties are broadly consistent with previous work on GSE, although the various selections of GSE in the literature yield significantly different samples \citep{Carillo24}. For example, \citet{Carillo24} found kinematic selections to have peak\ [Fe/H] (from Gaussian fits to the MDF) in the range $-1.18$ to $-1.24$, while a selection based on abundance ratios gave $-1.28$. The potential contamination of some GSE selections by Sagittarius stars with large distance errors (see above) may contribute to this variation. With our GSE selection, a mixture-model fit of two symmetric Gaussian yields $\langle \mathrm{[Fe/H]}\rangle_\mathrm{GSE,R} \simeq -1.17$ for the metal rich component, and $\langle \mathrm{[Fe/H]}\rangle_\mathrm{GSE,P} \simeq -1.51$ for the metal poor component. Fitting a single negatively-skewed Gaussian yields $\langle \mathrm{[Fe/H]}\rangle_\mathrm{GSE,skew} \simeq -1.30$ and a mode of $\approx-1.20$ (see section \ref{sec:gse_and_outer_halo}). Our result is therefore lower than the mass-weighted average of $-1.15^{+0.24}_{-0.33}$ reported by \citet{Naidu2020} based on a chemical evolution model fit, but within the range of the uncertainty and the variance among GSE selections. Although these differences may appear small (particularly given an intrinsic $\sim0.2$~dex uncertainty on our [Fe/H] measurements and the possibility of systematic offsets of similar size between surveys, e.g. \citealt{MWS_dr1}), they are significant to the extent that total mass estimates for the GSE progenitor are often made on the basis of the mass-metallicity relation \citep[e.g.][]{Naidu:2021aa,Carillo24} \footnote{Using the redshift-dependent stellar mass-metallicity relation of \citet{Ma2016_MZR}, a mean $\mathrm{[Fe/H]} = -1.30$ corresponds to $M_\star \sim 1.5 \times 10^8\,\msol$ at $z=0$ or $M_* \sim 1.7 \times 10^9 \,\msol$ at $z=2$. At either redshift, the range of  literature GSE mean metallicities, $-1.30 \lesssim \mathrm{[Fe/H]} \lesssim -1.17$, corresponds to a factor $\sim$ 2 difference in the estimated stellar mass.}.

We consider further the nature of GSE and its relationship to the residual stellar halo in section~\ref{sec:bulk}, after summarizing the remaining minor structures detected by \hdbscan{}.

\subsection{Minor structures}

Our experiments with sequential passes of the \hdbscan{} algorithm yielded two further detections of features previously reported but not associated with the major structures described above.

\subsubsection{Cluster A / Cetus-Palca}

\begin{figure*}
    \centering
    \includegraphics[width=\linewidth]{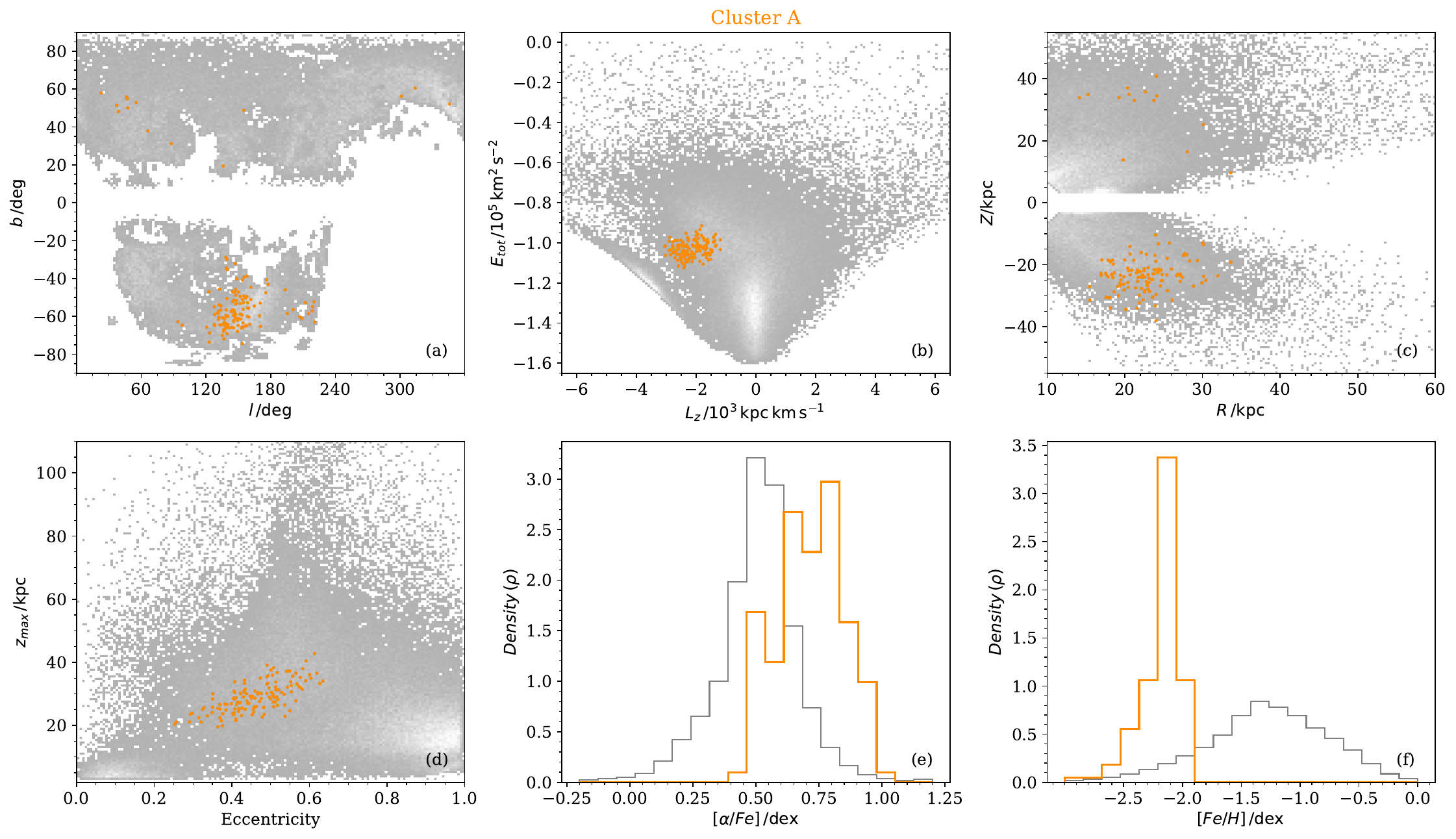}
    \caption{Summary plots for Cluster A, which we identify with the Cetus-Palca stream. Stars selected in this feature are shown in orange, and the whole sample as the grayscale background. The panels have the same meaning as those in Fig.~\ref{fig:fig15_cluster_3}.}
    \label{fig:fig16_cetus}
\end{figure*}

The kinematic and chemical properties of this cluster, found in the alternative second \hdbscan{} pass, are close to those of the Cetus-Palca stream \citep{Newberg2009_Cetus,GFT22_Cetus_Palca}. We identify 137 stars as part of this cluster and are shown in Fig.~\ref{fig:fig16_cetus}. In IoM space, this cluster lies near the northern section of the Sagittarius stream. However, most of its stars are in the southern Galactic hemisphere, along a diffuse track neighboring the southern Sagittarius stream and having a similar range of Galactocentric distance (24 to 47 kpc). The median eccentricity ($\sim 0.45$) and apocenter distance ($\sim 34$ kpc) of these stars are also very similar to the Sagittarius population.

However, Cetus-Palca is chemically distinct from Sagittarius: it has higher alpha enhancement (median $0.72$, with median [Mg/Fe] $\sim 0.40$) and lower metallicity (median $-2.13$; consistent with the mean of $-1.93$ found by \citealt{GFT22_Cetus_Palca}). The metallicity distribution is narrow. The proper motions of Cetus-Palca stars are also strongly clustered and apparently different from those in the Sagittarius stream. The distribution of this cluster in $L_{\perp}$ and action space suggests moderately eccentric prograde orbits in a well-defined plane. It resembles more closely the northern hemisphere section of Sagittarius in this respect, although all the stars in this cluster are in the southern hemisphere. Other work has speculated that this feature may be a companion or component of Sagittarius \citep{Yuan2019,Shipp2019,Chang2020}.

\subsection{Cluster B / Orphan-Chenab} \label{additional structures}

\begin{figure*}
    \centering
    \includegraphics[width=\linewidth]{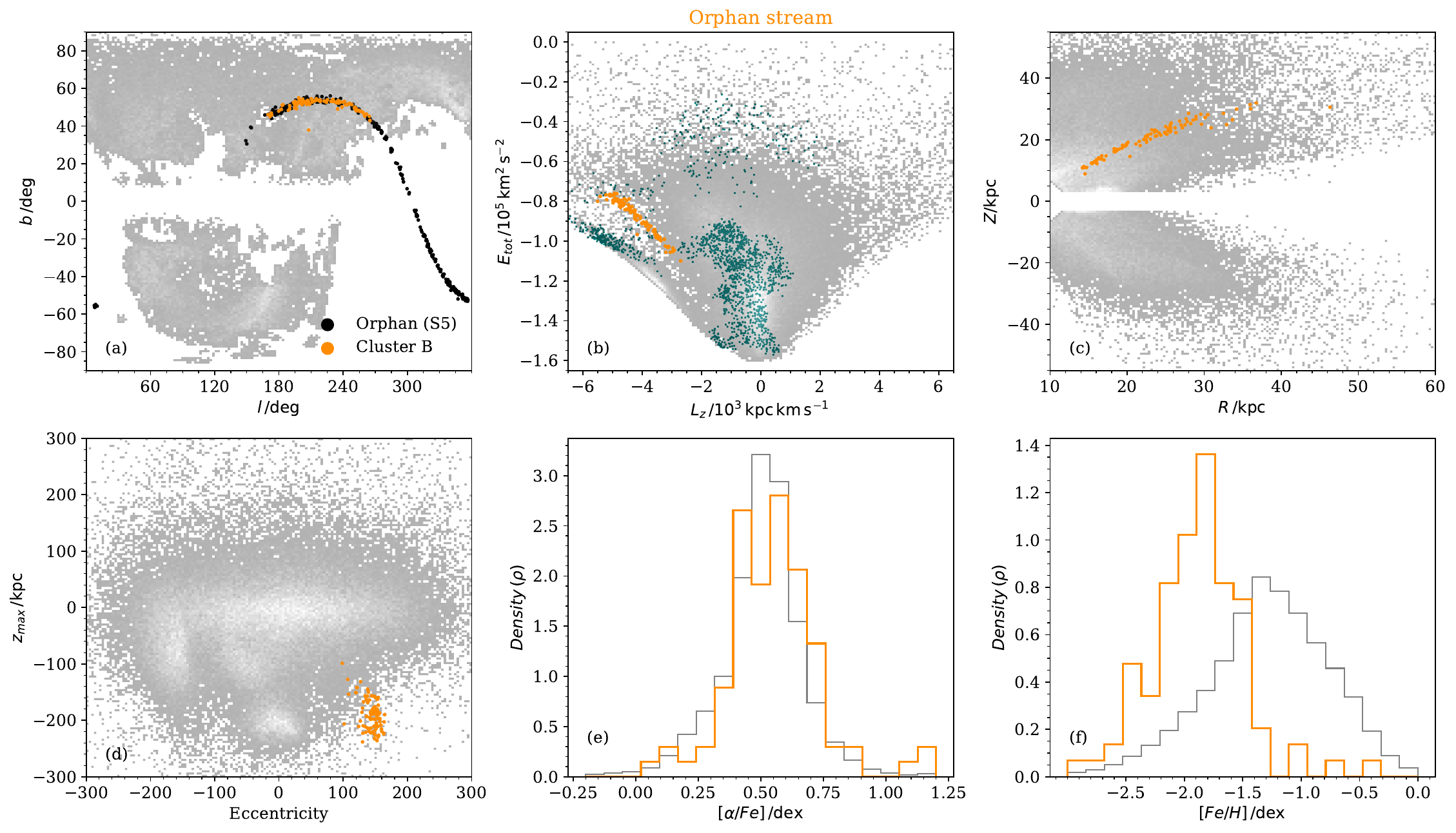}
    \caption{Summary plots for Cluster B, which we identify with the Orphan-Chenab stream. Stars selected in this feature are shown in orange, and the whole sample as the grayscale background. The panels have the same meaning as those in Fig.~\ref{fig:fig15_cluster_3}. The black points in panel (a) show the distribution of a spectroscopically confirmed sample of stars in the stream from \protect\citet{Koposov:2023aa}. The teal points in panel (b) show the other clusters found in the same \hdbscan{} pass, which we associated with the major halo substructures.}
    \label{fig:orphan}
\end{figure*}

To investigate whether our sequential clustering approach can identify structures with even fewer members, we ran \hdbscan{} with \texttt{[min\_cluster\_size, min\_samples]} $= [50,10]$ on those stars remaining in the catalog after removing all stars associated with a cluster in the second pass (we recall that the second pass used \texttt{[min\_cluster\_size, min\_samples]} $= [100,20]$). This identified a further 15 clusters. These are shown in Fig.~\ref{fig:orphan}b (teal-colored points). Most of these clusters apparently belong to low density regions of the clusters identified in the main pass. However, we clearly identified one new structure, with 95 stars, shown with orange points in Fig.~\ref{fig:orphan}. By comparing with data from \citet{Koposov:2023aa}, we confirmed that this is the northern (leading) section of the Orphan-Chenab stream. To our knowledge, this is the first clear `blind' detection of the stream in a spectroscopic survey \citep[although][identify $<10$ possible Orphan members in their FoF clustering analysis of the LAMOST K-giant catalog]{yang2019}. This demonstrates the sensitivity of the \hdbscan{} approach to features with very few stars in the DESI data.

Fig.~\ref{fig:orphan} shows the distribution of the Orphan-Chenab stream in different kinematic and orbital spaces. The black points in Fig.~\ref{fig:orphan}a show data from \citet{Koposov:2023aa}. The metallicity distribution agrees well with the mean of $\mathrm{[Fe/H]}=-1.9$ and dispersion $0.3$~dex reported by \citet{Koposov:2023aa}. The complex, extended distribution of this stream in $(E_{tot},L_z)$ space results from its interaction with the LMC \citep[see section 4.2 and figure 11 of][]{Koposov:2023aa}.

\section{The residual stellar halo} 
\label{sec:bulk}

Many stars in our catalog are not assigned to any structure, either by \hdbscan{} or our model-based selection of Sagittarius members. The large size of our sample enables us to quantify the bulk properties of these stars, and compare them to those of known structures. The residual (`smooth' or `phase-mixed') stellar halo may comprise stars that really are associated with known features but were not identified as such by our approach. Alternatively, the residual halo may comprise as-yet undiscovered large-scale substructure, or an underlying stratum of more diverse material from low-mass progenitors \citep[e.g.][]{Khoperskov2023,Pu2025}. The residual halo may not be intrinsically `smooth'. Substructure may simply be unresolved at the level of density and kinematic contrast (ratio of cluster member stars to non-member stars in the same region of feature space) accessible in our catalog with our \hdbscan{} method. This limit corresponds (at least) to features with the contrast of Orphan-Chenab. Another possibility is that the residual halo includes an `in situ' component, comprising stars formed directly on halo-like orbits or ejected from the disk \citep{Abadi:2006aa, Cooper:2015aa,Tau:2024aa,Thomas:2025aa}. A uniform phase-space distribution, on its own, is not conclusive evidence for in situ formation. Stars from massive accreted progenitors can spread out over large swaths of phase space \citep[e.g.][]{Jean-baptiste2017, Kizhuprakkat2024}, while simulations suggest the most likely source of weakly-bound in situ halo stars is gas stripped from satellites \citep[e.g.][]{Cooper:2015aa}.

\subsection{Residual halo sample and high angular momentum subsets}

We define a sample of K-giants in the `residual halo' as those remaining after we remove all stars in the clusters identified above (we use the model-based selection for Sagittarius members). To excise more thoroughly any remaining imprint of Sagittarius, we exclude all stars within a declination range of $\pm25$ degrees around the Sagittarius stream track\footnote{The Sagittarius stream's path is modeled using linear interpolation of its RA and DEC coordinates, creating a continuous function of stream's DEC as a function of its RA. For every star in the K-Giant catalog, the interpolated DEC of the stream at the star's RA is computed. A tolerance of 25 degrees is applied to define the region around the stream. In the northern hemisphere, stars with DEC $>$ (interpolated stream DEC + $25^\circ$) are classified as above the stream, and those with DEC $<$ (interpolated stream DEC - $25^\circ$) are considered as below the stream (opposite is true for the southern hemisphere). Stars within the $\pm25^\circ$ region are considered to be part of the stream, and are excluded.}. We note that this has the side-effect of removing many of the regions associated with previous detections of diffuse overdensities in the outer halo (see below) \footnote{As mentioned in Section \ref{sag model}, Sagittarius stars can also be selected by a cut in $(L_z, L_y)$ space. Although such a selection yields different numbers of prograde and retrograde stars in the residual halo, we do not find this has any significant effect on our results. }. Likewise, we remove the residual of Aleph by excluding all stars with circularity $|\varepsilon|>0.65$. For uniformity, we also remove high circularity stars from the retrograde halo. We further divide the residual halo sample into prograde and retrograde subsets at high angular momentum ($L_z < -1.5$ and $L_z > 1.5$), containing 1,895 and 2,043 stars respectively. This restriction to high $|L_z|$ excludes stars in the range of angular momentum associated with GSE. The `smooth' high-angular momentum halo in our sample thus shows a very weak bias towards retrograde orbits \footnote{\citet{Deason:2017aa_spin} used Gaia DR1 to recalibrate SDSS astrometry and compute proper motions for halo RR Lyrae stars, BHBs and K giants. After excluding Sagittarius members, they found a weak net \textit{prograde} spin of $\langle V_\phi \rangle \sim 5 -20\,\mathrm{km\,s^{-1}}$ for RR Lyrae stars, increasing slightly for more metal rich tracers.} ($\langle L_z \rangle = 0.015\,\mathrm{kpc\,km\,s^{-1}}$ and $\langle V_\phi \rangle = \ -0.53\,\mathrm{km\,s^{-1}}$).

\begin{figure*}
    \centering
    \includegraphics[width=\linewidth]{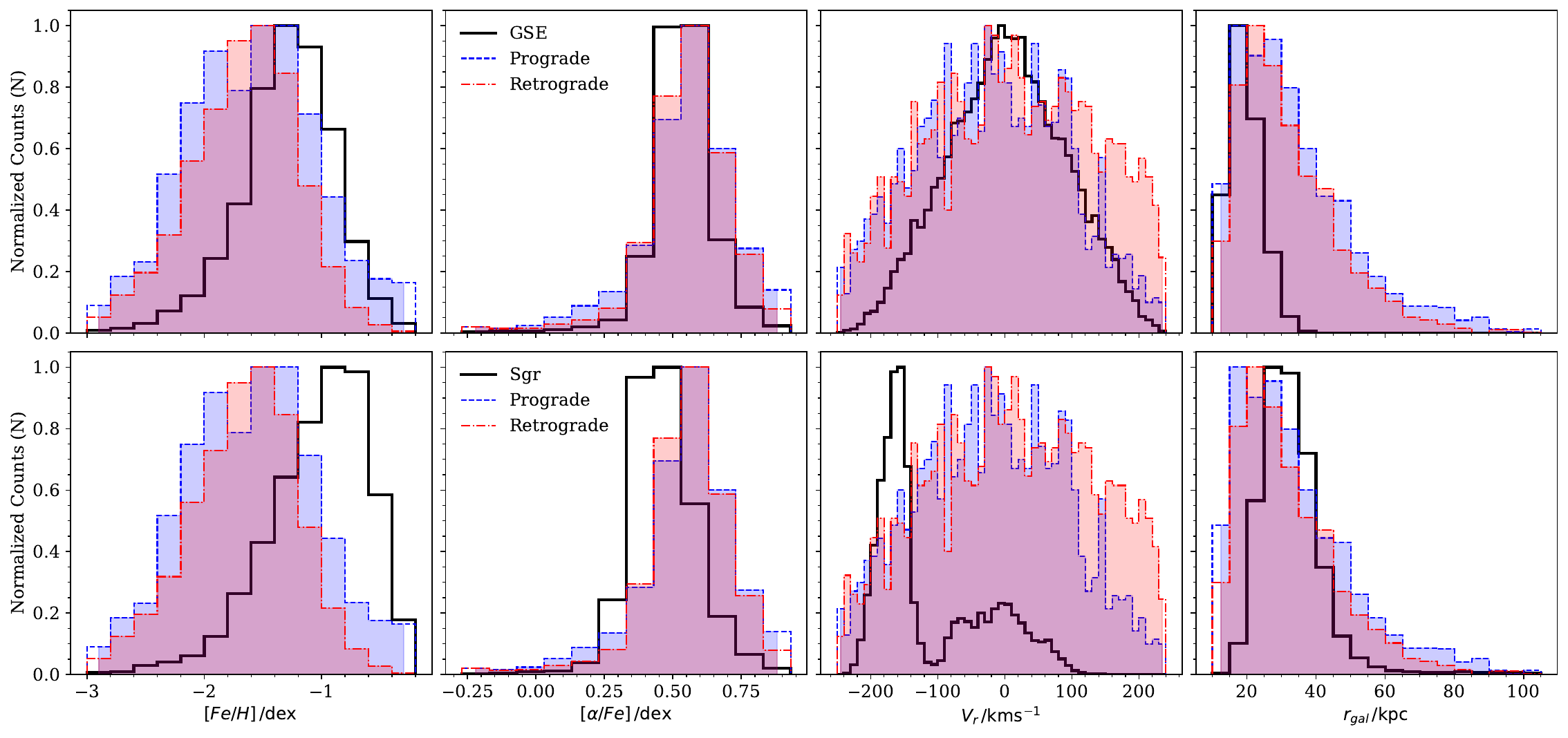}
    \caption{Distributions for stars not assigned to any \hdbscan{} cluster, divided into prograde ($L_z < -1.5$) and retrograde ($L_z > 1.5$) subsets. From left to right; metallicity; alpha abundance; Galactocentric radial velocity ($V_r$); and Galactocentric distance ($r_{gal}$). Blue shading represents the prograde subset, red represents the retrograde subset. The corresponding distributions for our GSE sample (top row) and Sagittarius sample (bottom row) are shown as black lines. The distributions in this figure are scaled to the same peak value, rather than the same area.}
    \label{fig:pro_retro_halo}
\end{figure*}

Fig.~\ref{fig:pro_retro_halo} shows distributions of the residual halo high-$L_z$ prograde and retrograde subsets in [Fe/H], [$\alpha$/Fe], radial velocity, and distance. We see a notable excess of retrograde stars with high radial velocity. The distributions of distance for the two subsets are similar. The high angular momentum prograde and retrograde residual halo also have very similar MDFs overall, with minor differences: the prograde MDF appears slightly broader and has more stars in the very metal-rich ([Fe/H]$\gtrsim0.5$) tail. The [$\alpha$/Fe] distribution shows a small excess in the prograde subset at low [$\alpha$/Fe], which may correspond to the excess at high [Fe/H].

Fig.~\ref{fig:pro_retro_halo} also compares the residual halo to our GSE and Sagittarius selections. As we discuss below, comparison to the stars identified with GSE can address the question of whether the diffuse material at high angular momentum may be a relic of an early phase of disruption of the GSE progenitor, or evidence of additional progenitors. The comparison to Sagittarius helps to determine how effectively we have removed its contribution, to the prograde residual halo subset in particular. We see that the Sagittarius distributions are substantially different from those of the residual halo, especially in [Fe/H] and radial velocity. This suggests a direct comparison between the prograde and retrograde residual halo subsets is unlikely to be dominated by the residual contribution of Sagittarius.

The small differences between the prograde and retrograde subsets of our residual halo sample contrast with the much more obvious differences between either subset and our GSE sample at comparable binding energy. The most notable differences are seen in the MDFs. The peak [Fe/H] and [$\alpha$/Fe] of both subsets are $\sim0.2$~dex lower than that of GSE. Their metal-rich tails are broadly similar to that of GSE. In contrast, the metal poor tails of the two subsets are very different to the metal-poor tail of GSE, containing a much larger fraction of stars with $\mathrm{[
Fe/H]}<-1.5$. A corresponding difference is apparent at high [$\alpha$/Fe], although the [$\alpha$/Fe] distributions are overall more similar to that of GSE. These statements regarding comparisons to GSE apply equally to the prograde and retrograde residual halo subsets.

\subsection{Implication of the residual halo metallicity distributions for the origin of GSE}

Following the discovery that the local stellar halo is dominated by stars associated with the GSE feature, various scenarios for the mass, arrival time and and initial orbit of the corresponding progenitor (or progenitors) have been proposed, based on a rapidly growing and somewhat complex body of chemodynamical observations. We now explore how our data can help to constrain the contribution of GSE debris to the outer halo.

\subsubsection{GSE progenitor scenarios}

To provide context for our results, we provide a brief summary of scenarios proposed to explain the observed local GSE feature, which may be contributed by either one or several progenitors. The debris of any given GSE progenitor may:
 \begin{enumerate}
      \item  \label{gse:one} Dominate \textit{either} the prograde or retrograde residual halo;
      \item \label{gse:both} Dominate \textit{both} the prograde and retrograde residual halo; 
      \item \label{gse:neither} Make a negligible or sub-dominant contribution to the residual halo at high angular momentum.
 \end{enumerate}

\citet{Naidu:2021aa} explore a single-progenitor example of case \ref{gse:one}, in which the progenitor has an initially retrograde orbit with moderate angular momentum. A fraction of stars are lost into the retrograde halo on early passages, and the bulk of the debris is subsequently stripped on highly eccentric orbits around the last few pericenters, after the progenitor orbit has been radialized by interaction with the host \citep[e.g.][]{Vasiliev:2022aa,Amarante2022}. A metallicity gradient in the progenitor is invoked to explain the lower metallicity of the debris with higher angular momentum. \citet{Naidu:2021aa} constrain the parameters of this scenario by comparing data from the H3 survey to many N-body realizations. They suggest that some of the apparent metal-poor structure in the retrograde halo \citep[specifically the `Arjuna' feature of][]{Naidu2020} corresponds to the first stages of stellar mass loss from the progenitor. Similarities in chemical abundances between retrograde residual halo stars and GSE have provided further support for this scenario \citep{Horta2023, Woody:2025aa}. 

In contrast, following early proposals of a retrograde origin for the Enceladus feature \citep{Helmi2018,Koppelman2019a}, \citet{Myeong2019} proposed a two-progenitor scenario. One progenitor, that of the `Sausage', arrives on an initially radial orbit, produces mostly radial or mildly prograde debris, and does not contribute any significant debris at high angular momentum (corresponding to our case \ref{gse:neither}). A second, less massive progenitor, `Sequoia', accounts for evidence of retrograde substructure $\lesssim10~\mathrm{kpc}$ from the Sun (case \ref{gse:one}). The catalog of potential substructures in this local region has grown with each data release from Gaia \citep[see, for example, the summaries in][]{Deason_belokurov2024,Kim2025} and it remains unclear to what extent any of the known local features contributes to the more distant halo \citep{Naidu:2021aa}. The local retrograde halo contains many small structures \citep[e.g.][]{Myeong2018_shards, Dodd:2023aa}, in contrast to the lack of large-scale retrograde structure in the more distant halo. An example is the Thamnos structure of \citet{Koppelman2019a} which is distinct from Sequoia chemically and dynamically.
The more distant ($\sim23$~kpc) Arjuna feature of \citet{Naidu2020} is also chemically distinct from Sequoia, and from I'itoi \citep[a feature associated with peak in the MDF of the retrograde halo, reported by][]{Naidu2020}. All of these features have distance and eccentricity (for example, respectively 23~kpc and $e\sim0.55$ for Arjuna) comparable to or lower than those of typical stars in our retrograde residual halo subset (median 28~kpc and $e\simeq0.59$, with many stars at much larger values).

As remarked on by \citet{Evans:2020aa}, the Gaia-Enceladus feature described by \citet{Helmi2018}, although retrograde on average, includes a substantial highly prograde component. Considering these stars as a single feature is an example of a single-progenitor scenario, akin to case \ref{gse:both}. \citet{Amarante2022} showed that a GSE-like merger event can produce both prograde and retrograde debris at higher energies, that are more metal-poor than the bulk of the GSE debris. A multi-progenitor GSE with several case \ref{gse:both} contributions is relatively common among the massive progenitors of cosmological simulations -- see, for example, figure 11 of \citet{Kizhuprakkat2024} and figures 3 and 4 of \citet{Khoperskov2023}, who note that this is due in part to the disturbance of the GSE-like debris by subsequent mergers \citep[c.f.][]{Jean-baptiste2017, Dillamore:2022aa}. Although features resembling GSE in integral-of-motion space are often seen in simulations \citep[e.g.][]{Fattahi2019, Dillamore:2022aa}, it is rare for them to be produced by only one early-accreted, fully phase-mixed progenitor \citep[][]{Orkney:2023aa,Rey:2023aa,Pu2025}.

\subsubsection{GSE and the outer halo MDF}
\label{sec:gse_and_outer_halo}
\begin{figure*}
    \centering
    \includegraphics[width=\linewidth]{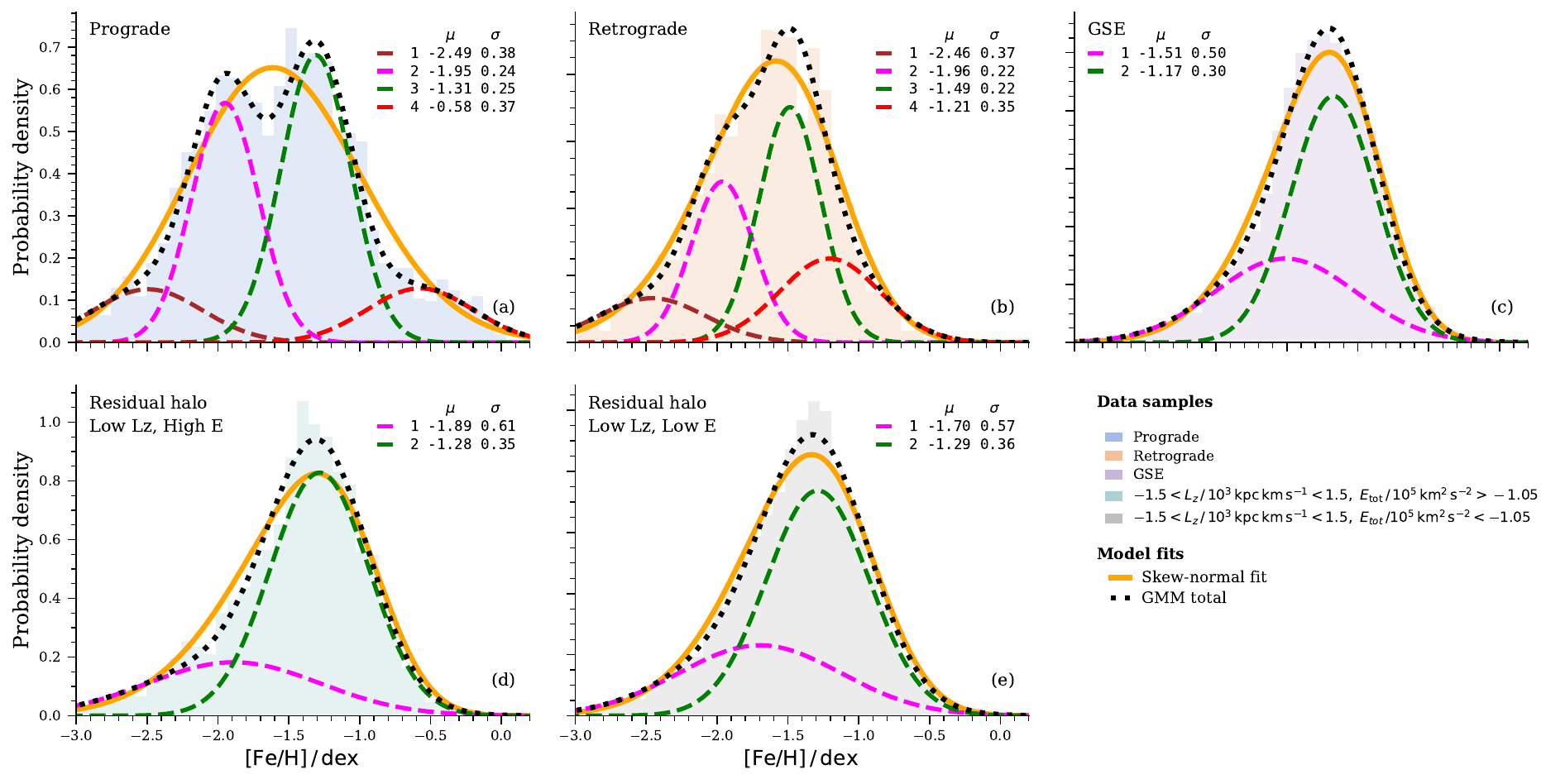}
    \caption{GMM fits to the MDFs of different data samples. In each panel, the histogram shows the metallicity distribution of the sample, the orange curve shows a skew-normal fit, the dotted black curve shows the total GMM fit and the dashed colored curves show the individual Gaussian components. Panels (a) and (b) use a 4-component GMM; panels (c) to (e) use a 2-component GMM. From top left, the samples are (a) prograde  residual halo, (b) retrograde resitual halo, (c) GSE (\hdbscan{} cluster 3), (d) the residual halo at low angular momentum and high energy, defined by $-1.5 < L_z < 1.5, E_\mathrm{tot} > -1.05$, (e) the residual halo at low angular momentum and low energy, defined by $-1.5 < L_z < 1.5, E_\mathrm{tot} < -1.05$. The mean ($\mu$) and standard deviation ($\sigma$) for each Gaussian component are given in the legend of each panel.}
    \label{fig:gmm_fits}
\end{figure*}

We now return to the interpretation of our results in Fig.~\ref{fig:pro_retro_halo}. We use Gaussian mixture model (GMM) fits of symmetric Gaussians to quantify the differences between the MDFs of the prograde high-$|L_z|$ residual halo, retrograde high-$|L_z|$ residual halo and GSE, as shown in the top left panel of Fig.~\ref{fig:pro_retro_halo}. Fig.~\ref{fig:gmm_fits} shows these fits; the mean metallicities of the GMM components are reported in Table~\ref{tab:metal_peaks}. We fit four components to the prograde and retrograde residual halo MDFs. The two components with highest weight represent distinct peaks at lower metallicity (component \#2) and higher metallicity (component \#3) that are present in both distributions, but most readily apparent in the prograde residual halo. The lower peak (for brevity we refer to this peak as P2 and R2, for the prograde and retrograde residual halo respectively) is $\sim0.1$ to $0.3$~dex below literature values for `Sequoia' \citep[e.g.][]{Naidu:2021aa,Horta2023}. The high peak (P3 and R3) is $\sim0.1$ to $0.2$~dex below the peak we find in the GSE MDF (see below). 

We do not consider the two further GMM components with lower weight (components \#1 and \#4) to be significant. These components (shown in italics in table~\ref{tab:metal_peaks}) effectively fit the extended tails of the MDFs. At least at the lowest metallicity, they likely arise from the intrinsically skew-normal shape of MDFs, as predicted by chemical evolution models. By including these somewhat artificial components, we obtain a better estimate of the metallicity peaks for GMMs restricted to symmetrical components. We note, however, that the most metal-rich component in the prograde residual halo (P4) does correspond to a visually distinct peak in this MDF (also visible in Fig.~\ref{fig:pro_retro_halo}), albeit with very low weight. The peak metallicity of this component is consistent with that of cluster 1 (Aleph), suggesting a small fraction of that component may remain in our prograde residual halo subset.

For GSE, we fit two GMM components, with the dominant component representing the peak of the MDF (GSE2). We have verified that the peak metallicity of a single skew-normal distribution fit is similar to the 2-component symmetrical GMM fit. 

For further comparison to GSE, we also fit the MDF of stars in the residual halo with low $|L_z|$, i.e.\ those that we do not \textit{not} assign to GSE or any other \hdbscan{} cluster. We separate these low $|L_z|$ stars into high energy (weakly bound) and low energy (tightly bound) subsets, at $E\approx-1.05\times10^{5}\,\mathrm{km^2\,s^{-2}}$, the highest energy we associate with GSE members. In both energy subsets, the MDF of the low-$|L_z|$ residual halo is unimodal and very similar in shape to the MDF of GSE, but approximately $0.1$~dex more metal poor. We note that the high energy, low-$|L_z|$ residual halo shows weak evidence of a second component around $\mathrm{[Fe/H]}\sim -2.0$.

We summarize our observations regarding the MDF fits in table~\ref{tab:metal_peaks} and Fig.~\ref{fig:gmm_fits} as follows:
\begin{itemize}
    \item The prograde and retrograde residual halo samples each appear to have two dominant progenitors with a clear separation in peak metallicity (P2, P3, R2, R3);
    \item The relatively metal poor P2 and R2 components have very similar peak metallicities and can be tentatively identified with the hypothetical metal-poor progenitor known as Sequoia;
    \item The relatively metal rich P3 and R3 components have slightly different peak metallicities, both more metal poor than GSE;
    \item The stars we identify with the GSE feature appears to have only one dominant progenitor;
    \item Residual halo stars with low angular momentum, but not identified with GSE, also appear to be dominated by one progenitor (LowLz2), with metallicity intermediate between GSE and P3/R3;
    \item We find tentative evidence for an additional very metal poor contributor to the low angular momentum halo (LowLz1), which is slightly clearer at the highest (weakly bound) energies.
\end{itemize}

There are many possible interpretations of these results, taken at face value. For example, it seems reasonable to associate both LowLz2 and at least one of P3 or R3 with GSE in a scenario similar to that of \citet{Naidu:2021aa}, invoking a metallicity gradient in the progenitor to explain a peak metallicity difference of $\sim0.3$. However, P3 and R3 have similar peak metallicity. The MDF peak is often used together with the assumption of a narrow and monotonic stellar mass-metallicity relation \citep[e.g.][]{Kirby2013_MZR,Ma2016_MZR} to estimate progenitor masses. Under such an assumption, components with similar peak metallicities must originate either from the same progenitor, or from two progenitors of similar mass. The MDF peaks of P3 and R3 are only slightly below that of GSE, implying a progenitor almost as massive as that of GSE. The K-giant counts in the prograde and retrograde subsets are similar. It therefore seems implausible (to us) that only one such progenitor could contribute to the GSE region -- if that were the case, it would beg the question of where the rest of the debris from other progenitor went. This suggests that either GSE is the superposition of two similar components, as argued for example by \citet{Myeong2019} and \citet{Donlon2020,Donlon2022} \citep[cf.][]{Pu2025} or else that it originates from a single progenitor, the debris of which is now spread across both the prograde and retrograde halos. Both scenarios would be incompatible with the retrograde-only model of \citet{Naidu:2021aa}.

Another possibility is an almost completely `radial' orbit for the GSE progenitor, with the prograde and retrograde residual halo then built up (for example) from one or two metal rich progenitors, and at least one metal poor progenitor. This would be close (though not identical) to the ``Sausage plus Sequoia'' hypothesis of \citet{Myeong:2019aa}. The identification of both PL and RL with Sequoia is challenged by the lack of a corresponding peak in the MDF of the low-$L_z$, weakly bound material. 

In all these cases, our results suggest to us that useful constraints are likely to follow from further analysis of residual halo components in MWS and other large spectroscopic surveys. Interpretations are unlikely to be straightforward or unique, due to the considerable spread and multi-modal nature of individual progenitor distributions in $(E,L_z)$ and other selection spaces, as demonstrated by (for example) the simulations of \citet{Khoperskov2023} and \citet{Kizhuprakkat2024}.

\begin{deluxetable}{lcccc}
\tablecaption{Representative GMM estimates of peaks in the [Fe/H] distributions of various halo star selections (equal to the mean for a symmetric Gaussian MDF). Entries in italics correspond to fit components that likely compensate for skew in the MDF. The typical errors on the metallicity of individual stars in our sample is $0.2$~dex \citep{MWS_dr1}. For Sequoia, we quote APOGEE measurements from \citet[][H23]{Horta2023} using the selections of \citet[][M19]{Myeong:2019aa} and \citet[][K19]{Koppelman:2019ab}, and the estimate made by \citet[][N21]{Naidu:2021aa} based on a peak in the retrograde halo MDF from the H3 survey.\label{tab:metal_peaks}}
\tablehead{ & \colhead{Poorest} & \colhead{Poor} & \colhead{Rich} & \colhead{Richest}}
    \startdata
    Prograde residual halo & $-$\textit{2.45} &  $-1.95$ & $-1.31$ & $-$\textit{0.58} \\
    Retrograde residual halo & $-$\textit{2.46} & $-1.96$ & $-1.49$ & $-$\textit{1.21} \\
    GSE (this work) & - & $-$\textit{1.51} & $-1.17$ & -  \\
    GSE (this work), skew & - & - & $-1.20$ & -  \\
    Low $|L_z|$, high $E$ residual halo & -  & $-$\textit{1.89} & $-1.28$ \\
    Low $|L_z|$, low  $E$ residual halo & -  & $-$\textit{1.70} & $-1.29$ & -\\
    \hline
    Sequoia H23/M19 & -  & $-1.41$ & - & - \\
    Sequoia H23/K19 & -  & $-1.53$ & - & - \\
    Sequoia N21 & - & $-1.60$ & - & - \\
    \enddata
\end{deluxetable}

Finally, we note that, if we fine-tune the width of the MDF mass bins, we can recover multiple peaks corresponding approximately to those associated with the retrograde Arjuna/Sequoia/I'itoi features in figure 1 of \citet{Naidu:2021aa}. With our wider fiducial bins (and greater energy range), their Sequoia and Arjuna peaks are blended into a single sub-GSE peak in Fig.~\ref{fig:pro_retro_halo}. I'itoi appears to correspond well to the metal poor tail of our retrograde distribution. However, we do not identify any clear spatial or kinematic overdensities associated with retrograde stars selected in any of these peaks.

\subsection{Diffuse overdensities in the residual halo}

\begin{figure*}
    \centering
    \includegraphics[width=\linewidth]{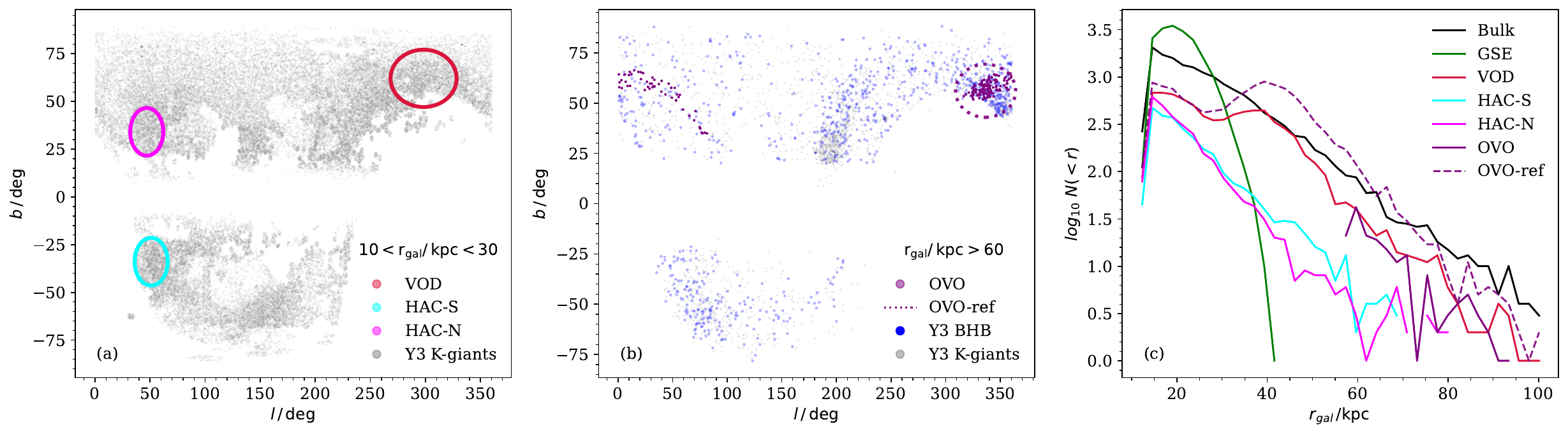}
    \caption{Distribution of overdensities identified in the Milky Way's stellar halo. Panel (a) shows the distribution of K-giants (gray) between $10 < r_{gal} < 30$~kpc in Galactic ($l,b$) coordinates. The red, cyan and magenta colored ellipses shows the apparent positions and extents of the structures VOD, HAC-S and HAC-N, respectively. Panel (b) shows the subset of K-giants (gray) and BHBs (blue) at $r_{gal} > 60$~kpc. Possible OVO members selected from our K-giant catalog are shown in purple. The dotted purple circle highlights a fiducial region around this overdensity. Panel (c) shows the logarithmic distribution of star counts with Galactocentric radius in these regions, with no correction for selection function.}
    \label{fig:vod_hac_ovo}
\end{figure*}

Finally, in addition to the chemodynamical structures we have discussed so far, we examine evidence in our data for several large-scale halo overdensities that have been identified in configuration space alone. Recent summaries of work on these features can be found (for example) in \citet{Chandra:2023aa} and \citet{Amarante:2024aa}. The Virgo Overdensity \citep[VOD;][]{Vivas:2001aa,Vivas:2016aa,Newberg:2002aa, Juric:2008aa} and the Hercules-Aquila clouds \citep[HAC-N and HAC-S;][]{Belokurov:2007ab} appear on opposite sides of the sky in the distance range $10<r_\mathrm{gal}<20$~kpc. They are aligned with an axis defined by the triaxial distribution of GSE members, and consequently have been interpreted as shell-like debris of the GSE merger \citep{Simion:2018aa,Simion:2019aa, Iorio:2019aa, Naidu:2021aa, Perottoni:2022aa, Han2022, Chandra:2023aa}. \citet{Donlon2020,Donlon2024} have argued that these features imply a merger time substantially later than the $\sim8$~Gyr age usually associated with the GSE event. Recently, \citet{Li2025_kg} measured radial density profiles for these regions using a sample of DESI K-giants. The Outer Virgo Overdensity at $r_\mathrm{gal}\sim80\,\mathrm{kpc}$ \citep[OVO;][]{Sesar:2017ab} has also been tentatively linked with GSE \citep{Chandra:2023aa}. 

Fig.~\ref{fig:vod_hac_ovo}a shows the distribution of K-giants in the range $10<r_\mathrm{gal}<30$~kpc, with the approximate positions and extents of VOD (red), HAC-N (magenta) and HAC-S (cyan) adopted from \citet{Perottoni:2022aa}:
\begin{align*}
    \mathrm{VOD:}\,& (270^\circ < l < 330^\circ)\,\cap (50^\circ < b < 75^\circ)\\
    \mathrm{HAC{\text -}N:}\,& (30^\circ < l < 60^\circ)\,\cap (20^\circ < b < 45^\circ) \\
    \mathrm{HAC{\text -}S:}\,& (30^\circ < l < 60^\circ)\,\cap (-45^\circ < b < -20^\circ)
\end{align*} 
Fig.~\ref{fig:vod_hac_ovo}b shows the projected density of stars from our catalog at $r_\mathrm{gal} > 60$~kpc in gray. Following \citet{Sesar:2017ab} we select OVO stars (purple points) according to their heliocentric distance and angle $\tilde{\Lambda}_{\odot}$ in the plane of the Sagittarius stream \citep{Vasiliev21_sag}:
\begin{equation*}
    \mathrm{OVO:}\, (r_\mathrm{helio} > 60~\mathrm{kpc})\, \cap (70^\circ < \tilde{\Lambda}_{\odot} < 80^\circ)
\end{equation*}
The cut on $\tilde{\Lambda}_{\odot}$ includes an apparent `tail' of stars extending beyond the visual OVO from $0 < l < 100^{\circ}$. We believe this tail is a selection artifact only. We find a typical [Fe/H] of $-1.0$ (median) or $-1.2$ (mean) for OVO members selected in this way, close to the value of $\sim-1.3$ found for OVO red giants by \citet{Chandra:2023aa}.

In Fig.~\ref{fig:vod_hac_ovo}b we overplot (in blue) the DESI BHB catalog of \citet{Bystrom:2025aa} in the same Galactocentric distance range ($r>60$~kpc). Similar overdensities are apparent in both K-giants and BHBs, with the OVO standing out most clearly in both tracers. Notably, the Sagittarius spur discussed in \ref{spur} is also prominent in the distribution of distant K-giants shown in Fig.~\ref{fig:vod_hac_ovo}b, around $(l,b)\approx(200,25)$. Unlike the OVO, the spur is much less obvious in the distribution of DESI BHBs.

It is not surprising that we do not detect any of these overdensities with \hdbscan{}. They are readily apparent only in configuration space, where they stand out particularly strongly in RR Lyrae star and BHB\footnote{We do not show the BHB sample in Fig.~\ref{fig:vod_hac_ovo}a because it would obscure the K-giant distribution. All three overdensities in this panel stand out more clearly in the BHB distribution than in K-giant distribution, in contrast to the OVO (which appears with similar clarity in the distribution of both tracers) and the Sagittarius spur (which is much more prominent in the distribution of K-giants).} counts \citep[e.g.][]{Sesar:2017ab}, but they are not easily detectable in kinematic or integral-of-motion spaces.

Fig.~\ref{fig:vod_hac_ovo}c shows the logarithmic profile of star counts with Galactocentric distance in each of circular approximations to the overdensity regions in panels (a) and (b). For the OVO, in addition to the stars selected as above, we show the profile of all stars in a circular patch similar to those used for the VOD and HAC; this OVO reference region partly overlaps with our VOD region. Both the VOD and OVO regions show two characteristic bumps at $\lesssim20$~kpc and $40$-$50$~kpc, corresponding to the VOD itself and Sagittarius, respectively. The density of GSE peaks over a similar radial range. These features are not seen in the profile of stars in our "residual halo" component (which excludes both Sagittarius and GSE). The HAC-N and HAC-S profiles are almost as featureless as the residual halo, but slightly steeper; a hint of Sagittarius is visible in the HAC-S profile. At larger distances, a small excess corresponding to the OVO is visible in the profile of the OVO members selected by $\tilde{\Lambda}_{\odot}$ and in the corresponding reference region. As in the case of the Sagittarius spur, the OVO feature (captured also by distant stars in the VOD region) has K-giant counts comparable to the entire residual halo sample at these distances.

\section{Conclusions}

\label{conclusion}

We present a K-giant catalog containing 88,959 stars from the DESI Y3 dataset. We selected these giants based on $T_\mathrm{eff}$, $\log g$ and proper motion, with an additional radial ($r_\mathrm{gal} > 12$ kpc) and vertical ($|Z| > 3$ kpc) cuts to restrict the sample to the stellar halo. We use the dynamical properties of these stars to associate them with stellar halo substructures, using the machine learning clustering algorithm \hdbscan{}. Since massive halo progenitors may have broad metallicity distributions resulting from extended star formation histories, we do not use metallicity or alpha abundance as a clustering feature.

The clusters we recover with \hdbscan{} using a six-dimensional feature space comprising IoM and spherical velocity ($E_{tot}, L_z, L_\perp, V_r, V_\phi, V_z$), correspond to well-known large-scale substructure in the stellar halo. With regard to each of these structures, we find the following:

\begin{itemize}

\item \textbf{Aleph:} We recover the Aleph structure reported by \citet{Naidu2020} (metal-rich stars on low-eccentricity orbits out to $r_\mathrm{gal}\gtrsim 20$~kpc) in our first pass of \hdbscan{}. We confirm that this structure extends to larger distance and lower latitude than reported in \citet{Naidu2020}. We also confirm a relatively narrow MDF for this feature, peaking at $\mathrm{[Fe/H]}\approx-0.5$. The structure shows a metallicity gradient, decreasing from $\mathrm{[Fe/H]}\simeq-0.47$ at $\sim 12\, \mathrm{kpc}$ to $-0.67$ at $\sim 30\, \mathrm{kpc}$. 
It is likely that the \hdbscan{} cluster we identify with Aleph includes some fraction of contamination associated with the thick disk and previously identified low-latitude structures such as the Monoceros ring and anticenter stream (ACS). Kinematically structures appear kinematically to be overdensities within a much more extensive but coherent system, although their chemical relationship is unclear. Aleph appears likely to have originated from disturbance to the outer disk. The broader view of Aleph from our K-giant catalog may help to constrain this scenario.

\item \textbf{Sagittarius:} In our first pass of \hdbscan{}, we find two large clusters associated with sections of the Sagittarius stream in the Northern and Southern Galactic hemispheres. Our second pass of \hdbscan{} recovers several additional clusters in the periphery of the first pass detections.
Through comparison to the Sagittarius model of \cite{Vasiliev21_sag}, we demonstrate that distance errors have a significant impact on the morphology of these clusters, and the overall recovery of Sagittarius stars across our footprint. We use the model to predict the regions of $(E,L_{z})$ space occupied by Sagittarius members after convolution with a fiducial distance error of 20\%. These regions are significantly different to those occupied by model stars in the absence of error, and included retrograde and lower-energy orbits, blended with the higher-energy periphery of GSE. It is possible that some selections for GSE in the literature may include this contamination from Sagittarius. We confirm that real Sagittarius members in our catalog are found in those regions, which \hdbscan{} did not associate with its two Sagittarius detections. By searching for DESI K-giants with sky coordinates and proper motions close to the predictions of the error-convolved model, we recover a more complete sample of Sagittarius stars, which traces the full extent of the stream across the sky. We consider this model-based selection to have higher completeness and purity than the \hdbscan{} clusters associated with Sagittarius. We find a small offset in peak metallicity ($\lesssim0.25$~dex) between the northern and southern components of the stream with the model-based selection.

Our sampling of Sagittarius at $r_\mathrm{gal} > 50~\mathrm{kpc}$ is sparse, and diluted further by the effect of distance errors. However, although the outermost portion of the Sagittarius stream predicted by the \cite{Vasiliev21_sag} model is not detected by \hdbscan{}, we find a strong signal of the `spur' associated with this region in configuration space \citep{Newberg:2003aa,Sesar:2017ab}. This overdensity is more clearly visible in K-giants than in BHB stars. The stars we identify in this feature largely agree with predictions for the outer wrap in the \cite{Vasiliev21_sag} model.

\item \textbf{GSE:} \hdbscan{} readily recovers the massive, centrally concentrated high-eccentricity component that dominates the inner stellar halo, known as GSE. The majority of stars associated with this are identified in the first \hdbscan{} pass, with a small additional contribution from the second. This GSE sample is consistent with previous studies of the feature, allowing for the variety of empirical definitions of GSE in the literature. We find a unimodal MDF for these stars, which is well described by a single skewed Gaussian and peaks . 

\item \textbf{Cetus-Palca:} Our `alternative' second pass of \hdbscan{} (restricted to metal poor stars) identified a structure spatially and kinematically closer to Sagittarius, but chemically distinct. Comparing with literature, we identified this structure as the southern hemisphere section of the Cetus-Palca stream. 

\item \textbf{Orphan-Chenab:} An additional pass of \hdbscan{} seeking to recover structures with very few member stars identified one clear system among many likely spurious clusters. We identify this as part of the Orphan-Chenab stream.

We did not identify as \hdbscan{} clusters any of the retrograde halo substructures for which evidence has been found in other surveys, such as Arjuna, I'itoi \citep{Naidu2020} or Sequoia \citep{Myeong2019}. This may be the result of their weaker coherence in velocity space, compared to IoM space. We also did not identify with \hdbscan{} any of the known `diffuse overdensities' prominent in BHB and RR Lyrae samples, although they are apparent in our dateset as overdensities in configuration space. The Outer Virgo Overdensity is particularly clear at distances $\gtrsim 60$~kpc.
\end{itemize}

The large size and distance range of our K-giant sample allows us to explore the chemical and kinematic structure of the apparently `smooth' residual halo, far from the solar neighborhood (section~\ref{sec:bulk}). This is particularly interesting in the context of models of the GSE accretion event, which make predictions for its contribution to the outer halo \citep[e.g.][]{Naidu:2021aa}. We find:

\begin{itemize}
    \item The MDFs of prograde and retrograde outer halo stars in the residual halo are similar to each other and dissimilar to the MDFs of Sagittarius and GSE. These MDFs are bimodal and do not resemble the MDFs of halo stars with lower angular momentum.
    \item The MDFs of low angular momentum stars broadly resemble the MDF of GSE, even at lower binding energy than GSE itself.
\end{itemize}

In Sec.~\ref{sec:gse_and_outer_halo} we speculate on the association of the metal-poor component of the residual halo with the hypothetical Sequoia progenitor, and the metal-rich component with either the outer structure of the GSE progenitor or a separate progenitor. At face value, our findings disfavor a scenario in which GSE originates from a single progenitor, initially on a more circular orbit, that deposits significant material into the retrograde halo alone. However, the distribution of stars from a given accretion $(E,L_z)$ can be broad, spread across the prograde and retrograde halos, and multi-modal (e.g.\ \citealt{Jean-baptiste2017, Amarante2022,Khoperskov2023}; many examples in the specific context of MWS are shown in \citealt{Kizhuprakkat2024}). This suggest that further efforts to characterize the residual outer halo may reveal a complex picture, but one that can still be useful for constraining the origin of GSE and the contributions of as-yet unknown progenitors.

Our catalog provides a baseline for halo studies based on DESI K-giants. The K-giant selection and distance determination are likely to be sensitive to pipeline systematics that have not yet been explored in detail. We hope to improve on these aspects in advance of the final DESI data release. Our approach to structure-finding is by no means unique, and it would be worthwhile to explore alternative approaches with the same underlying dataset. For this purpose, among others, the K-giant catalog constructed in this work will be made publicly available alongside the second DESI data release (DESI DR2). 

While the analysis in this paper is based solely on observational data, the interpretation of clustering performance and the subsequent results would benefit from a similar analysis on a dataset where the true progenitor information is known. The AuriDESI mock catalogs \citep{Kizhuprakkat2024} provide an opportunity to combine cosmologically derived stellar halo assembly histories with realistic DESI selection functions. By applying the same structure-finding procedure to AuriDESI, we will be able to quantify the completeness, purity and sensitivity as a function of progenitor mass and orbital properties. This detailed analysis will be presented in a future work.

\section*{Acknowledgments}

 NK and APC acknowledge support from Taiwan Ministry of Education Yushan Fellowship awarded to APC MOE-113-YSFMS-0002-001-P2 and the Taiwan National Science and Technology Council grants 112-2112-M-007-017, 113-2112-M-007-009, 114-2112-M-007-005 and 114-2811-M-007-079. We also thank Sy-Yun Pu for useful discussions regarding the implications of our results for the interpretation of GSE. This work used high-performance computing facilities operated by the Center for Informatics and Computation in Astronomy (CICA) at National Tsing Hua University. This equipment was funded by the Taiwan Ministry of Education, the Taiwan National Science and Technology Council, and National Tsing Hua University. C.A.P. is thankful for funding from the Spanish government through grants AYA2014-56359-P, AYA2017-86389-P and PID2020-117493GB-100. R.Z. acknowledges the funding from the European Research Council (ERC) under the European Union’s Horizon Europe research and innovation program (grant agreement No. 101221278, project OUTLIERS).

This material is based upon work supported by the U.S. Department of Energy (DOE), Office of Science, Office of High-Energy Physics, under Contract No. DE–AC02–05CH11231, and by the National Energy Research Scientific Computing Center, a DOE Office of Science User Facility under the same contract. Additional support for DESI was provided by the U.S. National Science Foundation (NSF), Division of Astronomical Sciences under Contract No. AST-0950945 to the NSF’s National Optical-Infrared Astronomy Research Laboratory; the Science and Technology Facilities Council of the United Kingdom; the Gordon and Betty Moore Foundation; the Heising-Simons Foundation; the French Alternative Energies and Atomic Energy Commission (CEA); the National Council of Humanities, Science and Technology of Mexico (CONAHCYT); the Ministry of Science, Innovation and Universities of Spain (MICIU/AEI/10.13039/501100011033), and by the DESI Member Institutions: \url{https://www.desi.lbl.gov/collaborating-institutions}. Any opinions, findings, and conclusions or recommendations expressed in this material are those of the author(s) and do not necessarily reflect the views of the U. S. National Science Foundation, the U. S. Department of Energy, or any of the listed funding agencies.

The authors are honored to be permitted to conduct scientific research on I'oligam Du'ag (Kitt Peak), a mountain with particular significance to the Tohono O’odham Nation.

For the purpose of open access, the author has applied a Creative Commons Attribution (CC BY) licence to any Author Accepted Manuscript version arising from this submission.
\vspace{5mm}

\subsubsection*{Data Availability}
The K-giant catalog constructed in this work will be made publicly available along with the second data release (DR2) of DESI. Data for reproducing all figures is available in a machine readable form on the Zenodo repository \url{https://doi.org/10.5281/zenodo.18873882}, which will be made accessible upon publication of the paper.

\facilities{Mayall 4-m, DECam, Gaia}.
\software{NumPy \citep{numpy}, SciPy \citep{scipy}, Astropy \citep{astropy:2013,astropy:2018,Astropy2022}, Matplotlib \citep{matplotlib}, Healpy \citep{healpy,healpy1}, HDBSCAN \citep{hdbscan_2013,hdbscan_McInnes2017}, scikit-learn \citep{scikit_learn}, Gala \citep{gala}, \texttt{AGAMA} \citep{Agama2019}.}

\appendix

\section{Distance calibration} \label{distance calibration}
The \texttt{rvsdistnn} method of Koposov et al.\ (\textit{in prep}) uses a neural network to map a combination of RVSpecfit stellar parameters ($T_\mathrm{eff}$, $\log\,g$, [Fe/H], [$\alpha$/Fe]), 
and photometric colors, to an absolute magnitude for each star, and hence to a distance modulus. The training loss function is based on Gaia parallax constraints; it makes use of the full MWS sample but is weighted towards nearby stars with smaller parallax errors. The process is explained briefly in \citet{Aganze2025}.

\begin{figure}
    \centering
    \includegraphics[width=\linewidth]{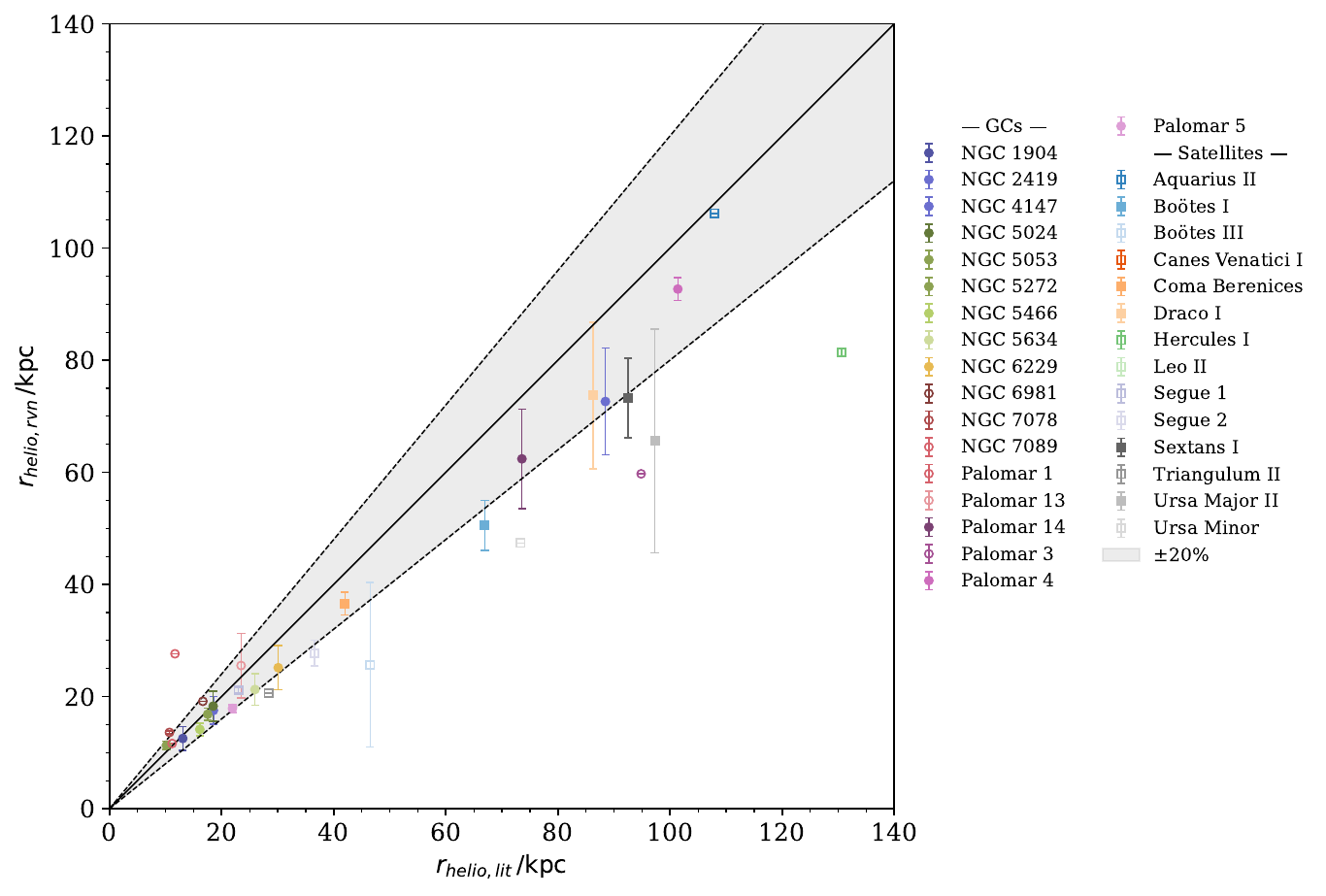}
    \caption{Median estimated distances for globular clusters and dwarf galaxies with likely members matched to our K-giant catalog, compared to their literature distances. The circle and square markers represent globular clusters and dwarf galaxies respectively. Hollow markers represent those structures contributing fewer than 3 stars to the K-giant catalog. The error bars represent the median absolute deviation of the estimated distances in a given structure. The solid black line shows the 1:1 relation, and the gray shaded region represents our fiducial uncertainty estimate of $\pm 20\%$. At distances $\gtrsim80~\mathrm{kpc}$, distances appear to be underestimated systematically by $\sim20\%$ (corresponding approximately to the lower boundary of the gray region).}
    \label{fig:distance_calib}
\end{figure}

Fig.~\ref{fig:distance_calib} compares distances estimated by the \texttt{rvsdistnn} method against literature values, for stars in our K-giant catalog with membership probability $> 0.9$ in known globular clusters \citep{Baumgardt2021} and dwarf galaxies \citep{Pace2024_lvd}. To $\sim 80$~kpc, the   estimates are within $\pm 20~\%$ of the literature values. A bias towards underestimated K-giant distances becomes more apparent at $\gtrsim 80$~kpc. This bias does not significantly affect the results in this paper, because almost all the structure we discuss is at (estimated) distances $< 80$~kpc. The detection of substructure at $\gtrsim 80$~kpc is fundamentally limited by the relatively small number of stars in our catalog at such distances, and their large distance uncertainties. We have verified that, in the case of the distant structure in the \citet{Vasiliev21_sag} model, the bias implied by Fig.~\ref{fig:distance_calib} has only a small effect on the inferred $(E,L_{z})$ distribution. The implied uncertainty (which has a large scatter among the different reference systems, but is consistent overall with our fiducial estimate of 20\%) has a much more significant effect.

\section{Sequential clustering} \label{sequential clustering}
This appendix provides further details of the successive passes of \hdbscan{} mentioned in section \ref{clustering_process}. 

\begin{figure*}
    \centering
    \includegraphics[width=\linewidth]{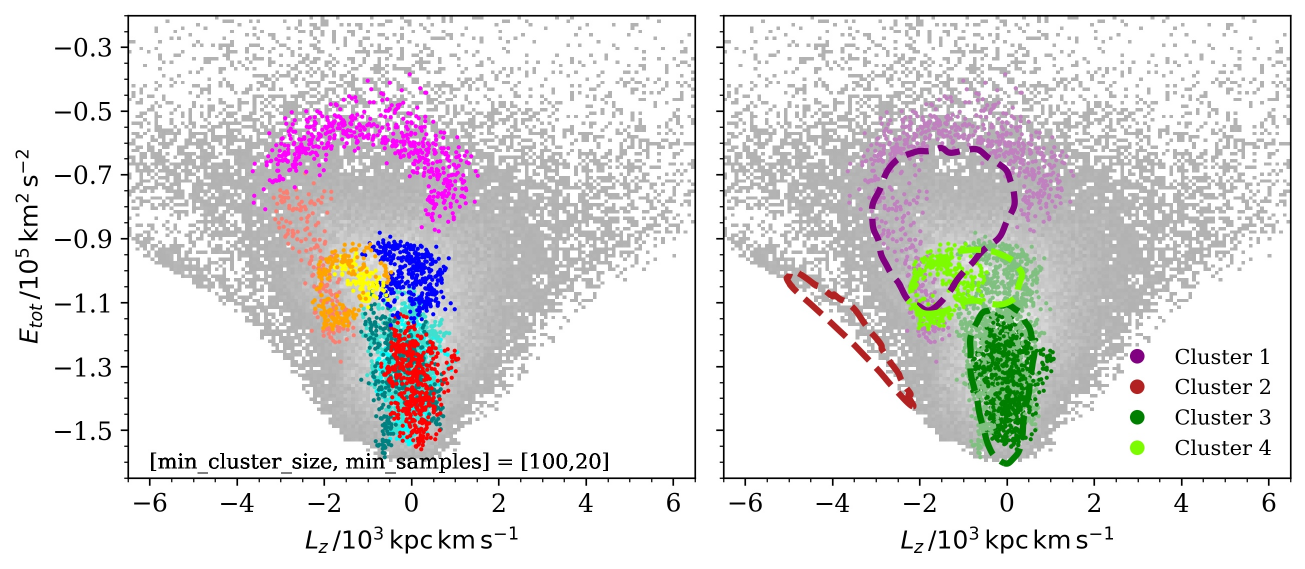}
    \caption{Results of the second pass of \hdbscan{} on the unclustered data from the first pass with \texttt{[min\_cluster\_size, min\_samples] = [100,20]}. The left panel shows the distribution of clusters identified in this pass on the $E_{tot} - L_z$ space. The right panel shows, with dashed contours, the regions corresponding to clusters from the first pass. Points, corresponding to second pass cluster members, are colored here according to their association with first pass clusters, based on the centroid method described in Section \ref{clustering_process}. A lighter hue indicates a larger distance (less likely association) between the second pass cluster centroid and the nearest first pass cluster centroid. The gray background on both panels shows the distribution of unclustered stars from the first pass.}
     \label{fig:fig20_hdbscan_run2}
\end{figure*}

The first pass of \hdbscan{} (see section \ref{hdbscan_intro}) identified four prominent clusters, all of which we could readily associate with well-known structures in the stellar halo. These clusters are relatively metal rich and, with the exception of cluster 3, mostly prograde. As described in the main text, we then applied \hdbscan{again} to the subset of stars not associated with any of these first-pass clusters, adjusting the parameters of \hdbscan{} to permit the recovery of smaller clusters.

Fig.~\ref{fig:fig20_hdbscan_run2} shows the results of this second \hdbscan{} pass in $E-L_z$ space. The gray background in both panels shows the distribution of stars that were not associated with clusters in the first pass. Colored points in the left panel correspond to clusters identified in the second pass. We do not label these individually because our next step is to group them into associations with each first pass cluster. We make these associations by hand guided by a simple estimate based on the distances between the cluster centroids, as described in section \ref{clustering_process}. The resulting associations are shown in the right panel. In this panel, different colors indicate the first pass cluster we associate with each star assigned to a cluster in the second pass. The contours show the regions occupied by the first pass clusters themselves. The dark (light) colored markers represent shorter (longer) distances between the centroid of the second pass cluster and its first-pass parent, measured in the scaled 6-dimensional feature space. 

\begin{figure*}
    \centering
    \includegraphics[width=\linewidth]{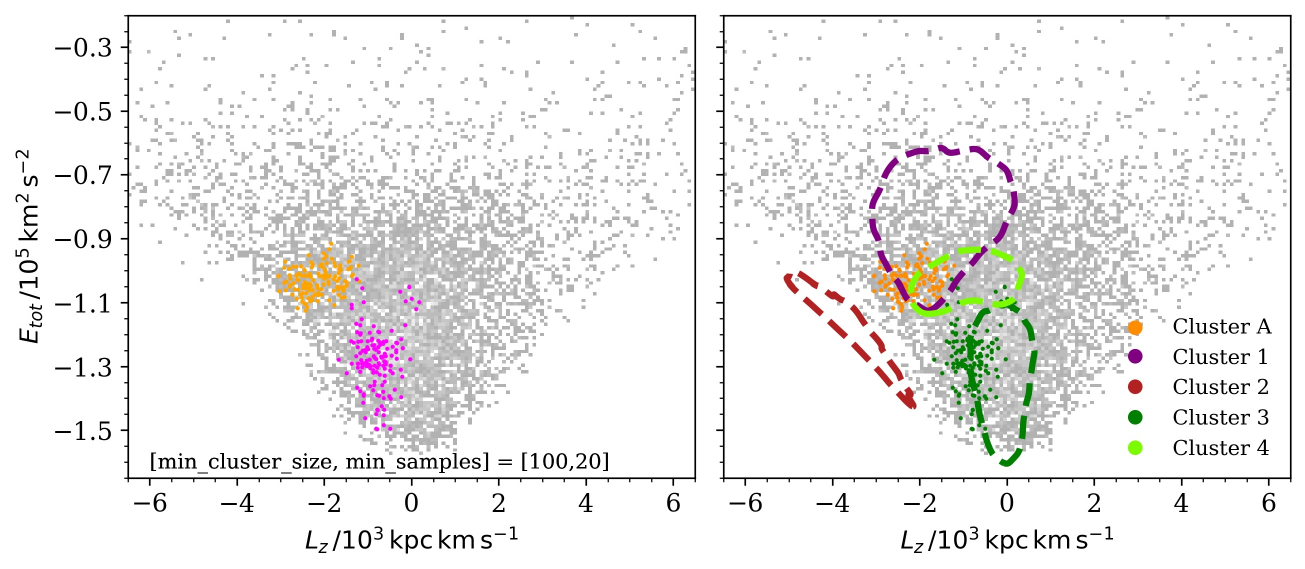}
    \caption{Results of an alternative second pass of \hdbscan{} on the unclustered data from the first pass, again with \texttt{[min\_cluster\_size, min\_samples] = [100,20]} as in Fig.~\ref{fig:fig20_hdbscan_run2}. In this alternative case, we use only the subset of the unclustered sample with $\mathrm{[Fe/H] < -2}$. The left panel shows the new clusters identified in this pass. The right panel shows the result of centroid-based matching (see sections~\ref{sequential clustering} and \ref{clustering_process}) to the first pass clusters (as in Fig.~\ref{fig:fig20_hdbscan_run2}). The dashed contours of different colors represent the first pass clusters. In this case, one new cluster is associated with first pass cluster 3 (dark green) while the other has no close match to a first pass cluster (orange). The grayscale background in both panels shows the distribution of metal-poor subset of the unclustered stars from the first pass.}
    \label{fig:fig21_hdbscan_run3}
\end{figure*}

The second pass did not identify any obviously new clusters, nor did it find any clusters in the retrograde halo or in the low-metallicity regime $(\mathrm{[Fe/H] \lesssim -2})$. We therefore ran an alternative metal-poor second pass, again based on the unclustered data from the first pass and using the same parameters as our fiducial second pass, but this time limited to K-giants with $\mathrm{[Fe/H] < -2}$.

Fig.~\ref{fig:fig21_hdbscan_run3} shows the two clusters found by \hdbscan{} in this metal-poor second pass. We associate the first of these with Cluster 3 from the first pass. The $(E,L_z)$ location and low metallicity are reminiscent of the \textit{LMS-1/Wukong} feature of \citet{Yuan:2020aa} and \citet{Naidu2020}. The second new cluster, shown with orange points, was associated to Cluster 4 by our centroid distance estimate. However, we believe this new cluster closely resembles a previously reported feature, Cetus-Palca \citep[][further details are given in Section \ref{hdbscan results}]{GFT22_Cetus_Palca,Zhen2022_cetus}. We therefore consider this a new cluster detected by the metal-poor second pass.

\section{Comparison to the SEGUE K-giant sample}
\label{appendix:segue}

We have compared our DESI K-giant sample with the \citet{Xue2014} sample of K-giants from the Sloan Extension for Galactic Understanding and Exploration (SEGUE) DR9 value added catalog. Based on measurements obtained by the SEGUE Stellar Parameter Pipeline (SSPP), \cite{Xue2014} used a probabilistic approach to calculate distances to 6036 K-giants, including 283 stars at distances greater than 50 kpc.

\begin{figure*}
    \centering
    \includegraphics[width=0.8\textwidth]{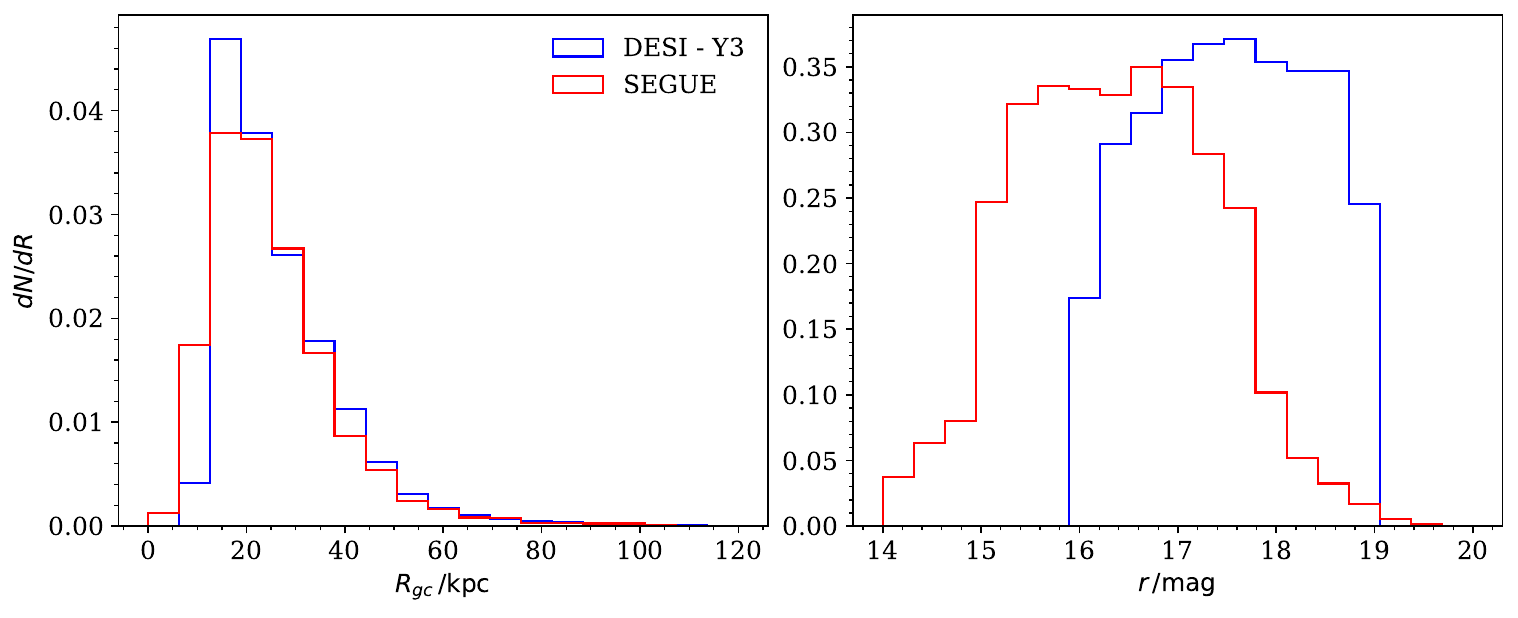}
    \caption{Comparison of Galactocentric distance (left) and apparent magnitude (right) distributions between the full \citet{Xue2014} SEGUE and DESI Y3 K-giant catalogs. The histograms are normalized to unit area.}
    \label{fig:fig22_segue_comparison_1}
\end{figure*}
\begin{figure*}
    \centering
    \includegraphics[width=\textwidth]{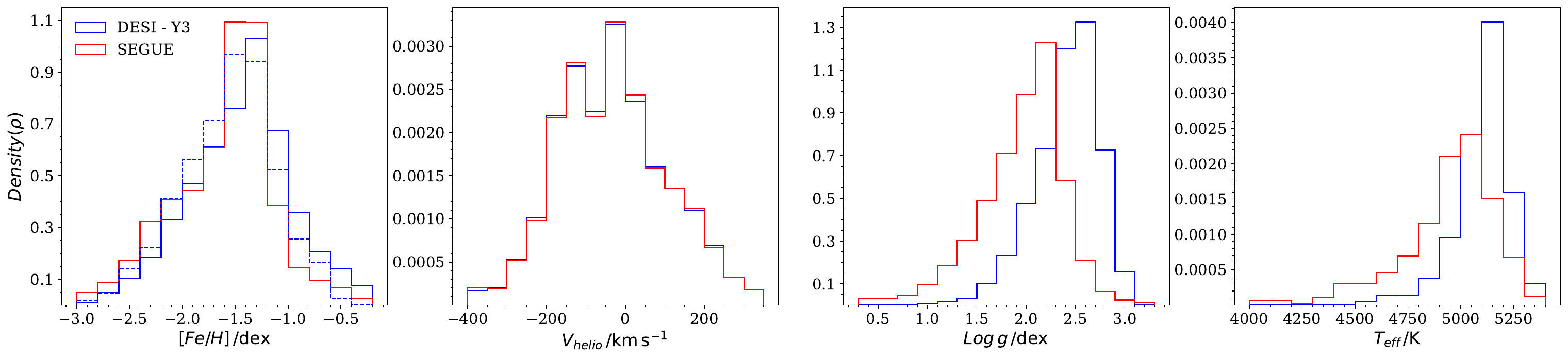}
    \caption{Comparison of atmospheric parameters measured by SEGUE (SSPP; blue) and MWS (RVS; red) for the subset of stars in both catalogs (matched on the sky to 0.5"). From left to right panels show [Fe/H], heliocentric velocity, $\log g$ and $T_\mathrm{eff}$. The dashed blue line in the first panel corresponds to the \protect\citet{MWS_dr1} recalibration of [Fe/H] from the RVS pipeline for DESI K-giants, to account for a systematic metallicity offset relative to other surveys.}
    \label{fig:fig23_segue_comparison_2}
\end{figure*}

Fig.~\ref{fig:fig22_segue_comparison_1} compares histograms of Galactocentric distance and $r$-band magnitude for the two K-giant samples in their entirety (i.e.\ all the stars in each sample, not only those stars common to both samples). Both datasets have a similar distance distribution, although SEGUE data are slightly more concentrated within inner 30 kpc. The right panel shows the distribution of $r$-band magnitude. SEGUE observes relatively brighter stars, up to $r\sim14$~mag, whereas DESI MWS has better sampling down to $r\simeq19$.

Fig.~\ref{fig:fig23_segue_comparison_2} compares the atmospheric parameters from MWS RVS pipeline for K-giants matched between SEGUE and MWS to within 0.5''. This yields 1,874 stars common to both catalogs. From left to right, we show distributions of [Fe/H], heliocentric radial velocity, $\log g$ and $T_\mathrm{eff}$. The radial velocity distributions are almost identical, which is unsurprising given the expected radial velocity precision of both surveys. A small shift ($\sim+0.2$ dex) of the DESI [Fe/H] distribution relative to SEGUE is apparent. \citet{MWS_dr1} have examined the metallicity offsets of the MWS RVS pipeline relative to other surveys. They find MWS metallicities for giants to be systematically high, relative to higher resolution spectroscopic surveys such as APOGEE, due to surface gravity and effective temperature dependent effects in the spectral modeling.The blue dashed line in the [Fe/H] distribution shows our K-giant [Fe/H] distribution calibrated using the quadratic function given by \citet{MWS_dr1}, which is in better agreement with the SEGUE distribution. The apparently systematic shifts of the $\log g$ and $T_\mathrm{eff}$ distributions likely reflect the different approaches and calibrations of the underlying spectroscopic pipelines. We conclude that these shifts do not lead to any significant systematic differences in distance or metallicity estimates.


\bibliography{sample631}{}
\bibliographystyle{aasjournal}



\end{document}